\documentclass[numbers=noenddot]{jfm}
\usepackage[latin9]{inputenc}

\usepackage{array, ragged2e}
\usepackage{float}
\usepackage{mathtools}
\usepackage{graphicx}
\usepackage{esint}
\usepackage{verbatim}
\usepackage{caption}
\usepackage{subcaption}
\renewcommand\[{\begin{equation}}
\renewcommand\]{\end{equation}}
\newcommand{\un}[1]{\,\mathrm{#1}}
\numberwithin{equation}{section}
\usepackage{amsmath} 
\newcommand{\beq}{\begin{equation}}
\newcommand{\eeq}{\end{equation}}
\newcommand{\lb}{\left(}
\newcommand{\rb}{\right)}
\usepackage{tikz}
\usepackage{pgfplots}
\usepackage{overpic}
\usepackage{todonotes}
\author{Graham P. Benham \aff{1,2}\corresp{\email{gpb35@cam.ac.uk}}, 
Mike J. Bickle \aff{1},  
Jerome A. Neufeld \aff{1,2,3}}
\affiliation{\aff{1} Department of Earth Sciences, University of Cambridge, Bullard Laboratories, Madingley Road, Cambridge CB3 0EZ, UK
\aff{2} BP Institute, University of Cambridge, Bullard Laboratories, Madingley Road, Cambridge CB3 0EZ, UK
\aff{3} Department of Applied Mathematics and Theoretical Physics, University of Cambridge, Wilberforce Road, Cambridge CB3 0WA, UK
}

\begin{document}

\title{Upscaling multiphase flow through heterogeneous porous media}

\maketitle

\abstract{
Upscaling the effect of heterogeneities in porous media is crucial for macroscopic flow predictions, with widespread applications in energy and environmental settings.
In this study, we derive expressions for the upscaled flow properties of a porous medium with a vertical heterogeneity, using a combination of asymptotic analysis and numerical simulations. Then, we use these upscaled expressions to describe the dynamic flooding of an aquifer, where the classic Buckley-Leverett formulation is modified to account for heterogeneities. 
In particular, we show that heterogeneities can modify flooding speeds significantly, and we discuss the implications of these results in the case of carbon dioxide sequestration. 
}

\section{Introduction}

The flow of immiscible fluids in heterogeneous porous media has widespread applications in energy and the environment. 
Nearly all subsurface rocks have a significant heterogeneous structure, often in the form of sedimentary layers, and it is well known that such heterogeneities play an important role in the resultant flow properties \citep{reynolds2015characterizing,jackson2018characterizing,nijjer2019stable}. For example, depending on the alignment of the sedimentary strata, flow of different fluid phases can be preferentially and significantly exacerbated or diminished, compared to the homogeneous case \citep{reynolds2015characterizing,krause2015accurate,rabinovich2016analytical}. 

One very topical application is the geological storage, or \textit{sequestration}, of carbon dioxide \citep{bickle2009geological,huppert2014fluid}. 
Currently one of the few proposed technological solutions to the global warming problem, this process involves trapping CO$_2$ emissions, either at power plants or industries, and pumping them several kilometres beneath the earth to be stored safely and securely \citep{szulczewski2012lifetime}. Possible sites for CO$_2$ storage include saline aquifers, depleted oil reservoirs and unprofitable coal seams. 
The CO$_2$, which is less dense than the ambient brine, rises gradually through the porous rock,  and is trapped as it migrates by a combination of impermeable cap rocks, by dissolution in the brine, or by residual trapping in the surrounding rock pores \citep{golding2011two,macminn2010co,macminn2011co,krevor2015capillary}. Due to the important relationship between flow speeds and residual trapping rates \citep{hesse2006scaling}, it is imperative to understand how the heterogeneities of the rock affect the large scale flow. Hence, we are motivated to develop a macroscopic model for the migration of immiscible fluids in an aquifer with an underlying heterogeneity.

Flow in porous rocks is generally a multi-scale phenomenon, with relevant length scales varying from the pore size ($\sim\mathcal{O}(1\mathrm{mm})$) up to the aquifer size ($\sim\mathcal{O}(10\mathrm{km})$). 
Due to the large computational cost involved in simulating flow in heterogeneous resevoirs, it is largely desirable to avoid modelling all of these scales. 
In porous media flow, it is common to neglect much of the small scale details, and instead attempt to describe their bulk effect on the macroscopic scale, which is often referred to as \textit{upscaling}.
Whilst there are many studies which focus on upscaling from the pore scale \citep{krevor2015capillary}, here we focus on length scales between the size of the rock heterogeneities (layers) and the size of the aquifer. 

Heterogeneities, of which there are many varieties, refer to spatial variations in rock features such as pore size, pore geometry, faults and fractures, as well as variations in rock type itself (e.g. sandstone, clay, \ldots). These heterogeneities often play a strong role on multiphase fluid flow by means of small scale capillary forces acting on the phases. For example, in two-phase flow, the non-wetting phase tends to be preferentially drawn to regions of larger pore space by capillary forces, resulting in a pronounced non-uniform flow.  The effect of the heterogeneities also depends on how they are distributed. Perhaps the most common type of heterogeneity is sedimentary layering in a particular orientation (e.g. parallel or perpendicular to the flow), though elsewhere distributions may be arranged over some correlation length scale (e.g. in the horizontal), or they may be largely randomly distributed. In pressure driven flows, heterogeneities frequently result in unstable displacement of phases (so long as capillary forces are large enough to overcome the driving pressure), and fingering \citep{dawe1992experimental,dawe2011immiscible}. Hence, an analogy can be drawn between the capillary-driven mixing of immiscible fluids, and the classic diffusion/dispersion-driven mixing of miscible fluids \citep{tchelepi1993dispersion,nijjer2019stable}. 
However, for this study we focus on the case of immiscible fluid flow in a layered porous medium.


The role of heterogeneities is often characterised by the non-dimensional capillary number, which is given as the ratio between typical horizontal pressure gradients $\Delta p/L$ (over length scale $L$), and typical vertical gradients in the pore entry pressure $\Delta p_e/H$ (over length scale $H$), giving
\beq
\mathrm{N}_{c}=\frac{\Delta p}{\Delta p_e}\frac{H}{L}.\label{capdef}
\eeq
At small N$_c$, the background flow is sufficiently weak that the flow of fluid phases is largely dominated by the heterogeneity-driven capillary forces, whereas at large N$_c$, the background flow dominates, such that heterogeneities can be largely ignored. Hence, the limit N$_c\rightarrow 0$ is known as the \textit{capillary limit} and N$_c\rightarrow \infty$ is known as the \textit{viscous limit}. To model the flow in any case which is far away from the viscous limit, one needs detailed knowledge of the structure of the heterogeneities to describe the flow, which presents a significant challenge.

Recently, there has been strong emphasis on attempting to upscale the effect of heterogeneities in porous media \citep{reynolds2015characterizing,boon2017observations,jackson2018characterizing}.  
One of the key difficulties lies in the sheer number of measurements, either experimental or numerical, needed to characterise the effect of rock layers across a broad range of flow conditions. For example, in the case of immiscible flow of wetting and non-wetting phases, the effect of the heterogeneities not only depends on the capillary number, as described above, but also on the fractional flow of either phase \citep{woods2015flow}. Furthermore, since each type of rock heterogeneity is different, it is difficult to transpose results without performing experiments and simulations for each specific case.

One successful approach involves using X-ray CT scans of flow in layered rocks, in conjunction with detailed numerical simulations.
The recent study by \citet{jackson2018characterizing} presents a systematic approach to estimate the global effect of rock layers on the flow. A set of CT scan experiments is first performed at high capillary number to determine the \textit{intrinsic} properties of the flow, such as the relative permeabilities and capillary pressure (which are both typically functions of the saturation). Then a similar set of experiments is performed at low capillary number to characterise the heterogeneity of the rock by means of fitting a set of capillary pressure scaling factors (one for every scanned voxel) to match numerical simulations to the CT scans. 
Having performed this two-stage analysis, Jackson \textit{et al.} then use the fitted numerical model to describe the flow at intermediate capillary numbers, thereby enabling a systematic upscaling of the heterogeneities.
In this way, relationships for the \textit{equivalent} properties of the flow are derived, such as equivalent relative permeability, which are particularly useful when employed in conjunction with flow simulators to make predictions in the field.
However, without being able to perform CT scans of flow in the rock samples, such analysis is impossible. 
Furthermore, there exists no general upscaled theory for the flow regime in between the viscous and capillary limits.

The objectives of the current study are to develop a simple theoretical tool that can be used to upscale the effect of heterogeneities in arbitrary flow conditions, where the heterogeneity can be given as a model input. 
The ultimate goal is to be able to study a vast range of scenarios, to provide ensemble forecasts for the migration of immiscible fluids in porous media.
Such a tool can be used not only to pinpoint optimal sites and predict trapping efficiencies for CO$_2$ sequestration, for example, but also for inverse modelling of rock heterogeneities given field measurements.

In the present study, we restrict our attention to a layered porous medium, with heterogeneity varying in the vertical direction and flow driven in the horizontal direction only. Furthermore, we focus on drainage flows, where a non-wetting phase drives out a wetting phase, though the analysis can easily be extended to imbibition flows.
Using a combination of asymptotic analysis and numerical simulations of steady-state flow conditions, similar to \citet{ekrann2000steady}, we derive relationships for the equivalent relative permeabilities and capillary pressure relationships that are valid across all capillary numbers and saturations. 
We then use the upscaled properties to describe the dynamic flooding of an aquifer with small scale heterogeneities. The latter is an extension of the classic model of \citet{buckley1942mechanism}, where a one-dimensional system is used to model the displacement of immiscible fluids in a long thin porous medium. 
In the original study, fluid displacement is characterised by an advective velocity that depends on the relative permeabilities of the two phases, and often results in shock behaviour. Here, we calculate advection speeds using the upscaled equivalent relative permeabilities, which derive from the underlying rock heterogeneity. In particular, we demonstrate that the capillary number may vary significantly along the aquifer, such that different regions may lie within the capillary, viscous, or intermediate regimes simultaneously. Using our extension to the Buckley-Leverett problem, we illustrate the effects heterogeneities have on flows in aquifers, and we discuss the implications in the context of CO$_2$ sequestration.

Section 2 describes the heterogeneous system we consider, and derives relationships for the upscaled flow properties in the viscous and capillary limits. In the case of intermediate capillary numbers, numerical simulations are used to characterise the viscous-capillary transition. Then Section 3 uses the upscaled flow properties to study flooding dynamics via the Buckley-Leverett problem, extended to heterogeneous media. In Section 4 we compare our upscaling predictions with the experimental measurements of other authors, and finally we close by summarising the results.

\section{Upscaling heterogeneities}
\label{upscale}

The general approach taken here is as follows: We start by summarising the governing equations and boundary conditions for two-phase flow in a layered porous medium; we define upscaled quantities, such as the equivalent relative permeabilities; we then derive expressions for these upscaled quantities in each of the two limiting \textit{viscous} and \textit{capillary} cases, using some simple examples for illustration; finally, we use numerical simulations to calculate the upscaled quantities for intermediate capillary numbers, showing how to incorporate all regimes using some simple parameterisations.

\subsection{Immiscible two-phase flow in porous media}

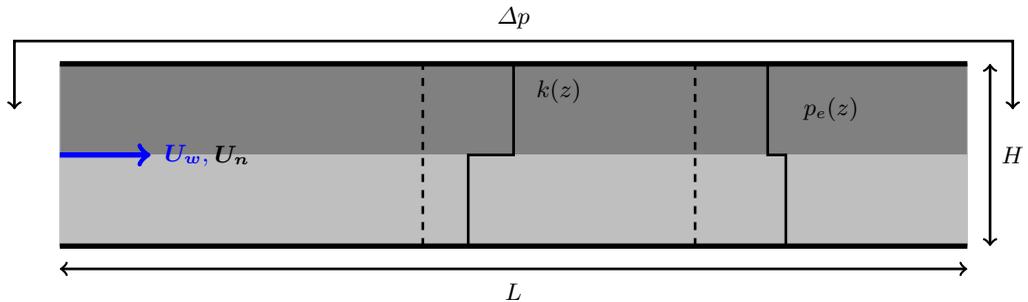
\begin{figure}
\centering
\begin{tikzpicture}[scale=1.2]
\draw[line width=0,lightgray,fill=lightgray]  (0,0) rectangle  (10,1);
\draw[line width=0,lightgray,fill=gray]  (0,1) rectangle  (10,2);
\draw[line width=2] (0,0) -- (10,0);
\draw[line width=2] (0,2) -- (10,2);
\draw[line width=2,blue,->] (0,1) -- (1,1);
\node[blue] at (1.4,1) {$\boldsymbol{U_w},$};
\node[black] at (1.9,1) {$\boldsymbol{U_n}$};
\draw[line width=1,<->] (0,-0.25) -- (10,-0.25);
\draw[line width=1,<->] (10.25,0) -- (10.25,2);
\node at (5,-0.5) {$L$};
\node at (10.5,1) {$H$};
\draw[line width=1,dashed] (4,0) -- (4,2);
\draw[line width=1] (5,2) -- (5,1) -- (4.5,1) -- (4.5,0)  ;
\draw[line width=1,dashed] (7,0) -- (7,2);
\draw[line width=1] (7.8,2) -- (7.8,1) -- (8,1) -- (8,0) ;
\node at (5.5,1.7) {$k(z)$};
\node at (8.5,1.5) {$p_e(z)$};
\draw[line width=1,<->] (-0.5,1.5) -- (-0.5,2.25) -- (10.5,2.25) -- (10.5,1.5);
\node at (5,2.5) {$\Delta p$};
\end{tikzpicture}
\caption{Schematic diagram of a long, thin two-dimensional aquifer with steady, pressure-driven flow of wetting and non-wetting phases. Vertical heterogeneity is given by variation in the pore entry pressure $p_e(z)$ and permeability $k(z)$, which here is illustrated in the case of a two-layered system. \label{fig1}}
\end{figure}

We consider the flow of a non-wetting phase driving out a wetting phase (e.g. carbon dioxide driving out water) in a two-dimensional aquifer of length $L$, height $H$, and whose intrinsic properties (e.g. porosity $\phi$, permeability $k$, pore entry pressure $p_e$) vary in the vertical direction $z$ (see figure \ref{fig1}). 
We model the flow behaviour at the continuum scale (but below the scale of the heterogeneities) using conservation of mass and the multiphase extension to Darcy's law under gravity \citep{bear2013dynamics}.
Hence, the governing equations for the flow are
\begin{align}
\phi(z) \frac{\partial S_i}{\partial t}+{\nabla}\cdot\boldsymbol{u}_i&=0,\quad &i=n,w\label{ge1},\\
\boldsymbol{u}_i&=-\frac{k(z)k_{ri}(S_i)}{\mu_i}{\nabla}\, \lb p_i -\rho_i g \boldsymbol{z} \rb,\quad &i=n,w\label{ge2},
\end{align}
where subscripts $n$ and $w$ indicate non-wetting and wetting phases, and we require the fluids to fill the pore spaces $S_n+S_w=1$. The parameters $\mu_i$ and $\rho_i$ are the viscosities and densities of either phase, $k_{ri}(S_i)$ are the relative permeabilities, and $p_i$ are the pressures of each phase, which differ by an amount
\beq
p_n-p_w=p_c(S_i)\label{cappres},
\eeq
where $p_c$ is known as the capillary pressure since it is associated with the micro-scale capillary forces between phases. 
Although $k_{ri}$ and $p_c$ depend on many factors in general, they are often assumed to be functions of the saturation alone \citep{golding2011two}. 
A simple, commonly used empirical relationship for the capillary pressure is that proposed by \citet{brooks1964hydrau}, 
\beq
p_c=p_e(z)(1-{s})^{-1/\lambda},
\eeq
where $p_e(z)$ is the pore entry pressure, $\lambda\geq1$ represents the pore size distribution, and 
\beq
s=\frac{S_n}{1-S_{wi}}\label{recaleds}
\eeq
is the rescaled saturation.
The irreducible wetting phase saturation $S_{wi}$ represents the amount of wetting phase that cannot be removed, and is therefore always trapped in the pores by capillary forces. Using this notation, the rescaled saturation $s$ varies between 0 and 1. 
The pore entry pressure $p_e$ describes the minimum pressure required to allow any non-wetting phase into the pore spaces. For $p_n-p_w=p_e$, only the largest pore spaces are filled with non-wetting phase, and for $p_n-p_w>p_e$, smaller and smaller pore sizes are invaded. Clearly, the pore entry pressure depends on the porosity and geometry of the pores, as does the permeability, and we assume these vary in the vertical direction. Therefore, in this study, heterogeneities are defined solely by $\phi(z)$, $p_e(z)$ and $k(z)$. It is often assumed that $p_e(z)$ and $k(z)$ depend on the porosity under some power law that reflects the geometry of the pore spaces \citep{leverett1941capillary}. Hence, we have $p_e\propto \phi^{-a},\,k\propto \phi^b$, for parameters $a,b$. Therefore, the pore entry pressure and permeability are related according to 
\beq
p_e=p_{e_0}\lb \frac{k}{k_0} \rb ^{-B},\label{powerlaw}
\eeq
where $p_{e_0}$ and $k_0$ are typical dimensional scalings, and $B=a/b>0$ is a positive constant, since larger pore spaces should correspond to lower pore entry pressure. 
It has long been argued that such power law relationships do not apply generally \citep{cloud1941effects}, but specific power laws are often used for particular rock types (e.g. see \citet{nelson1994permeability}). For example, using $b=2$ and the scaling proposed by \citet{leverett1941capillary}, where $p_e\sim (\phi/k)^{1/2}$, gives a value of $B=1/4$.

There are a vast number of different empirical relationships for the relative permeabilities $k_{ri}$ which have been proposed by various authors \citep{krevor2012relative}, and the appropriate choice depends on the specific rock type and fluid phases. 
The relative permeabilities are monotonic functions of their respective phase saturations, and lie between 0 and 1. In the limiting case where the flow becomes single phase, the relative permeability of that phase should be 1 (and 0 for the other phase). But as we have already discussed, there may be an irreducible  wetting phase saturation, and hence we have $k_{rn}(s=1)=k_{rn0}$, for some $0\leq k_{rn0} \leq1$.
In this paper, we propose a general framework which is not limited by a specific choice of empirical relationship. However, we make comparisons with several commonly used laws, including those proposed by \citet{corey1954interrelation} and \citet{chierici1984novel}, which we give explicitly in Appendix \ref{appA}.

Finally, to complete the model, we require a set of boundary conditions. There are many possible choices of boundary conditions for such flow, as discussed by \citet{krause2012modeling}. We note that after some simple rearranging, it is possible to convert \eqref{ge1}-\eqref{cappres} to equations for the pressure and saturation of one of the phases only. Therefore, without loss of generality, we formulate our model focussing on the non-wetting phase, and we consider a pressure driven flow, resulting in the boundary condition
\beq
{p_n}|_{x=0}-{p_n}|_{x=L}=\Delta p\label{bc1}
\eeq
for some overarching pressure drop $\Delta p\geq 0$. We assume the flow at the inlet is well-mixed, and hence we fix the saturation to a constant value
\beq
s|_{x=0}=s_i.\label{bc2}
\eeq
In addition, we assume that the aquifer is sufficiently long that saturation gradients are negligible at the outlet, 
\beq
\left.\frac{\partial s}{\partial x}\right|_{x=L}=0.\label{bc3}
\eeq
Finally, we impose impermeability conditions at the top and bottom boundaries, such that
\begin{align}
\left.\frac{\partial p_n}{\partial z}\right|_{z=0,H}=0,\\
\left.\lb\frac{\mathrm{d} p_e}{\mathrm{d} z}+\frac{p_e}{\lambda(1-s)}\frac{\partial s}{\partial z}\rb\right|_{z=0,H}=0.\label{imperm}
\end{align}
Note that \eqref{bc1} determines the flow rate of non-wetting phase at the inlet. Similarly, \eqref{bc2}-\eqref{bc3} determine the flow rate of the wetting phase (or equivalently the pressure drop of the wetting phase). Hence, it is often useful to replace \eqref{bc1}-\eqref{bc3} by flow conditions
\beq
{u}_i|_{x=0}=U_i,\quad i=n,w,\label{bcu}
\eeq
where the inflow parameters $U_n, U_w$ are related to $s_i$ and $\Delta p$ by the multiphase flow model. 
To summarise, the model consists of the governing equations \eqref{ge1}-\eqref{cappres}, as well as boundary conditions \eqref{bc1}-\eqref{imperm}, and some initial conditions for $p_n$ and $s$. The heterogeneity is characterised by $\phi(z)$, $k(z)$, and $p_e(z)$, which are related by \eqref{powerlaw}.

\subsection{Upscaling}

As discussed by numerous authors \citep{reynolds2015characterizing,krause2015accurate,rabinovich2016analytical}, heterogeneities have the capability of changing the overall flow properties of porous media. 
In particular, in the presence of heterogeneities the empirical relative permeability relationships discussed earlier tend to become wholly inaccurate as we deviate away from the classic homogeneous or \textit{viscous} limiting case. 
Typically, parallel layering (as studied here) tends to segregate phases in such a way as to increase the overall flow of non-wetting phase, and decrease the flow of wetting phase \citep{krause2015accurate}.
For this reason, and as a method of reducing the requirement to resolve individual heterogeneities, it is useful to define so-called \textit{equivalent} properties instead which give a description of the flow that upscales the effects of these heterogeneities.

For the purpose of upscaling, we restrict our attention to the steady-state case. 
Therefore, similarly to \citet{jackson2018characterizing}, we define the equivalent relative permeabilities as
\beq
k_{ri_\mathrm{eq}}=\frac{\left<u_i\right>\mu_i L}{k_0\left<\Delta p_i\right>},\quad i=n,w,\label{eqrelperm}
\eeq
where the pressure changes $\Delta p_i$ refer to the difference between inlet and outlet for each respective phase, and the operator $\left<\cdot\right>$ refers to a type of spatial averaging, which we leave in general terms for now but discuss later in Sections \ref{seccap}, \ref{secvisc} and \ref{secint}.
Similarly, we define the equivalent capillary pressure as
\beq
p_{c_\mathrm{eq}}=\left<\frac{p_c}{p_e}\right>,\label{eqpc}
\eeq
which is a dimensionless quantity. As discussed earlier, the effect of heterogeneities is often characterised by the so-called \textit{capillary number} N$_c$ \eqref{capdef}, which is given as the ratio between typical horizontal pressure gradients, and typical vertical gradients in the pore entry pressure.
For the horizontal pressure change in \eqref{capdef}, we choose the constant non-wetting pressure difference \eqref{bc1}, though we could equally choose the wetting pressure, or some kind of combination. As we will discuss later, this choice is satisfactory for our purposes. For the characteristic vertical pore entry pressure change $\Delta p_e$, we choose the maximum difference 
\beq
\Delta p_e=\max_{z\in[0,H]} p_e(z)- \min_{z\in[0,H]} p_e(z).
\eeq
The equivalent properties \eqref{eqrelperm}-\eqref{eqpc}, which are the main focus of this paper, depend on the following different quantities:
\begin{enumerate}
\item The underlying heterogeneity of the rock, characterised by $p_e(z)$ and $k(z)$ via \eqref{powerlaw}.
\item The flow-driving pressure drop across the aquifer $\Delta p$.
\item The aspect ratio of the domain $\delta$.
\item The inlet conditions of the saturation $s_i$.
\end{enumerate}
Note, the capillary number N$_c$ contains all of (i)-(iii), but has no notion of (iv). Furthermore, it doesn't describe the spatial variation of the heterogeneity, only the typical variation scale $\Delta p_e$. In addition, the definition of N$_c$ depends on the choice of length scales $H$ and $L$, which are not necessarily well-defined in real applications. Therefore, even though N$_c$ is not sufficient on its own to characterise the complete flow picture, we use it primarily as a metric for describing the type of flow regime (horizontal pressure-driven flow versus vertical capillary-driven flow), to which it lends itself well.

\subsection{Non-dimensionalisation and asymptotic analysis}

Before we address each of the viscous and capillary limits it is useful to convert to dimensionless variables. Let us attribute the following scalings to each variable
\beq
\begin{split}
x=L\hat{x},\quad & z=H\hat{z},\quad \lb u_i,w_i\rb=\frac{k_0\Delta p}{\mu_n L}\lb\hat{u}_i,\delta\hat{w}_i\rb,\\
 & p_e=p_{e_0}+\Delta p_e \hat{p}_e,\quad p_i=\Delta p \,\hat{p}_i,
\end{split}
\eeq
where $\delta=H/L$ is the aspect ratio, which we assume to be small, and $w_i$ is the vertical velocity component of each phase. Written in terms of these new non-dimensional variables, the governing equations \eqref{ge1}-\eqref{cappres} (in the steady state) become
\begin{align}
{\hat{\nabla}}\cdot{\hat{\boldsymbol{u}}}_n&=0,\label{nondim1}\\
{\hat{\nabla}}\cdot{\hat{\boldsymbol{u}}}_w&=0,\label{nondim11}\\
\hat{u}_n&=-\hat{k}(\hat{z})  k_{rn}(s)\frac{\partial \hat{p}_n}{\partial \hat{x}},\label{nondim2}\\
\delta^2\hat{w}_n&=-\hat{k}(\hat{z})  k_{rn}(s)\lb \frac{\partial \hat{p}_n}{\partial \hat{z}}-\psi_n \rb,\label{nondim3}\\
 M  \hat{u}_w&=-\hat{k}(\hat{z})  k_{rw}(s)\frac{\partial \hat{p}_w}{\partial \hat{x}},\label{nondim4}\\
 M  \delta^2\hat{w}_w&=-\hat{k}(\hat{z})  k_{rw}(s)\lb \frac{\partial \hat{p}_w}{\partial \hat{z}}-\psi_w\rb,\label{nondim5}\\
\hat{p}_n-\hat{p}_w&=\frac{1}{\sigma_P\tilde{\mathrm{N}}_c}\lb 1+\sigma_P\hat{p}_e(\hat{z})\rb(1-s)^{-1/\lambda},\label{capnondim}
\end{align}
where we have introduced the non-dimensional variables $M=\mu_w/\mu_n$ (mobility ratio), $\sigma_P=\Delta p_{e}/p_{e_0}$, $\psi_i=\rho_i g H/\Delta p$, and $\tilde{\mathrm{N}}_c=\Delta p/\Delta p_e=\mathrm{N}_c/\delta$ is the reduced capillary number. For this study, we restrict our attention to thin aquifers $\psi_i\ll1$, in which gravity can be neglected, similarly to the core flooding experiments of \citet{jackson2018characterizing}. The boundary conditions \eqref{bc1}-\eqref{imperm} become
\begin{align}
\hat{p}_n|_{\hat{x}=0}-\hat{p}_n|_{\hat{x}=1}&=1,\label{nondimbcp}\\
s|_{\hat{x}=0}&=s_i,\label{nondimbcinlet}\\
\left.\frac{\partial s}{\partial \hat{x}}\right|_{\hat{x}=1}&=0,\label{nondimbc3}\\
\left.\frac{\partial \hat{p}_n}{\partial \hat{z}}\right|_{\hat{z}=0,1}&=0,\\
\left.\lb\sigma_P \frac{\mathrm{d} \hat{p}_e}{\mathrm{d} \hat{z}}+\frac{\lb 1+\sigma_P\hat{p}_e\rb}{\lambda(1-s)}\frac{\partial s}{\partial \hat{z}}\rb\right|_{\hat{z}=0,1}&=0.\label{nondimimperm}
\end{align}
Likewise, the inflow of each phase is given by
\begin{align}
\hat{u}_n|_{\hat{x}=0}&=U,\label{nondimbc1}\\
\hat{u}_w|_{\hat{x}=0}&=f_0 U ,\label{nondimbc2}
\end{align} 
where we have introduced the two non-dimensional flow parameters 
\begin{align}
 U &=\frac{U_n\mu_n L}{k_0 \Delta p},\\
 f_0 &=\frac{U_w  }{U_n },
\end{align}
which represent the flow of non-wetting phase and the flow fraction, respectively.
Finally, the power law describing the scaling between permeability and pore entry pressure, \eqref{powerlaw}, becomes 
\beq
1+\sigma_P \hat{p}_e=\hat{k}^{-B}.\label{powerlawnondim}
\eeq
We choose the dimensional scaling $k_0$ as the vertical average of the permeability, such that $\hat{k}$ averages to unity but note that $1+\sigma_P \hat{p}_e$ may not.

\subsection{Capillary limit}
\label{seccap}

To find solutions in the capillary limit, we consider an asymptotic expansion in the scaled capillary number $\tilde{\mathrm{N}}_c\ll1$. 
We assume that the statistical properties of the heterogeneity are fixed, such that $\sigma_P$ remains order $\mathcal{O}(1)$ (i.e. we consider a weak overarching pressure gradient that is independent of the rock properties). 
In addition, we restrict our attention to the case where the aspect ratio is much smaller than the flow perturbation, such that  $\delta\ll\tilde{\mathrm{N}}_c\ll1$.

From the capillary pressure equation \eqref{capnondim}, it is clear that both wetting and non-wetting pressure should scale like $\hat{p}_i\sim 1/\tilde{\mathrm{N}}_c$. Therefore, the variables $s$, $\hat{p}_n$ and $\hat{p}_w$ are expanded in $\tilde{\mathrm{N}}_c$ as
\begin{align}
s&=s_0+\tilde{\mathrm{N}}_c s_{1}+\ldots,\\
\hat{p}_n&=\tilde{\mathrm{N}}_c^{-1}\hat{p}_{n_{-1}}+ \hat{p}_{n_0}+\ldots,\\
\hat{p}_w&=\tilde{\mathrm{N}}_c^{-1}\hat{p}_{w_{-1}}+ \hat{p}_{w_0}+\ldots.
\end{align}
Hence, \eqref{nondim2}-\eqref{nondim5} indicate that the pressures in both phases must be constant to leading order, such that $\hat{p}_{n_{-1}}-\hat{p}_{w_{-1}}=\gamma$, for some value of $\gamma$. This is consistent with the definition of capillary limit given by other authors \citep{ekrann2000steady,rabinovich2016analytical}. 
From \eqref{capnondim} we therefore derive a leading order expression for the saturation
\beq
s_0=1-\lb \frac{ \hat{P}_e(\hat{z}) }{\gamma\sigma_P}\rb^\lambda,\label{leads}
\eeq
where we write $\hat{P}_e=1+ \sigma_P\hat{p}_e$ for convenience. Given the form of \eqref{eqrelperm}-\eqref{eqpc}, we would like to express \eqref{leads} in terms of the averaged saturation. Since, to leading order, the capillary limit solution only depends on $\hat{z}$, we select our averaging operator here as the vertical average $\left< \cdot \right>= \int_0^1 \cdot \,\mathrm{d}\hat{z}$. In this way, \eqref{leads} becomes
\beq
s_0=1-\frac{ \hat{P}_e(\hat{z})^\lambda}{\overline{  \hat{P}_e^\lambda}}(1-\overline{s}).\label{leads2}
\eeq 
Note that the solution \eqref{leads2} also satisfies the outlet condition \eqref{nondimbc3} and the impermeability condition \eqref{nondimimperm}. The inlet condition \eqref{nondimbcinlet} is not satisfied, which will lead to a boundary layer over which the saturation transitions to the outlet state, as we discuss later.

To calculate the equivalent relative permeabilites \eqref{eqrelperm}, we first need the averaged Darcy velocities, which only appear at first order. These are obtained by vertically integrating \eqref{nondim2},\eqref{nondim4} and using \eqref{nondimbc1},\eqref{nondimbc2}, to give
\begin{align}
 U &=- \frac{\mathrm{d} \hat{p}_{n_0}}{\mathrm{d} \hat{x}}  \overline{\hat{k}(\hat{z})k_{rn}(s_0(\hat{z}))},\label{leadp1}\\
 f_0  M   U &=- \frac{\mathrm{d} \hat{p}_{w_0}}{\mathrm{d} \hat{x}}  \overline{ \hat{k}(\hat{z})k_{rw}(s_0(\hat{z})) }.\label{leadp2}
\end{align}
By integrating \eqref{leadp1}-\eqref{leadp2} across the channel length, we arrive at expressions for the total changes in pressure across the channel, which we then insert into \eqref{eqrelperm} to finally arrive at the capillary limit for the equivalent relative permeabilities
\begin{align}
k_{rn_\mathrm{cap}}=  \quad\,\, \frac{ U }{ U /\overline{\hat{k}k_{rn}(s_0)}}&\quad=\overline{\hat{k}k_{rn}}(\bar{s}),\label{caplim1}\\
k_{rw_\mathrm{cap}}=\frac{ f_0  M   U }{ f_0  M   U /\overline{\hat{k}k_{rw}(s_0)}}&\quad=\overline{\hat{k}k_{rw}}(\bar{s}).\label{caplim2}
\end{align}
The expressions \eqref{caplim1}-\eqref{caplim2} are a generalisation of the arithmetic mean expressions derived by \citet{rabinovich2016analytical} in the case where the heterogeneity consists of a set of layers.
The equivalent capillary pressure is found by inserting \eqref{leads2} into \eqref{eqpc}, giving
\beq
p_{c_\mathrm{cap}}=\overline{\hat{P}_e^{-1}}\,\,\overline{\hat{P}_e^{\lambda}}^{1/\lambda} (1-\bar{s})^{-1/\lambda}.\label{leadpc}
\eeq

It should be noted that the capillary limit solution \eqref{leads2} may lead to negative saturation values for
\beq
\bar{s}<1-{\overline{ \hat{P}_e^\lambda}}/{\max_{\hat{z}\in[0,1]}\{\hat{P}_e(\hat{z})^\lambda\}},\label{scriteria}
\eeq
which is clearly unphysical. 
In such situations, the saturation profile is instead given by
\beq
s_0=\max \{1-(\hat{P}_e(\hat{z})/\gamma\sigma_P)^\lambda,0\},\label{leads2nonneg}
\eeq
and consequently there are regions of space devoid of non-wetting phase, a phenomenon associated with very strong heterogeneities. In this case, it is less straightforward to relate the capillary pressure constant $\gamma$ to the mean saturation analytically. However, a nonlinear relationship can be established numerically instead.
Note that we could go to higher order in the asymptotic expansions to capture near-capillary-limit behaviour. However, for the purposes of understanding the leading order impact of capillary heterogeneity on the flow, we find leading order solutions sufficient.

\subsection{Viscous limit}
\label{secvisc}

In contrast to the capillary limit, the viscous limit relates to the regime where the flow-driving pressure gradient is much larger than the capillary forces, such that the heterogeneities do not affect the flow. Therefore, to address this limit we consider a small capillary correction $\Delta p_e/\Delta p=\tilde{\mathrm{N}}_c^{-1}\ll1.$
Note that the pore entry pressure is related to the scaled capillary number via the parameter $\sigma_P=C\tilde{\mathrm{N}}_c^{-1}$, where $C=\Delta p/p_{e_0}$. 
For this analysis, we assume that the overarching pressure gradient is fixed, such that $C$ remains order $\mathcal{O}(1)$ (i.e. we consider a weak heterogeneity $\Delta p_e$ independently of the pressure gradient). Furthermore, we assume that the aspect ratio is much smaller than the heterogeneity perturbation, such that $\delta\ll\tilde{\mathrm{N}}_c^{-1}\ll1$. 
Given the power law relationship \eqref{powerlawnondim}, we also have
\beq
\hat{k}=1-BC \hat{p}_e(\hat{z}) \tilde{\mathrm{N}}_c^{-1}+\ldots. 
\eeq
Similarly to the capillary limit, here we seek an asymptotic solution, except now this is given in terms of powers of $\tilde{\mathrm{N}}_c^{-1}$, such that
\begin{align}
s&=s_0+\tilde{\mathrm{N}}_c^{-1} s_{1}+\ldots,\\
\hat{p}_n&=\hat{p}_{n_{0}}+\tilde{\mathrm{N}}_c^{-1} \hat{p}_{n_1}+\ldots,\\
\hat{p}_w&=\hat{p}_{w_{0}}+ \tilde{\mathrm{N}}_c^{-1}\hat{p}_{w_1}+\ldots.
\end{align}
In this way, \eqref{nondim2},\eqref{nondim4} indicate that there are no leading order vertical pressure gradients $\partial \hat{p}_{n_0}/\partial \hat{z}=\partial \hat{p}_{w_0}/\partial \hat{z}=0$. 
Furthermore, \eqref{capnondim} indicates that to leading order
\beq
\hat{p}_{n_0}-\hat{p}_{w_0}=C^{-1}(1-s_0)^{-1/\lambda},\label{capnondim2}
\eeq
which implies that $s_0$ must also be independent of $\hat{z}$. This also ensures that the impermeability condition \eqref{nondimimperm} is satisfied at leading order.

The Darcy velocities are obtained by vertically integrating the system \eqref{nondim1}-\eqref{capnondim} and using \eqref{nondimbc1},\eqref{nondimbc2}, to give
\begin{align}
 U &=- \frac{\mathrm{d} \hat{p}_{n_0}}{\mathrm{d} \hat{x}}  k_{rn}(s_0(\hat{x})),\label{leadp3}\\
 f_0  M   U &=- \left[\frac{\mathrm{d} \hat{p}_{n_0}}{\mathrm{d} \hat{x}}-(C\lambda)^{-1}(1-s_0)^{-1/\lambda-1}\frac{\mathrm{d} s_{0}}{\mathrm{d} \hat{x}}\right]   k_{rw}(s_0(\hat{x})) .\label{leadp4}
\end{align}
Due to \eqref{leadp4}, the zero gradient boundary condition \eqref{nondimbc3} can only be satisfied if $s_0$ is constant. This is equivalent to the condition
\beq
 f_0  M   k_{rn}(s_0)=k_{rw}(s_0),
\eeq
which enforces a relationship between the flow fraction $ f_0 $ and the saturation $s_0$. 
Therefore, since the viscous limit solution is constant to leading order, the averaging operator in \eqref{eqrelperm}-\eqref{eqpc} is trivial.
With this taken into account, the viscous limit expressions for the equivalent relative permeabilities are
\begin{align}
k_{rn_\mathrm{visc}}&=\frac{ U }{ U /k_{rn}(s_0)}\quad=k_{rn}(\bar{s}),\label{visclim1}\\
k_{rw_\mathrm{visc}}&=\frac{ f_0  M   U }{ f_0  M   U /k_{rw}(s_0)}\quad=k_{rw}(\bar{s}).\label{visclim2}
\end{align}
Furthermore, the equivalent capillary pressure is given by
\beq
p_{c_\mathrm{visc}}=(1-\bar{s})^{-1/\lambda}.\label{visclim3}
\eeq
The viscous limit expressions \eqref{visclim1}-\eqref{visclim3} are identical to the original expressions for relative permeability and capillary pressure, which is expected in the limit of vanishing heterogeneity. 
Note that this analysis can be extended to higher order terms to approximate the case of a large but finite capillary number. However, we find a leading order analysis satisfactory for our purposes.

\subsection{Types of heterogeneity}

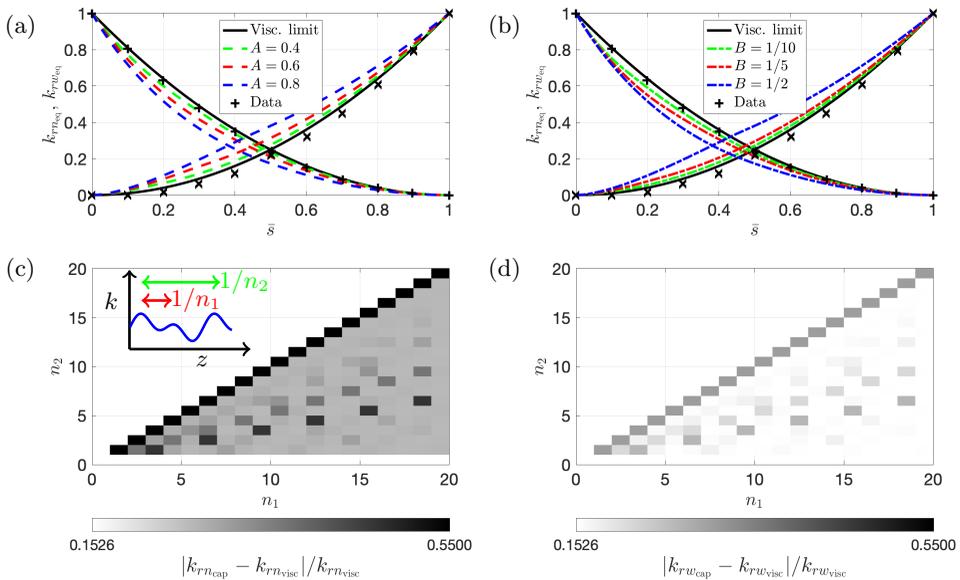
\begin{figure}
\centering
\begin{tikzpicture}[scale=0.8]
\node at (0,0) {\includegraphics[width=0.45\textwidth]{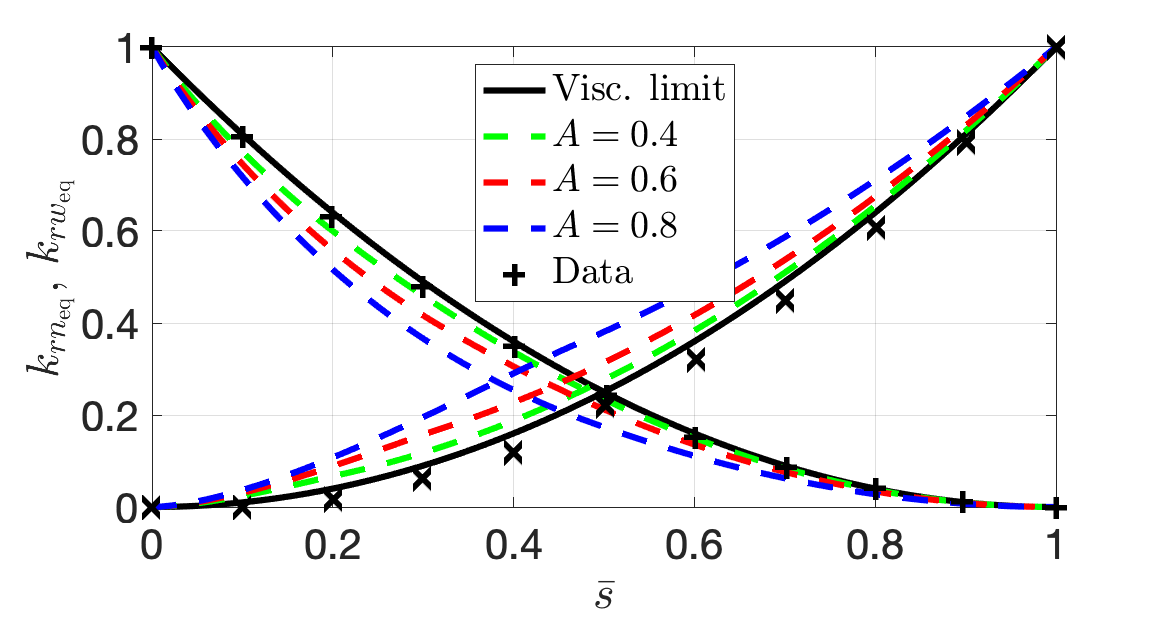}};
\node at (8,0) {\includegraphics[width=0.45\textwidth]{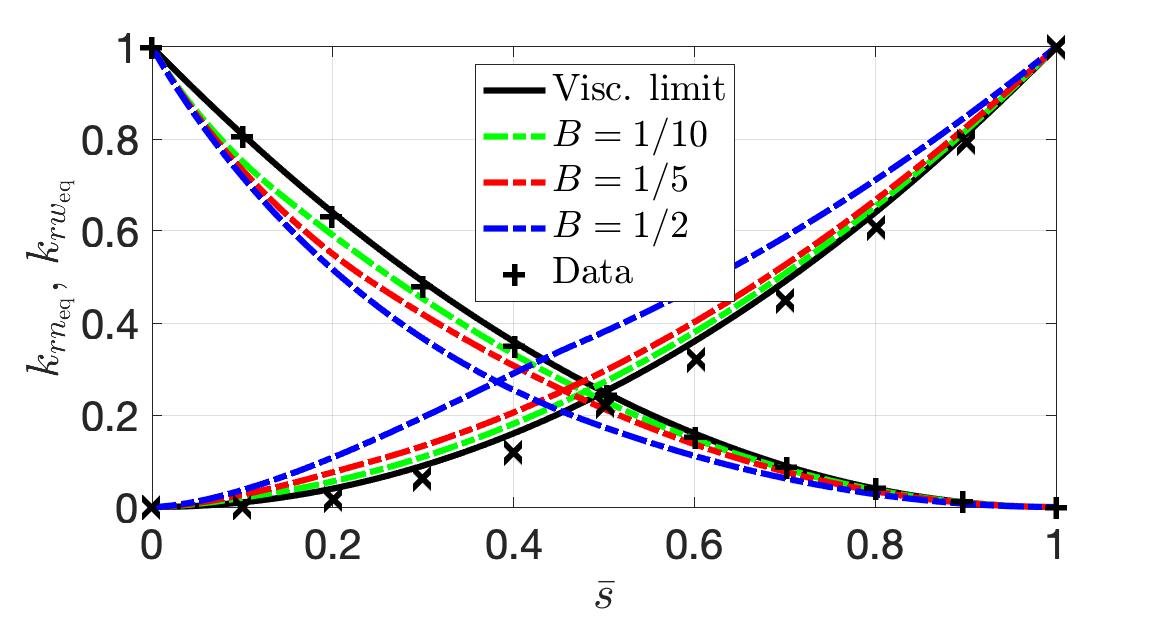}};
\node at (0,-5) {\includegraphics[width=0.45\textwidth]{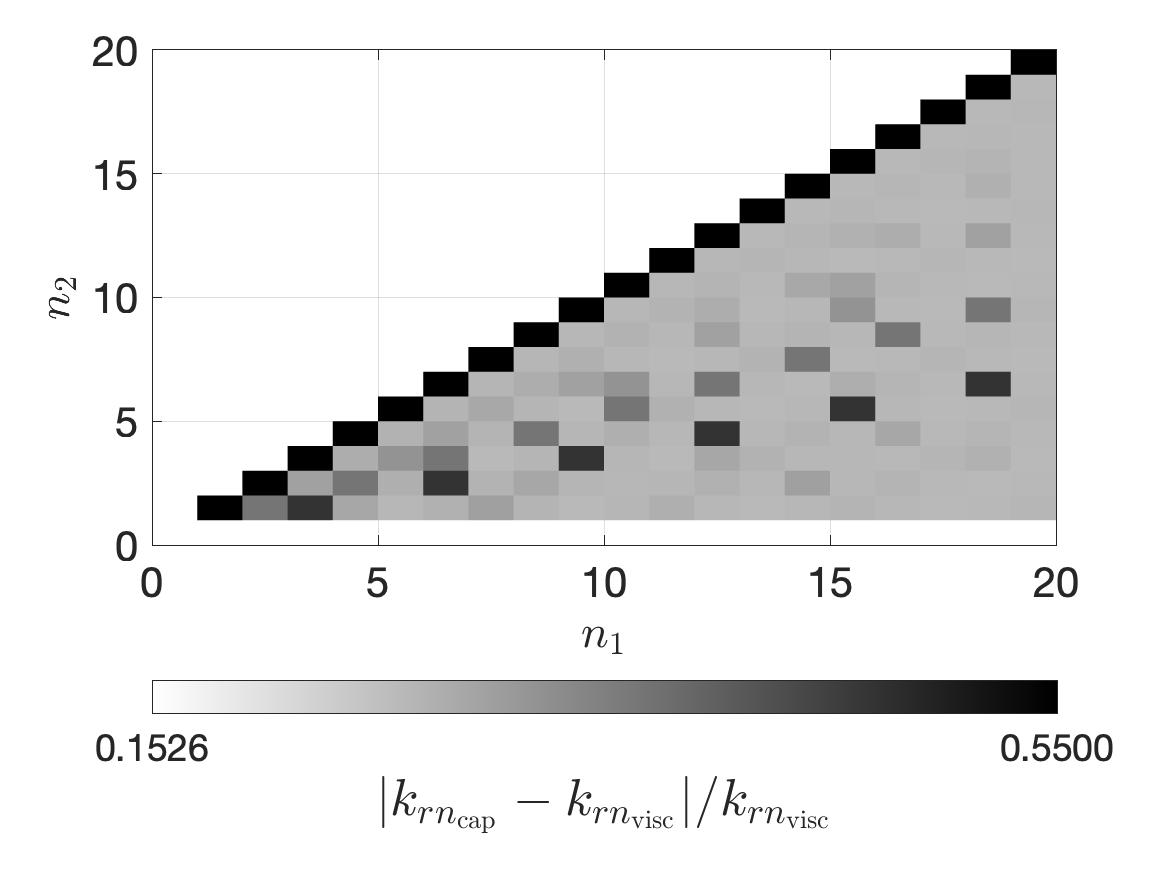}};
\node at (8,-5) {\includegraphics[width=0.45\textwidth]{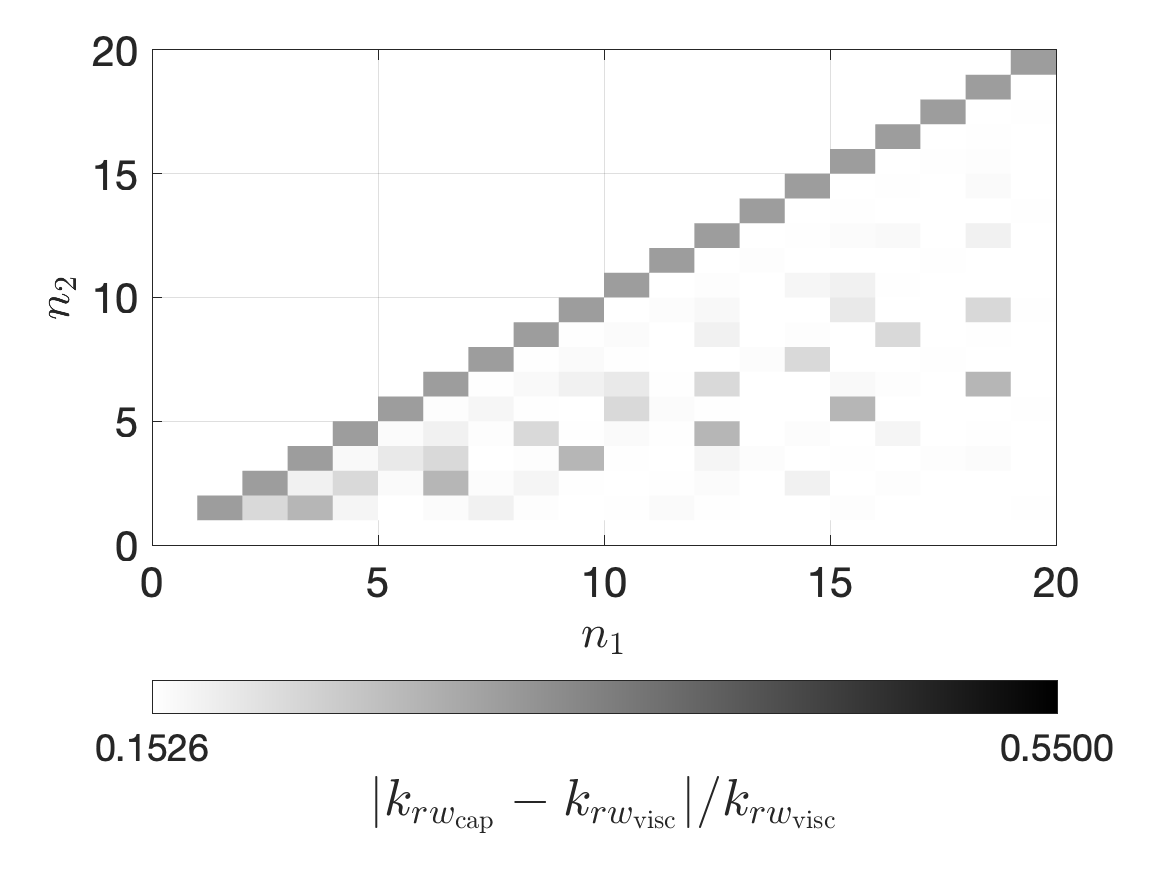}};
\draw[line width=1,->] (-2.2,-3.8) -- (-2.2,-2.5);
\draw[line width=1,->] (-2.2,-3.8) -- (-0.2,-3.8);
\draw[line width=1,<->,red] (-2,-3.) -- (-1.5,-3.);
\draw[line width=1,<->,green] (-2,-2.7) -- (-0.7,-2.7);
\node[red] at (-1.1,-3.) {$1/n_1$};
\node[green] at (-0.3,-2.7) {$1/n_2$};
\node at (-2.5,-3) {$k$};
\node at (-1,-4) {$z$};
\begin{axis}[domain=0:50,hide axis, scale only axis, width=0.15\textwidth,
     height=0.04\textwidth, samples=100, at={(-0.175\textwidth,-0.275\textwidth)}]
      \addplot[mark=none,color=blue,very thick]{sin(20*x)+sin(10*x)};
    \end{axis}
\node at (-4,1.5) {(a)};
\node at (4,1.5) {(b)};
\node at (-4,-2.5) {(c)};
\node at (4,-2.5) {(d)};
\end{tikzpicture}
\caption{Viscous and capillary limits of equivalent relative permeability \eqref{eqrelperm} (note the non-wetting relative permeability is normalised by $k_{rn_0}=0.116$) for a sinusoidal heterogeneity \eqref{sink} and a power law relationship for the pore entry pressure \eqref{powerlawnondim}. The capillary limit is shown for different values of the heterogeneity amplitude $A$ (fixing $B=1/2$) (a) and power law $B$ (fixing $A=0.8$) (b). Experimental data taken from \citet{bennion2005relative} in the viscous limit. (c,d) Greyscale maps of the percentage difference between viscous and capillary limit predictions for a heterogeneity with two wavenumbers $n_1$, $n_2$ \eqref{sink2}.
\label{capvisc}}
\end{figure}

Whilst the above analysis applies for any given vertical heterogeneity and empirical relative permeability relationships $k_{rn},k_{rw}$, we shall now discuss how our predictions manifest in an example scenario.
We choose a simple background heterogeneity which consists of a sinusoidal perturbation on a uniform permeability profile
\beq
\hat{k}= 1+A\sin {2n\pi \hat{z}} ,\label{sink}
\eeq
for some amplitude $A$ and wavenumber $n\in \mathbb{N}$. Meanwhile, the pore entry pressure is given by \eqref{powerlawnondim}, in terms of some power $B$.  For the intrinsic relative permeabilities $k_{ri}$, we use the classic empirical power law of \citet{corey1954interrelation}, which is given by \eqref{corey1}-\eqref{corey2}, with a quadratic power law. A full list of parameter values is found in Appendix \ref{appA}.

In figure \ref{capvisc} we plot the viscous limit (which is independent of heterogeneity) and the capillary limit for different values of $A$ and $B$ (for a fixed value of $n=1$). 
The plots confirm that heterogeneity has the overall effect of lowering the flow of the wetting phase, and raising the flow of non-wetting phase. This can be explained by \eqref{leads2}, which indicates that $s$ is larger in places where the pore entry pressure is smaller, and hence in regions of larger pore space. Hence, capillary pressure forces the non-wetting saturation to preferentially segregate to regions of larger space, where it is easier to flow. Increasing the amplitude $A$ accentuates this effect, since this corresponds to stronger heterogeneity. It is also accentuated by increasing the power law $B$, since this increases the strength of the pore entry pressure heterogeneity.

Note in some cases it is possible to derive analytical formulae for the equivalent relative permeabilities in the capillary limit. For example, in the simple case where $B=1$, the resulting expressions are
\begin{align}
k_{rn_\mathrm{cap}}&=1+\sqrt{1-A^2}\lb \bar{s}^2 - 1\rb,\label{anal1}\\
k_{rw_\mathrm{cap}}&=\sqrt{1-A^2}\lb 1-\bar{s}\rb^2.\label{anal2}
\end{align}
The expressions \eqref{anal1}-\eqref{anal2} are valid for amplitudes $A<1$, though only for values of $\bar{s}$ large enough so that \eqref{leads2} doesn't have $s=0$ anywhere (or according to \eqref{scriteria}, for $\bar{s}>1-\sqrt{(1-A)/(1+A)}$). In situations where there are regions of zero saturation, an analytical formula is still possible, though the expressions are more complicated so we do not display them here.

In contrast to $A$ and $B$, varying the wavenumber of the perturbation $n\in \mathbb{N}$ does not have a significant effect on $k_{rn_\mathrm{cap}},k_{rw_\mathrm{cap}}$. However, more interesting effects are observed when two different wavelengths are introduced, such that the permeability
\beq
\hat{k}= 1+\frac{A F}{2} \lb \sin {2n_1\pi \hat{z}}+\sin {2n_2\pi \hat{z}} \rb,\label{sink2}
\eeq
where the factor $F$ is chosen such that the difference between the maximum and minimum perturbation (and hence the capillary number) is kept the same. In figure \ref{capvisc}c,d we display greyscale plots of the percentage difference in equivalent relative permeability between the viscous and capillary limits, for different values of $n_1$ and $n_2$. 
Since the plots are symmetric about $n_1\leftrightarrow n_2$, we only display half of the phase space. Clearly, the maximum difference occurs when $n_2=n_1$ (at constant values of $55\%$ and $31\%$), but there are also streaks near $n_2= n_1/3$, $n_2=n_1/2$, $n_2=n_1/4$, and so on (in descending order of magnitude).

Whilst these heterogeneities are idealised, this simple investigation serves as an illustration for the different types of permeability and pore entry pressure one might encounter in the field. In particular, we have indicated how upscaled quantities depend on model parameters in the two limiting viscous and capillary limits, which will be useful throughout the paper. Next, we move on to model situations which are not in either of these two limits, but instead lie somewhere in between.

\subsection{Intermediate capillary number}
\label{secint}

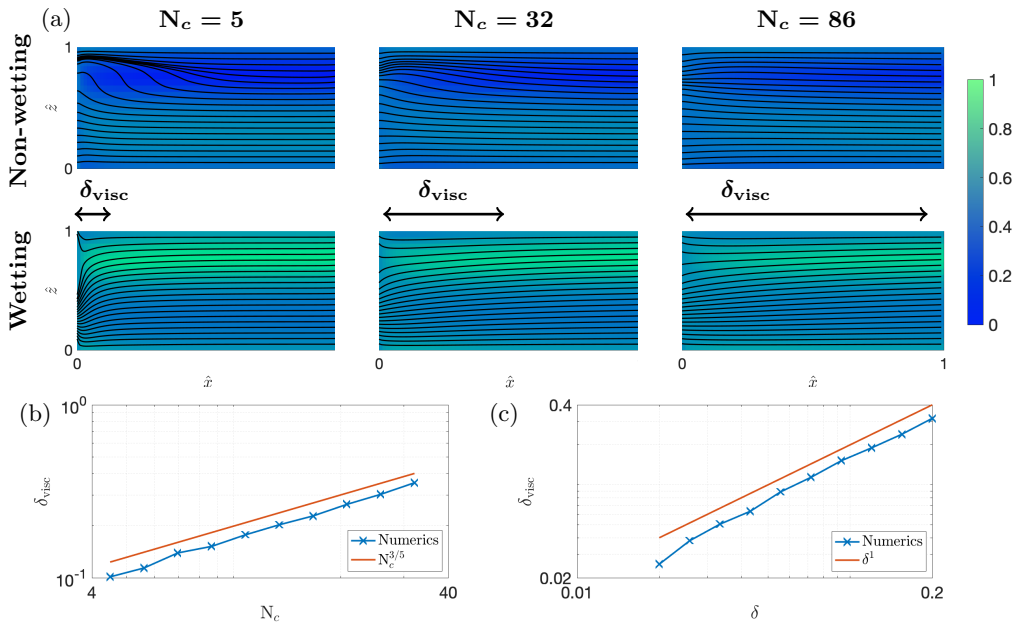
\begin{figure}
\centering
\begin{tikzpicture}[scale=0.8]
\node at (0,0) {\includegraphics[width=0.33\textwidth]{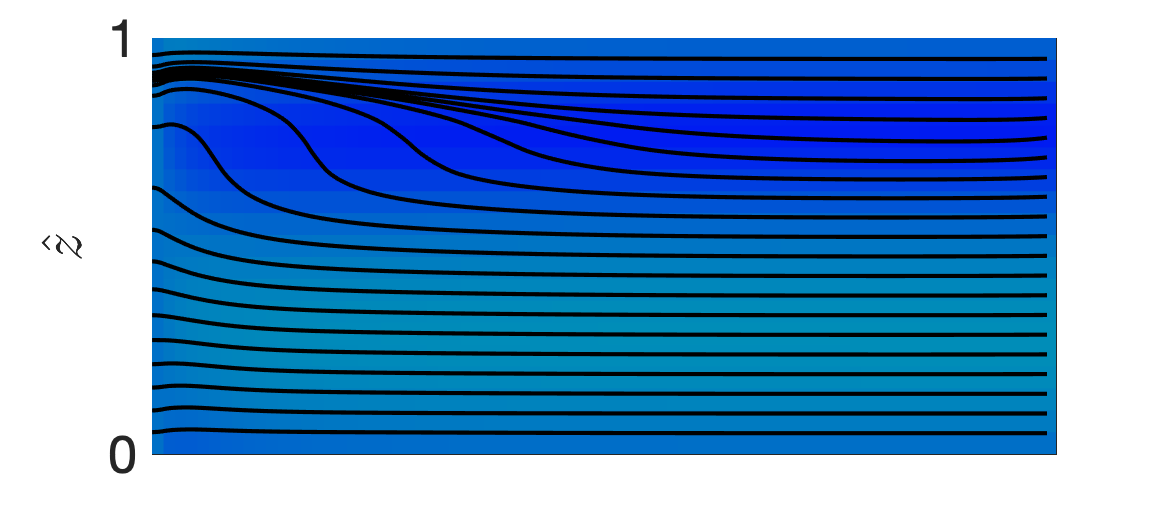}};
\node at (0,-3.2) {\includegraphics[width=0.33\textwidth]{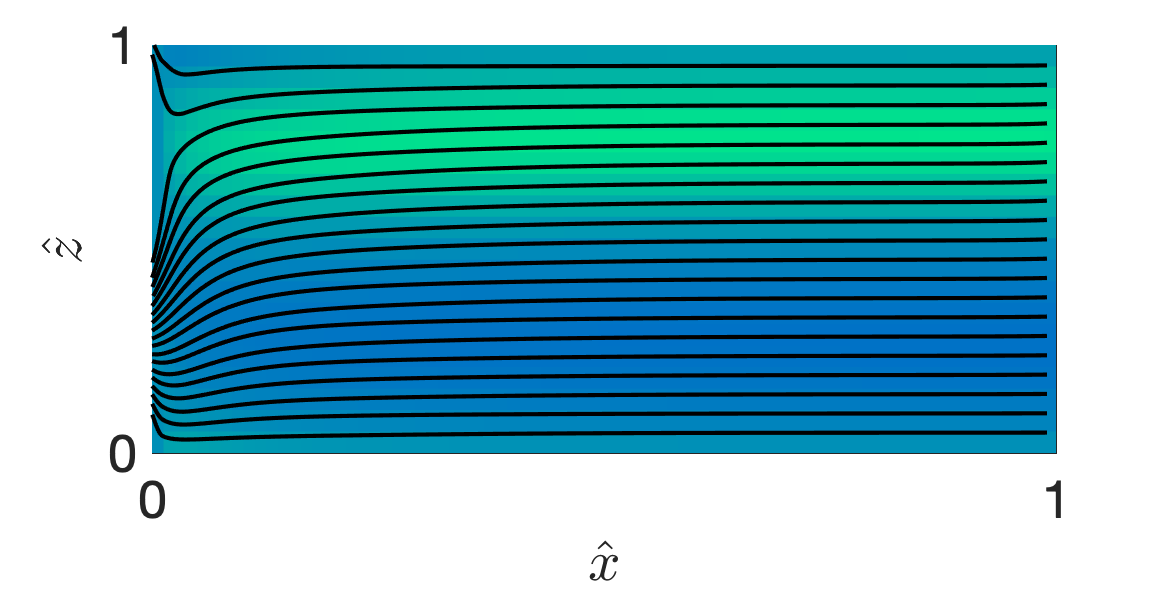}};
\node at (5.0,0) {\includegraphics[width=0.33\textwidth]{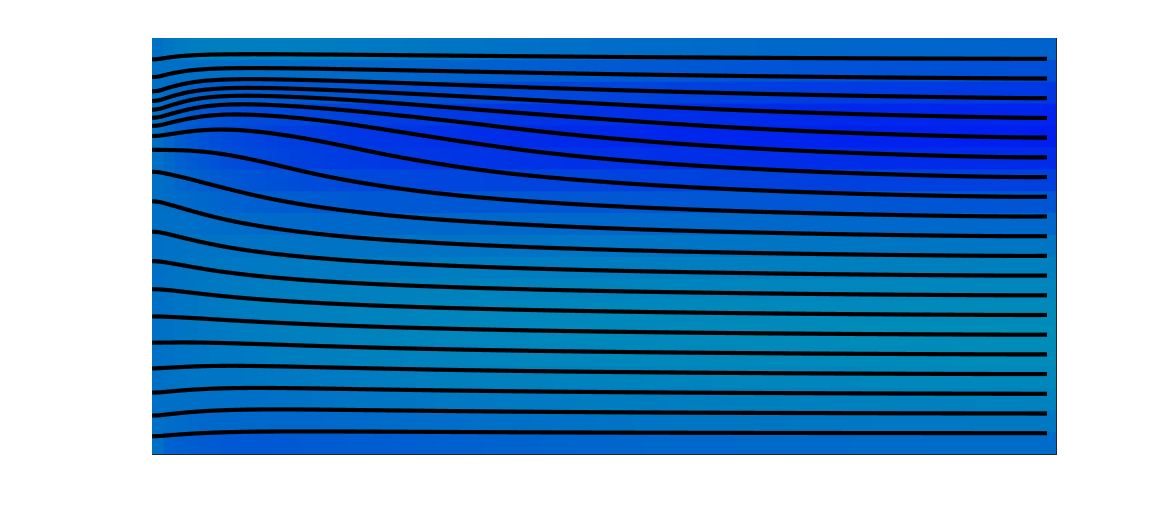}};
\node at (5.0,-3.2) {\includegraphics[width=0.33\textwidth]{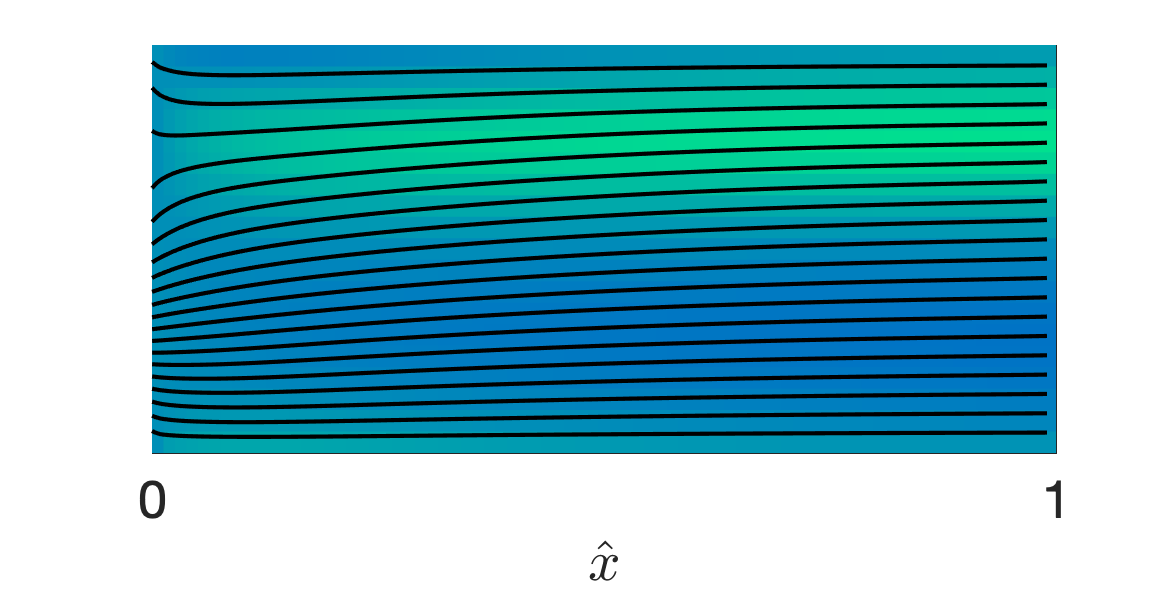}};
\node at (10,0) {\includegraphics[width=0.33\textwidth]{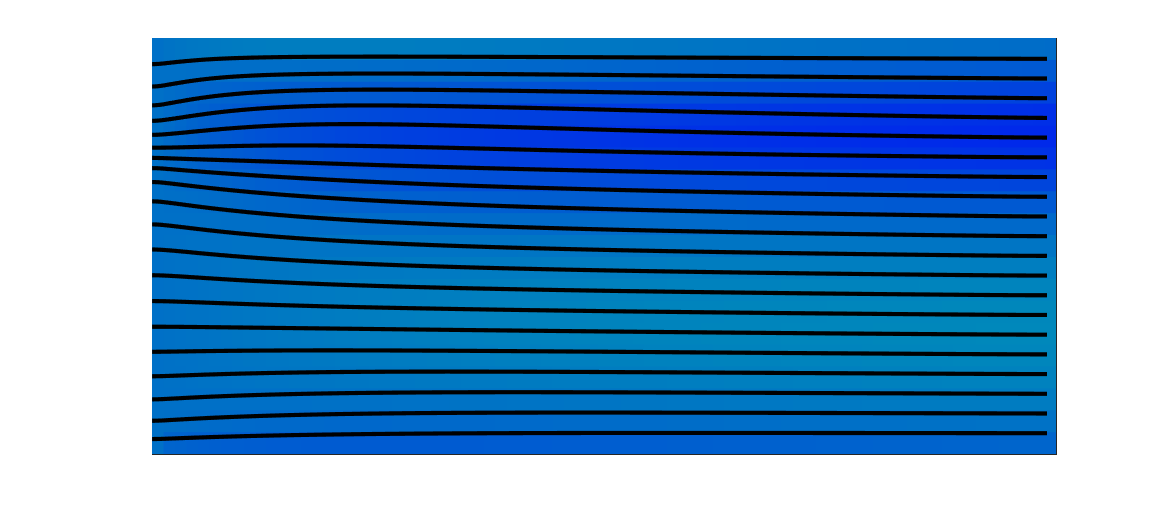}};
\node at (10,-3.2) {\includegraphics[width=0.33\textwidth]{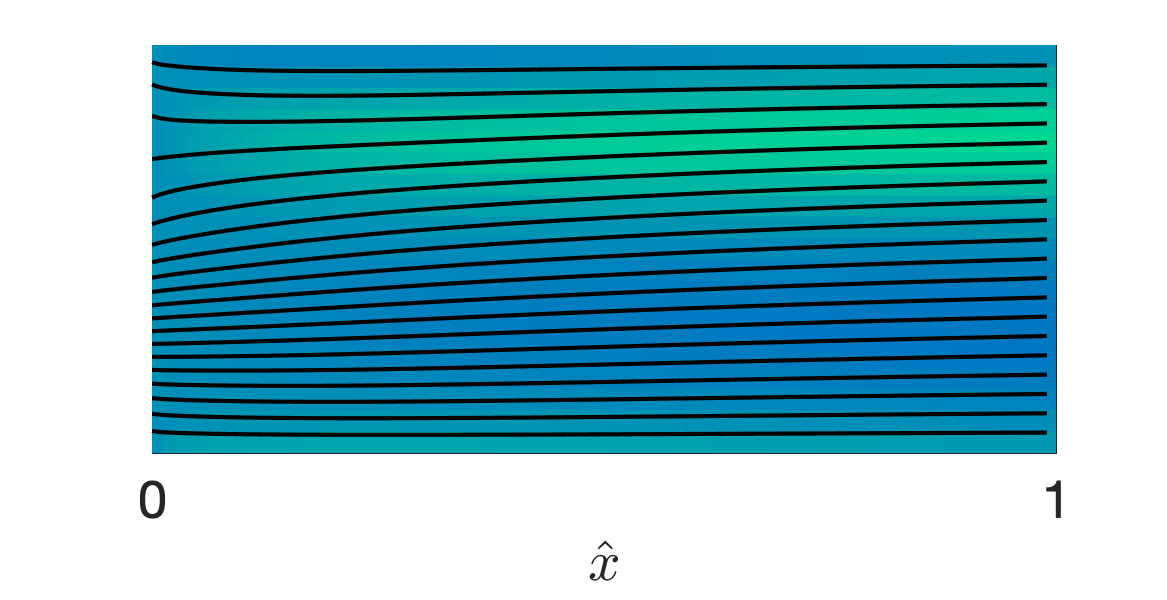}};
\draw[line width=1,<-> ] (-2.1,-1.7) -- (-1.5,-1.7);
\draw[line width=1,<-> ] (3.,-1.7) -- (5,-1.7);
\draw[line width=1,<-> ] (8.,-1.7) -- (12.,-1.7);
\node at (-1.6,-1.3) {\bf $\boldsymbol{\delta_\mathrm{visc}}$};
\node at (4,-1.3) {\bf $\boldsymbol{\delta_\mathrm{visc}}$};
\node at (9.,-1.3) {\bf $\boldsymbol{\delta_\mathrm{visc}}$};
\node at (0,1.5) {\bf N$\boldsymbol{_c=5}$};
\node at (5.,1.5) {\bf N$\boldsymbol{_c=32}$};
\node at (10.,1.5) {\bf N$\boldsymbol{_c=86}$};
\node at (-3,0) {\rotatebox{90}{\bf Non-wetting}};
\node at (-3,-2.8) {\rotatebox{90}{\bf Wetting}};
\node at (1,-6.5) {\includegraphics[width=0.45\textwidth]{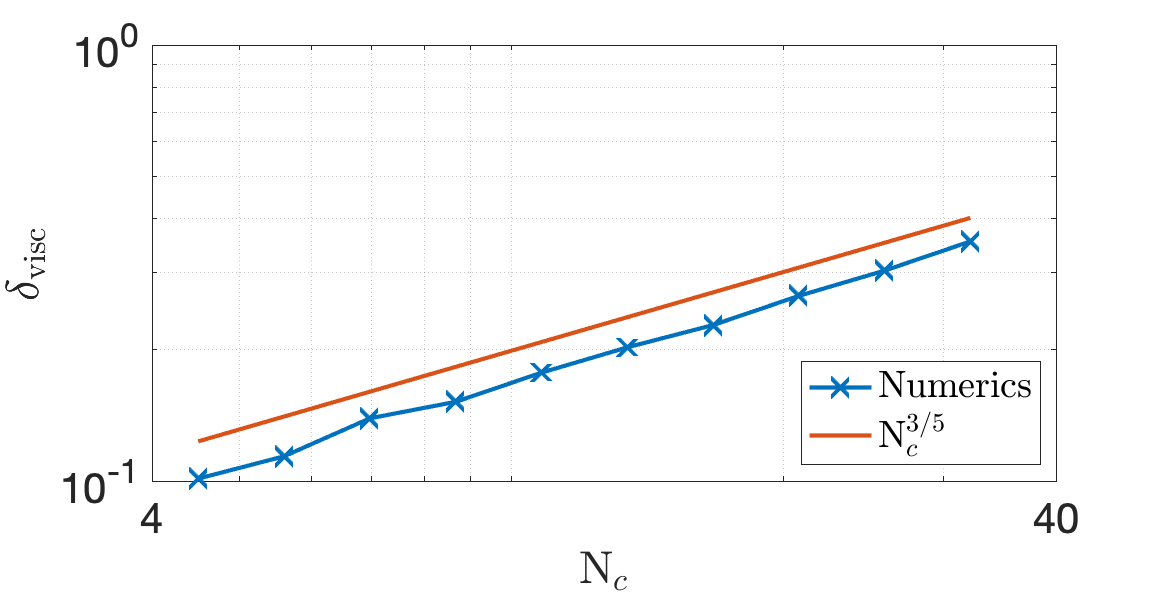}};
\node at (9,-6.5) {\includegraphics[width=0.45\textwidth]{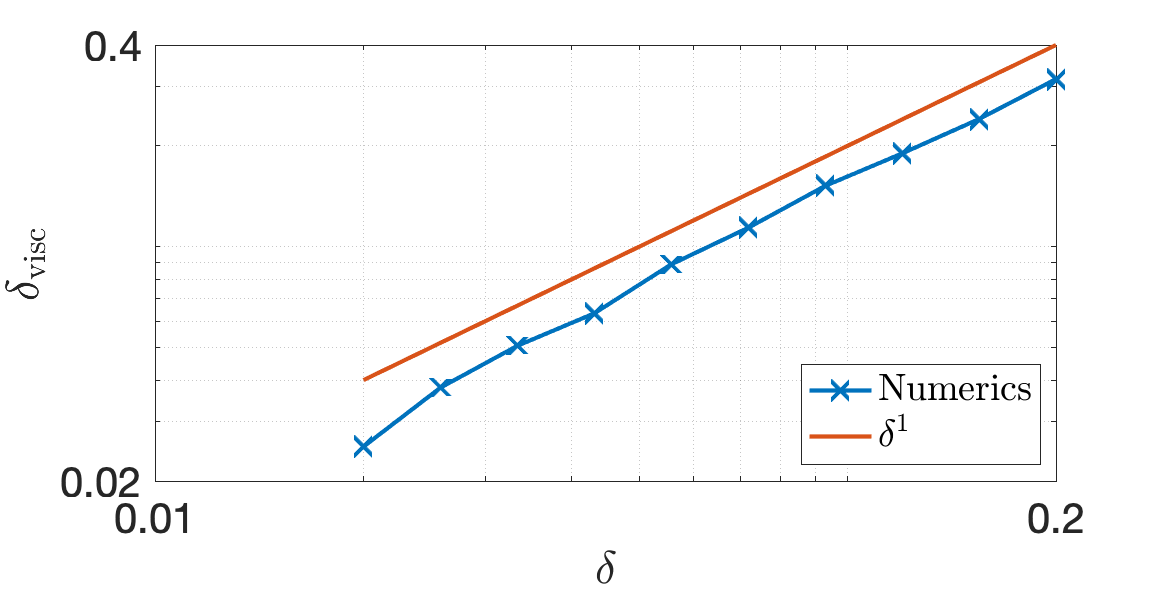}};
\node at (13,-1.5) {\includegraphics[width=0.05\textwidth]{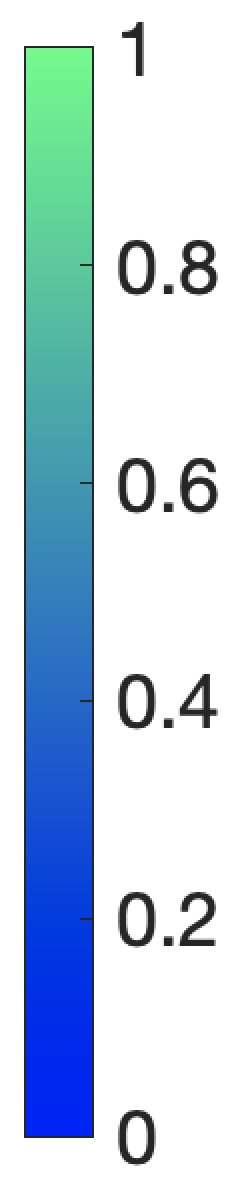}};
\node at (-2.4,1.5) {(a)};
\node at (-2.75,-5) {(b)};
\node at (5,-5) {(c)};
\end{tikzpicture}
\caption{(a) Steady numerical solutions of the saturation of non-wetting $s$ and wetting $1-s$ phases across a range of capillary numbers (where N$_c$ \eqref{capdef} is given in terms of non-wetting pressure change). Streamlines of the Darcy velocity fields $\hat{\boldsymbol{u}}_{n}$ and $\hat{\boldsymbol{u}}_{w}$ are overlaid on each plot. Boundary layer thickness $\delta_\mathrm{visc}$ plotted against capillary number (holding $\delta=0.1$ fixed) (b) and against aspect ratio (holding N$_c=8$ fixed) (c), using logarithmic scales. \label{numcol}}
\end{figure}

In the case of intermediate capillary number, there are two possible approaches: Either we can perform numerical simulations of steady Darcy flow \eqref{nondim1}-\eqref{capnondim} with boundary conditions \eqref{nondimbcp}-\eqref{nondimimperm} and then calculate the equivalent properties \eqref{eqrelperm}; or we can go to higher order terms in the asymptotic expansion of each of the viscous limit or the capillary limit. We prefer to use the numerical approach here, similarly to \citet{virnovsky2004steady},  since it gives a complete description that is valid across all capillary numbers, and this is more convenient than patching together asymptotic solutions from different regimes. Although the previous analysis related to the scaled capillary number $\tilde{\mathrm{N}}_c$, here we keep everything in terms of the original capillary number N$_c$, since this is more common in the literature, and therefore makes our results more accessible.

We have calculated numerical solutions for capillary number N$_c$ between $1$ and $10^4$ and a heterogeneity \eqref{sink} with amplitude $A=0.6$ and power law $B=1/2$. In addition, we set the aspect ratio as $\delta=0.1$. 
The numerical solutions are calculated using a $4^{\mathrm{th}}$ order central difference scheme in space (with $80\times20$ grid points in the ($x,z$) directions) and a pseudo-time-stepping method that converges iteratively. We use the method of continuation to advance quickly through several orders of magnitude of the capillary number.

In figure \ref{numcol}a we display colour plots of both the wetting and non-wetting saturations, overlaid with streamlines given by the Darcy velocities $\hat{\boldsymbol{u}}_i$ for three different values of the capillary number.
For small capillary numbers, the flow segregates into two separate streams, where all the non-wetting phase moves to the more permeable regions, and vice versa. 
There is a small region of strong transverse flow of wetting phase near the inlet due to sharp saturation gradients.
For larger capillary numbers, the saturation profile is more uniform throughout. The segregation of phases is less pronounced, and there is little transverse flow near the inlet.

There is a kind of horizontal boundary layer in saturation distribution that exists near the inlet, over which the saturation transitions from the constant inflow value $s_i$ to the capillary limit solution downstream. 
The boundary layer thickness, which we denote $\delta_\mathrm{visc}$, grows with capillary number. By defining $\delta_\mathrm{visc}$ as the distance needed to reach the capillary limit solution \eqref{leads2} to $90\%$ accuracy, we can plot the variation with capillary number, as can be seen in figure \ref{numcol}b. Hence, we find that the boundary layer thickness $\delta_\mathrm{visc}$ is approximately proportional to N$_c^{3/5}$.

Note that if we were to extend the aquifer sufficiently, all cases would eventually reach the capillary limit. This is evident by noticing that the only solution to \eqref{nondim1}-\eqref{nondimimperm} which is independent of $\hat{x}$ is the capillary limit solution ($p_c=$constant). Therefore, in the transition between the viscous and capillary limits, the inlet condition $s_i$ is of critical importance. Indeed, if we were to choose the inlet profile as \eqref{leads2}, then any capillary number would result in the capillary limit solution. To mitigate this, we have chosen $s_i$ as a constant value so that both viscous and capillary limits can be recovered in the limit of large and small capillary number, respectively. In addition to the capillary number, the boundary layer thickness must clearly depend on the aspect ratio $\delta$, and we have plotted this dependence in figure \ref{numcol}c, holding the capillary number fixed at N$_c=8$. In this case, we see that $\delta_\mathrm{visc}$ grows linearly with aspect ratio. This is expected due to a uniform stretching of the domain. Clearly, the choice of the domain dimensions for upscaling has a significant impact on the resulting upscaled quantities, presenting a challenge for creating a general theory of upscaling. Later in Section \ref{seclength} we discuss how varying the choice of domain size may affect predictions.

To calculate equivalent properties of the flow, it is necessary to choose an appropriate averaging operator $\left<\cdot\right>$ in \eqref{eqrelperm},\eqref{eqpc}. 
We are dissuaded from choosing a core average, since undesirable boundary layer effects
from the inlet make it impossible to recover the capillary limit solution \eqref{caplim1}-\eqref{caplim2} as we decrease N$_c$.
Instead, we find the most convenient choice is a vertical average at the aquifer outlet $\left<\cdot\right>=\int_0^1 \cdot \,\mathrm{d}\hat{z}|_{\hat{x}=1}$. Since we have chosen zero gradient conditions \eqref{bc3}, this removes boundary effects from the averaging process as much as possible. In the case of the pressure drop in \eqref{eqrelperm}, we use an average of the non-dimensional pressure gradient $\Delta \hat{p}_i=\overline{\partial \hat{p}_i/\partial \hat{x}}$. Using this averaging method allows the solution to converge to both capillary and viscous limit solutions consistently.

\begin{figure}
\centering
\begin{tikzpicture}[scale=0.8]
\node at (0,5) {\includegraphics[width=0.45\textwidth]{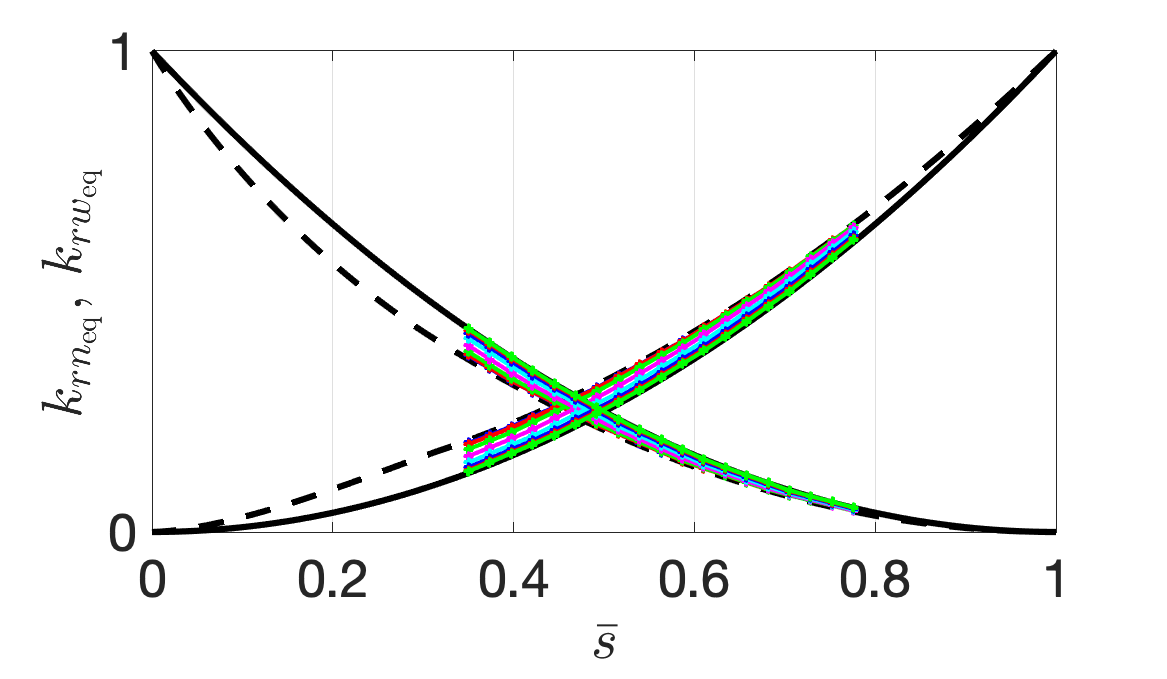}};
\node at (8,5) {\includegraphics[width=0.45\textwidth]{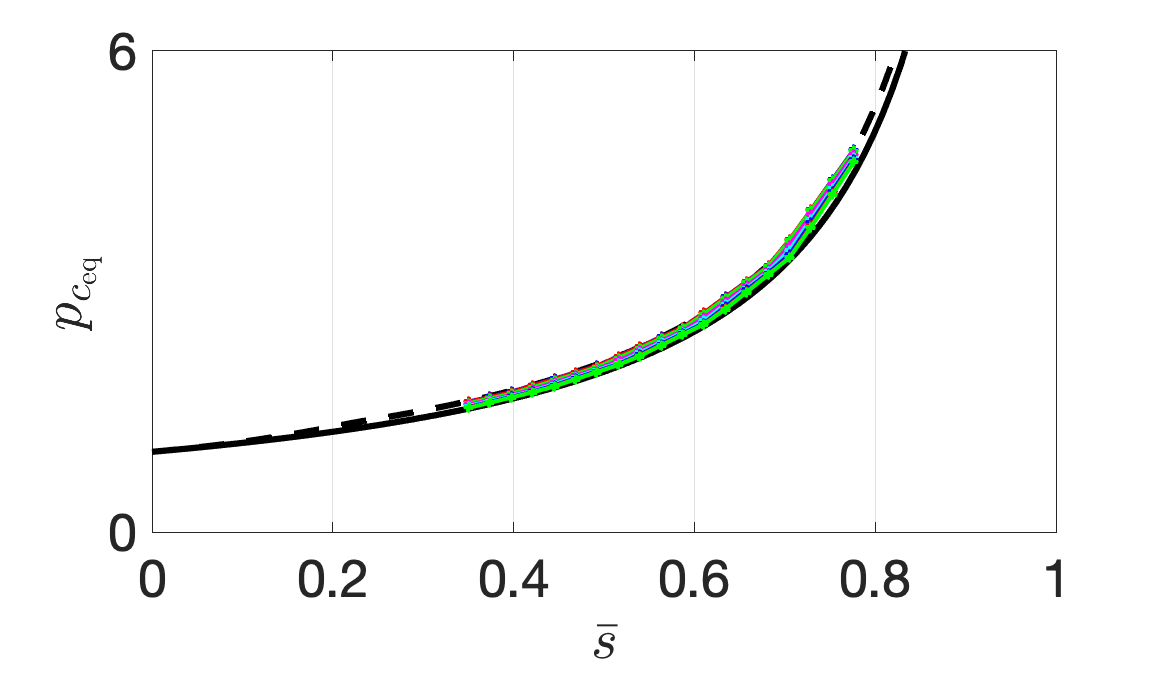}};
\node at (0,0) {\includegraphics[width=0.45\textwidth]{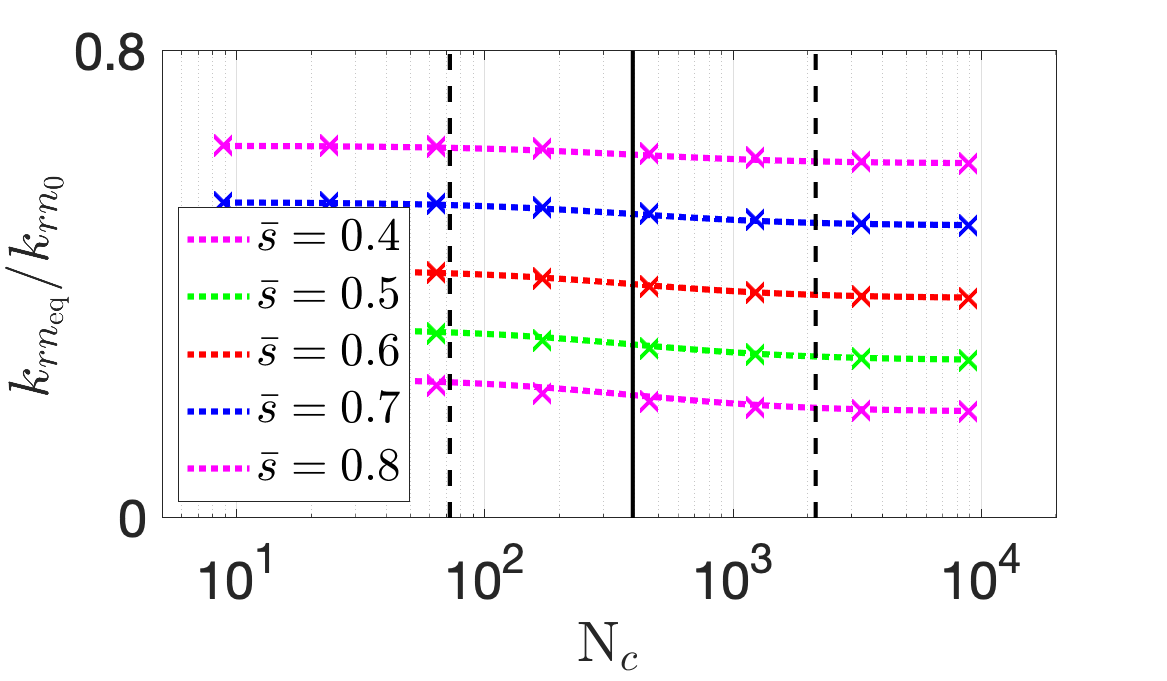}};
\node at (8,0) {\includegraphics[width=0.45\textwidth]{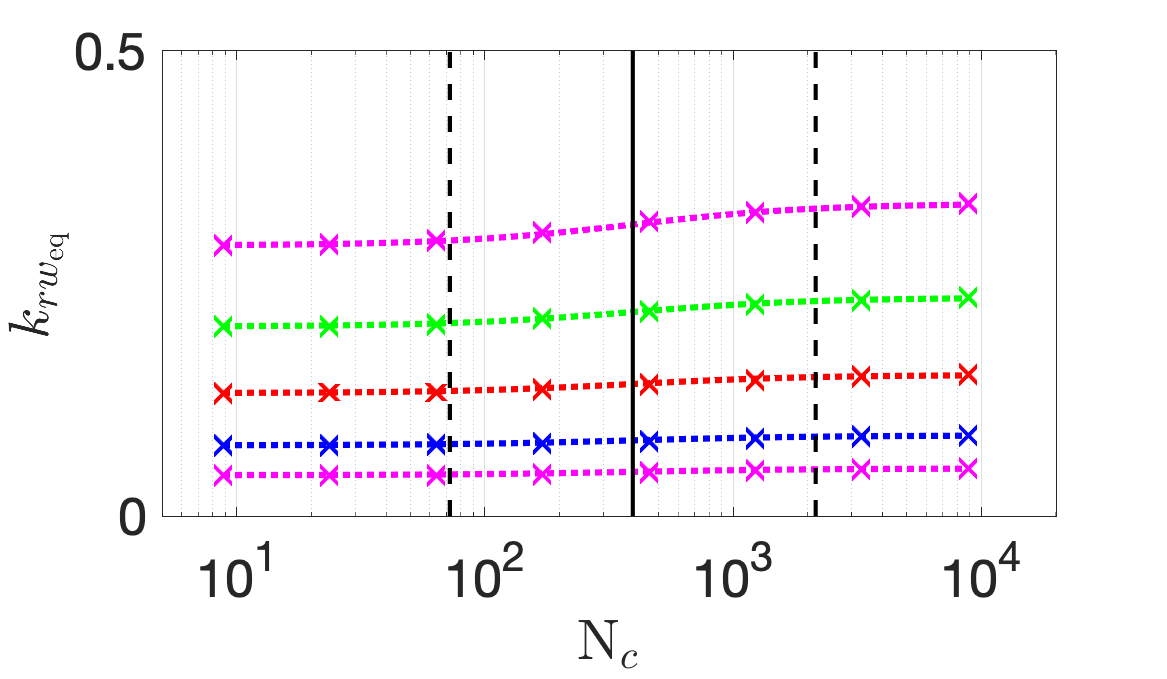}};
\draw[line width=1,-> ] (1,4.5) -- (1,5.5);
\node at (0.5,6) {Decreasing N$_c$};
\node at (0.3,2.4) { N$_{c_t}\times\Delta^{\pm1}$};
\draw[line width=1,<-> ] (-0.9,2.1) -- (1.45,2.1);
\node at (-4,1.5) {(c)};
\node at (4,1.5) {(d)};
\node at (-4,6.5) {(a)};
\node at (4,6.5) {(b)};
\end{tikzpicture}
\caption{  (a) Equivalent relative permeabilities \eqref{eqrelperm} (note the non-wetting relative permeability is normalised by $k_{rn_0}=0.116$), and equivalent capillary pressure \eqref{eqpc} (b), calculated with numerical simulations across a range of capillary numbers. (c,d)  Best fit of composite hyperbolic tangent function \eqref{transfun}, modelling the transition between capillary and viscous limits, illustrating the fitted parameter N$_{c_t}$ and one folding scale $\Delta$ on either side. \label{transfig}}
\end{figure}
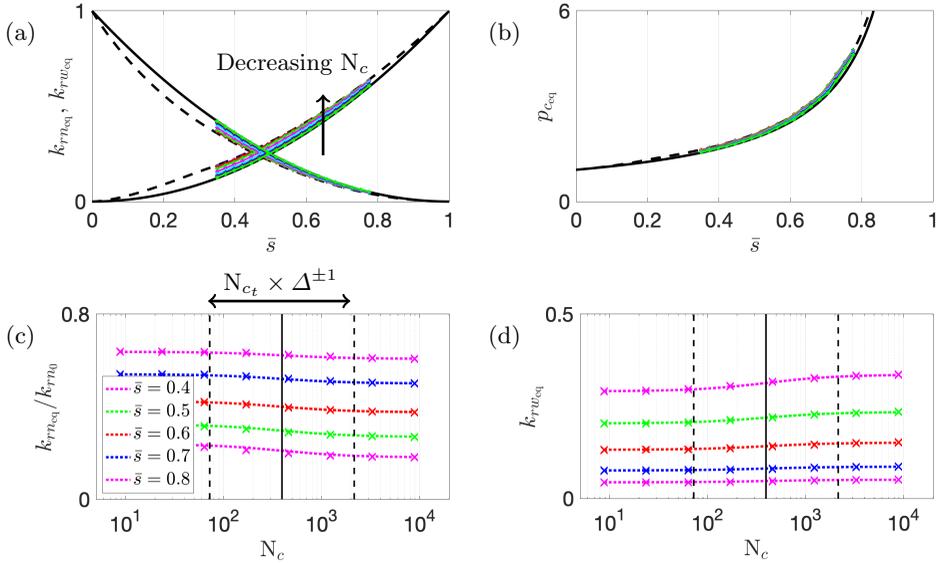

The equivalent relative permeabilities and capillary pressure are shown in figure \ref{transfig}a,b. Each coloured line on the plot has the same  capillary number and different values of the inlet saturation $s_i$ (or equivalently the flow fraction $ f_0 =U_w/U_n$). 
In this way, it is possible to observe how the equivalent relative permeabilities vary over both saturation and capillary number, as illustrated in figure \ref{transfig}c,d. As indicated in the plots, the equivalent relative permeabilities are very well approximated by the transition function
\beq
k_{ri_\mathrm{eq}}=\frac{1}{2}\left[ k_{ri_-}(\bar{s}) \tanh \lb\frac{\log{\mathrm{N}_c}-\log{\mathrm{N}_{c_t}}}{\log \Delta}\rb+k_{ri_+}(\bar{s}) \right] ,\quad i=n,w,\label{transfun}
\eeq
with parameter values N$_{c_t}= 394$, $\Delta=5.5$, and $k_{ri_\pm}=k_{ri_\mathrm{visc}}\pm k_{ri_\mathrm{cap}}$, where the viscous and capillary limits are given by \eqref{caplim1},\eqref{caplim2},\eqref{visclim1},\eqref{visclim2}. 
The composite expression \eqref{transfun} captures the numerical results with mean relative error of around $\sim1\%$.
Although an even better fit can be attained by allowing N$_{c_t}$ and $\Delta$ to vary with saturation and capillary number, we take them as constants here for the sake of simplicity. 

The transition capillary number N$_{c_t}$ represents the capillary number that lies logarithmically as a midpoint between the viscous and capillary regimes. The parameter $\Delta$ represents one logarithmic folding scale. As we can see in figure \ref{transfig}c,d, the viscous and capillary limits are little more than one folding scale away from the transition capillary number on either side. These two parameters N$_{c_t}$ and $\delta$ fully characterise the flow regime for intermediate capillary numbers, and they are subtly related to the boundary layer thickness discussed earlier. Hence, they are not universal for every scenario, since we have shown that the boundary layer thickness depends on the choice of domain aspect ratio and inlet conditions $s_i$. Therefore, great care must be taken when choosing the domain for upscaling, as we discuss later in Section \ref{seclength}.

Note that, we could have equally fit the data to the capillary number defined in terms of the wetting pressure change (see \eqref{capdef}). However, we observe that the ratio of these pressure changes is
\beq
\frac{\Delta p_n}{\Delta p_w}=\frac{1}{ M   f_0 }\frac{k_{rw_\mathrm{eq}}}{k_{rn_\mathrm{eq}}}.
\eeq
Hence, the two definitions are not independent, and would just result in a different form of \eqref{transfun}. Therefore, without loss of generality, we keep the capillary number defined in terms of non-wetting pressure difference.

Variation in the equivalent capillary pressure \eqref{eqpc} is much less significant, since $p_{c_\mathrm{cap}}/p_{c_\mathrm{visc}}=1.06$. This can be seen in figure \ref{transfig}b, where the capillary and viscous limit curves lie almost on top of each other. Therefore, there is not a great need to model the transition behaviour, and it is sufficient to assume the viscous limit everywhere
\beq
p_{c_\mathrm{eq}}=(1-\bar{s})^{-1/\lambda}.
\eeq
In the next part of the study, we use the equivalent properties derived here to study dynamic flooding in an aquifer.

\section{The Buckley-Leverett problem for heterogeneous media}
\label{bucksec}
\subsection{Problem summary}

Now that we have analytical expressions for the equivalent relative permeabilities in the viscous and capillary limits \eqref{caplim1},\eqref{caplim2},\eqref{visclim1}, \eqref{visclim2}, and a composite expression \eqref{transfun} for intermediate capillary numbers fitted against numerical data, we have a full description of the equivalent properties across all flow conditions. Next, following the classic study of \citet{buckley1942mechanism} for the displacement of immiscible flows in a long-thin aquifer, we extend this to the case of heterogeneous media, using our upscaled equivalent properties.

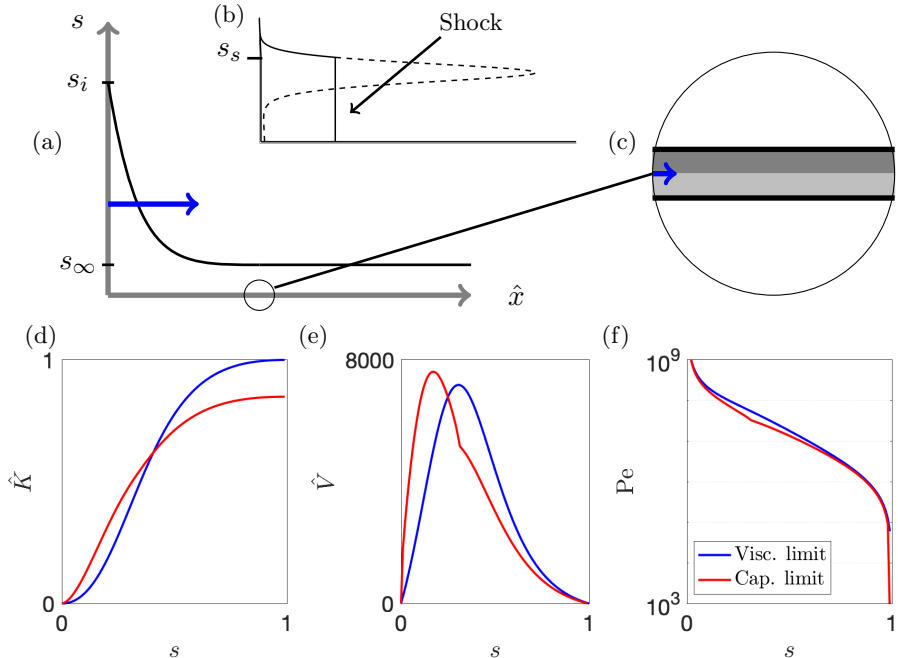
\begin{figure}
\centering
\begin{tikzpicture}[scale=0.4]
\draw[line width=2,->,gray] (0,-1) -- (0,8);
\draw[line width=2,->,gray] (0,-1) -- (12,-1);
\node at (-1,8) {\large ${s}$};
\node at (-1,6) {\large $s_i$};
\node at (-1,0) {\large $s_\infty$};
\draw[line width=1] (-0.2,0) -- (0.2,0);
\draw[line width=1] (-0.2,6) -- (0.2,6);
\node at (13.5,-1) {\large $\hat{x}$};
\draw[line width=1, black]  (0,6) .. controls (1,0) and (2,0) .. (5,0);
\draw[line width=1, black]  (5,0) -- (12,0);
\node at (22,3) {\begin{tikzpicture}[scale=0.32]
\draw[line width=0,lightgray,fill=lightgray]  (0,0) rectangle  (10,1);
\draw[line width=0,lightgray,fill=gray]  (0,1) rectangle  (10,2);
\draw[line width=2] (0,0) -- (10,0);
\draw[line width=2] (0,2) -- (10,2);
\draw[line width=2,blue,->] (0,1) -- (1,1);
\end{tikzpicture}};
\draw (22,3) circle (4cm);
\draw (5,-1) circle (0.5cm);
\draw[line width=1] (5.5,-0.75) -- (18,3);
\draw[line width=2,blue,->] (0,2) -- (3,2);
\node at (10,6) {\includegraphics[width=0.4\textwidth]{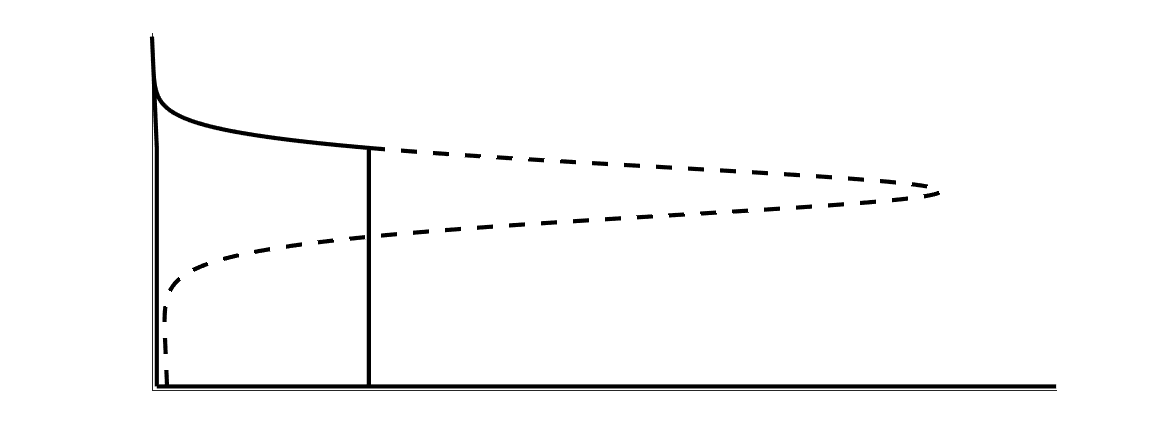}};
\node at (12,8) {Shock};
\draw[line width=1,->] (11,7.5) -- (8,5);
\node at (4,7) {\large $s_s$};
\draw[line width=1] (4.6,6.8) -- (5.1,6.8);
\node at (-2,4) {(a)};
\node at (4,8.2) {(b)};
\node at (17,4) {(c)};
\end{tikzpicture}
\begin{tikzpicture}[scale=0.8]
\node at (0,0.5) {\includegraphics[width=0.3\textwidth]{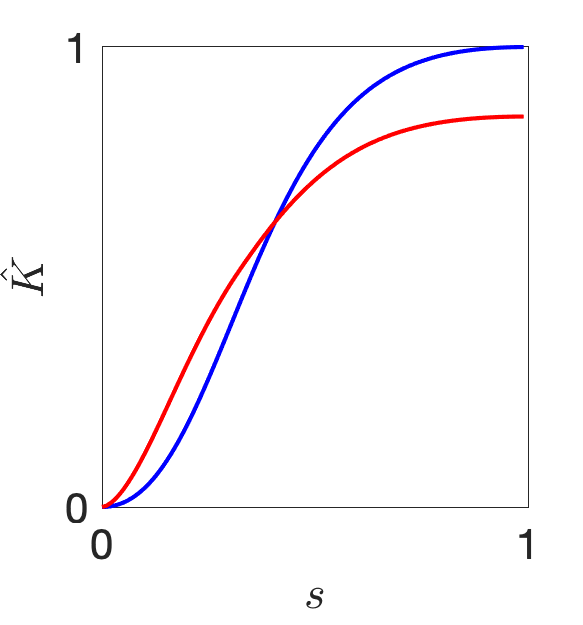}};
\node at (5,0.5) {\includegraphics[width=0.3\textwidth]{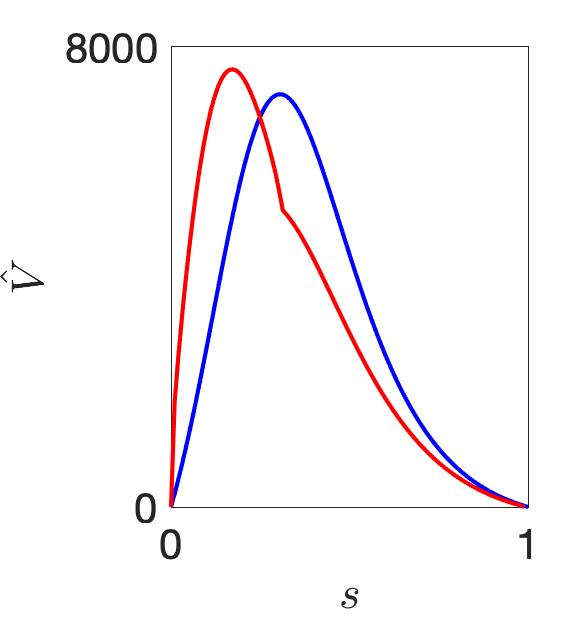}};
\node at (10,0.5) {\includegraphics[width=0.3\textwidth]{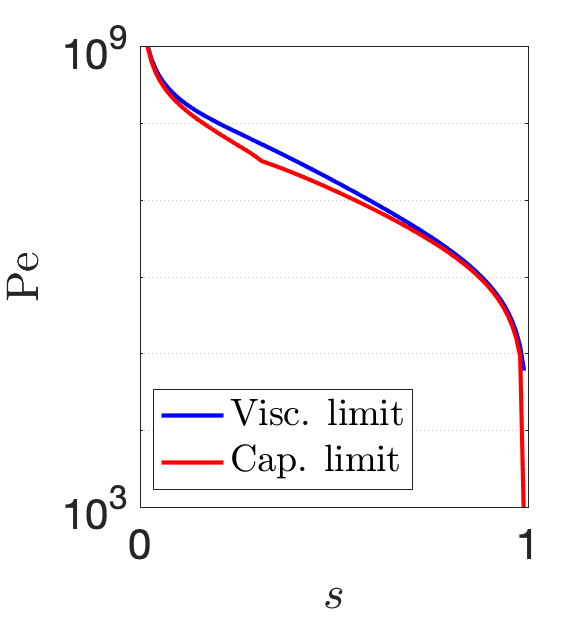}};
\node at (-2,3.2) {(d)};
\node at (2.5,3.2) {(e)};
\node at (7.5,3.2) {(f)};
\end{tikzpicture}
\caption{(a) Illustration of flooding a long, thin aquifer with saturation $s_i$, where the initial saturation was $s_\infty$ (Buckley-Leverett problem). (b) When a multi-valued distribution of saturation develops, a shock forms at saturation $s_s$. (c) Illustration of the underlying heterogeneity in the aquifer. (d,e) Plots of the non-dimensional diffusion and advection coefficients $\hat{K}(s),\hat{V}(s)$ for the capillary and viscous limits. (f) Peclet number Pe$=\hat{V}/\hat{K}$.\label{fig6}}
\end{figure}

In the classic Buckley-Leverett problem, a one-dimensional porous medium, initially filled with a base level saturation $s_\infty$, is flooded with a saturation $s_i$ at the inlet $x=0$ (see figure \ref{fig6}a). Unlike our previous flow study, this problem is time-dependent. However, we make the key assumption that the equivalent properties derived earlier still apply even when the flow is unsteady, which is similar to the approach taken in industrial applications. Our analysis here can be interpreted as the macroscopic flow picture of an aquifer with an underlying heterogeneity, where the length scale of the heterogeneity is much smaller than the flow length scale (see figure \ref{fig6}c).

A complete discussion of the Buckley-Leverett problem can be found in any standard porous media textbook, such as \citep{bear2013dynamics,woods2015flow} for example. Here, we simply summarise the problem and describe how it can be extended to heterogeneous media. In the original problem formulation (for homogeneous media), the governing dimensional equation for the saturation is
\beq
\frac{\partial s}{\partial t}+V(s)\frac{\partial s}{\partial x} = \frac{\partial }{\partial x}\lb K(s)\frac{\partial s}{\partial x}\rb,\label{sgoveq}
\eeq
where the advective and diffusive terms are given by
\begin{align}
V&=V_{tot}\frac{\partial}{\partial s} \left[ \frac{ M   k_{rn}}{ M   k_{rn}+k_{rw}} \right],\label{advective}\\
K&=\frac{k_0 p_{e_0}}{\mu_w}\left[\frac{ M   k_{rn}k_{rw}}{ M   k_{rn}+k_{rw}}\right]\frac{\partial }{\partial s} \lb\frac{p_c}{p_e}\rb,\label{diffusive}
\end{align}
which can be derived by combining \eqref{ge1}-\eqref{ge2}, where $V_{tot}=u_n+u_w$ is the total Darcy flow. Note that we have rescaled time in \eqref{sgoveq} by a factor of $\phi(1-S_{wi})$ for convenience. To extend to heterogeneous media, we replace the relative permeabilities and capillary pressure in \eqref{advective}-\eqref{diffusive} by their equivalent counterparts derived earlier, and the saturation $s$ is interpreted as an upscaled saturation\footnote{Note that in the case where relative permeability depends on the capillary number \eqref{transfun} the advective velocity \eqref{advective} contains a partial derivative with respect to N$_c$. However, due to the logarithmic dependence this contribution is very small (e.g. $\mathcal{O}(10^{-9})-\mathcal{O}(10^{-3})$ for typical parameter values) and so we ignore it. }. Hence, this extension to the Buckley-Leverett problem, though it is one-dimensional, contains information about the vertical variation in the rock and flow properties. Furthermore, the rock heterogeneities only manifest in these upscaled quantities and their typical scalings ($\phi_0,p_{e_0},k_0$).

In figure \ref{fig6}d,e,f we plot the advective and diffusive components, given in non-dimensional terms $\hat{V}=V L\mu_w/k_0p_{e_0}$, $\hat{K}=K\mu_w/k_0p_{e_0}$, for both the capillary and viscous limits. We also plot the nonlinear Peclet number Pe$=\hat{V}/\hat{K}$.
For the purposes of this comparison we define a non-dimensional flow rate 
\beq
\mathcal{U}=\frac{V_{tot}L\mu_w}{k_0p_{e_0}},
\eeq
and we use typical parameter values, giving $\mathcal{U}=3167$ and a viscosity ratio of $M  =30$. A full list of dimensional parameters is given in Table \ref{table1} (taken from the Salt Creek case study, which we discuss later). 

Several observations can be made immediately. Firstly, for these typical parameter values the diffusive term is much smaller than the advective term (indicated by the Peclet number), indicating that the diffusive term can be neglected, except perhaps when saturation gradients are very large (e.g. for shock solutions \citep{woods2015flow}), or when $s$ is very close to 1. 
Secondly, the faster limit (between viscous and capillary) depends on the saturation value.
Finally, the slight kink in the capillary limit advection velocity curve in figure \ref{fig6}e is due to non-smooth changes in saturation distribution due to \eqref{leads2nonneg}.

It is well known that the non-monotone behaviour of $V$ can result in multi-valued saturation distributions, as illustrated in figure \ref{fig6}b. This is often dealt with by introducing a shock at some intermediary saturation $s_s$, where the saturation value is found by solving the equation
\beq
V(s_s)=\frac{J(s_s)-J(s_\infty)}{s_s-s_\infty},\label{shockspeed}
\eeq
in terms of the advective flux $J=\int V \,\mathrm{d}s$ and the initial saturation $s_\infty$. The shock equation \eqref{shockspeed} can be derived by a conservation of mass balance across the shock \citep{woods2015flow}. A typical shock solution is illustrated in figure \ref{fig6}b, where the original multi-valued solution is overlaid as a dashed line. In reality, the steep saturation gradients present in such a shock solution would be softened by the diffusive term \eqref{diffusive} over a growing length scale $\ell\propto(t/$Pe$)^{1/2}$. For typical situations, this results in a diffusive boundary layer of around $1-5\%$ of the total aquifer length.

\begin{figure}
\centering
\begin{tikzpicture}[scale=0.4]
\node at (2,-2.3) {\includegraphics[width=0.5\textwidth]{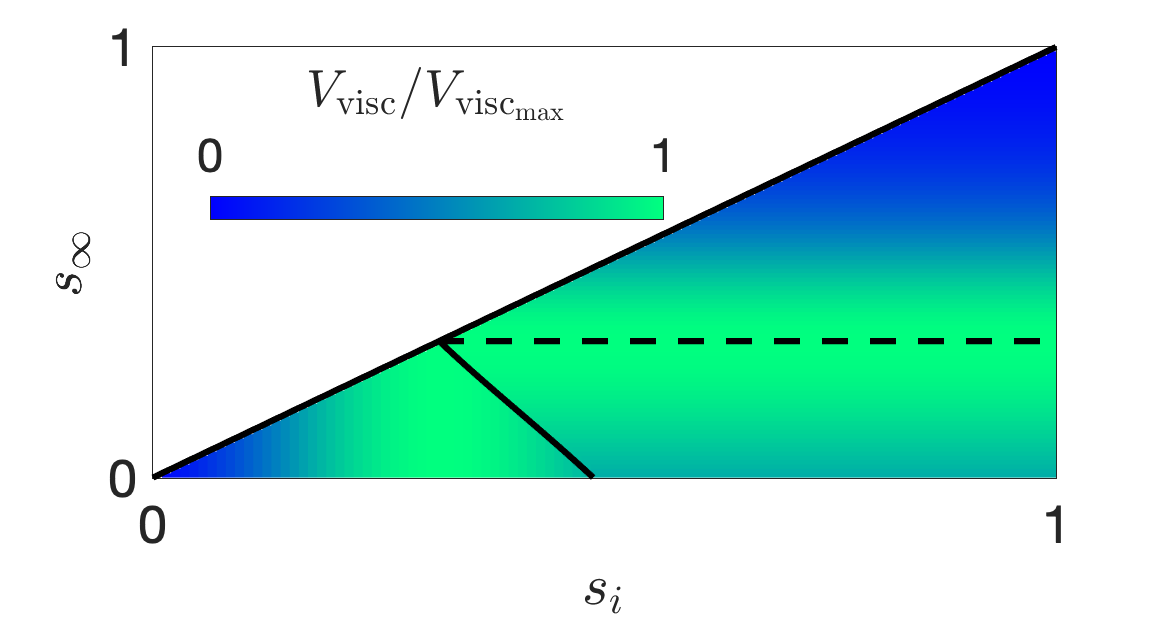}};
\node at (18,-2.3) {\includegraphics[width=0.5\textwidth]{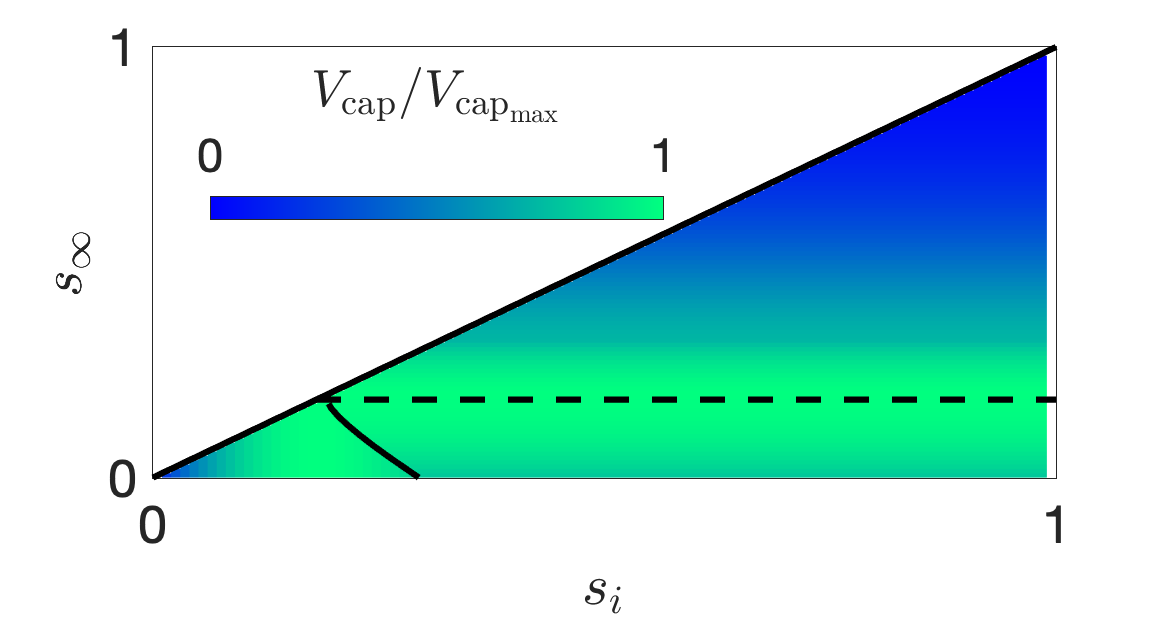}};
\node at (2,-11.5) {\includegraphics[width=0.5\textwidth]{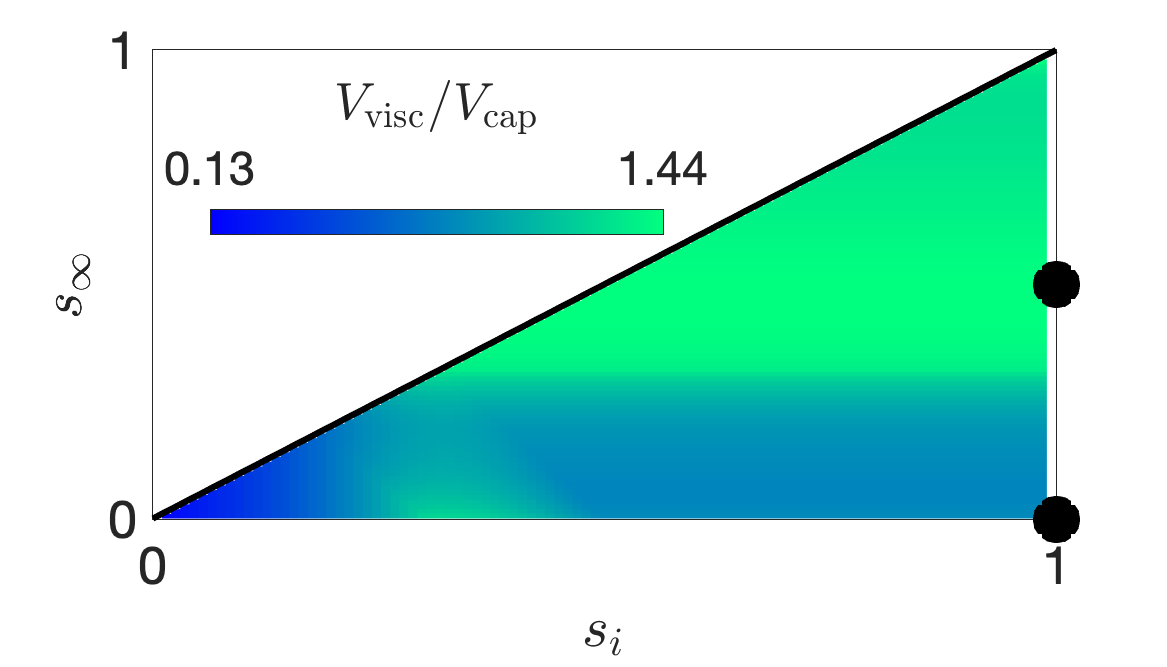}};
\node at (18,-11.5) {\includegraphics[width=0.5\textwidth]{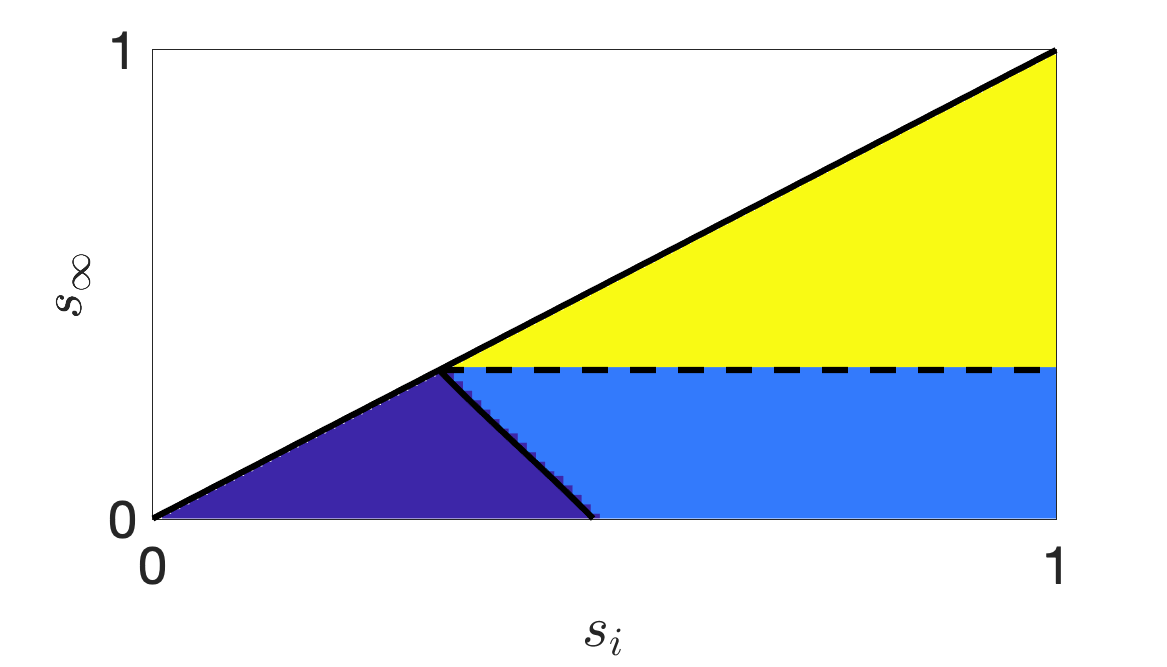}};
\node at (21,-10) {\footnotesize \rotatebox{30}{\bf No shocks}};
\node[blue] at (21.9,-10.5) {\footnotesize \rotatebox{30}{$\boldsymbol{V=V(s_\infty)}$}};
\node[white] at (21.5,-13.7) {\footnotesize $\boldsymbol{V=V(s_s)}$};
\node[white] at (15.5,-13.7) {\footnotesize $\boldsymbol{V=V(s_i)}$};
\node[black] at (25.5,-4.) {${s_m}$};
\node[black] at (9.5,-3.) {${s_m}$};
\node[black] at (25.5,-12) {${s_m}$};
\node[black] at (7,-10.5) {\ref{fig8}a,c,e};
\node[black] at (7,-13.5) {\ref{fig8}b,d,f};
\node at (18.5,-12.5) {\footnotesize \bf Shocks};
\node at (-5.5,2.5) {(a)};
\node at (10,2.5) {(b)};
\node at (-5.5,-6.5) {(c)};
\node at (10,-6.5) {(d)};
\end{tikzpicture}
\caption{ Advection coefficient colormaps for the extended Buckley-Leverett problem in the viscous (a) and capillary (b) limits for all possible values of inlet and initial saturation $s_i$, $s_\infty$ (normalised by their maximum value). (c) Ratio between advection coefficients in viscous and capillary limits (note the different colour scale). Two markers indicate the solutions in figure \ref{fig8}. (d)  Phase diagram illustrating different parameter regimes, indicating front speed definition, the stationary point $s_m$, and where shocks occur.\label{fig7}}
\end{figure}
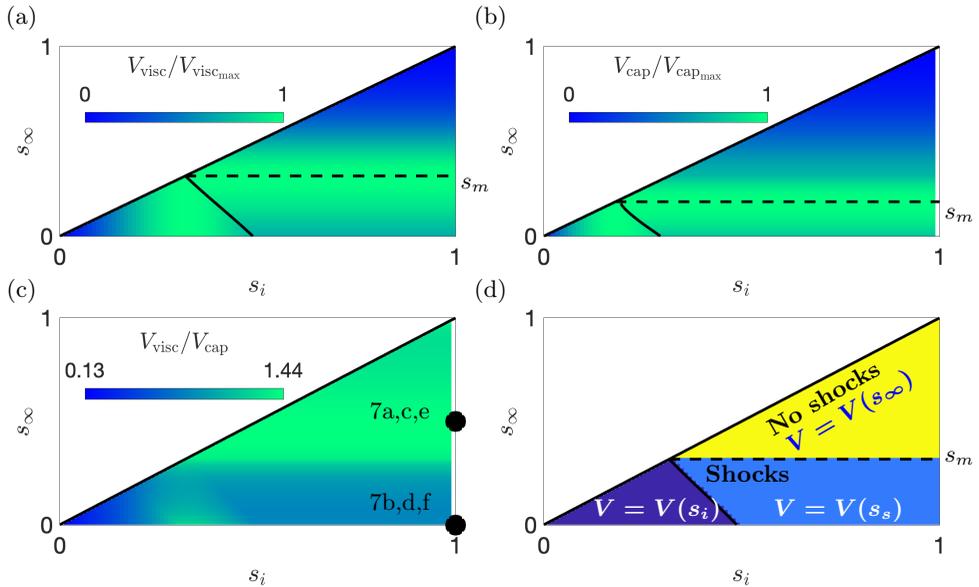

The solution behaviour of the Buckley-Leverett problem is characterised by several saturation values: the inlet saturation $s_i$, the initial far-field saturation $s_\infty$ and, should a shock develop, the shock saturation $s_s$. 
Since we restrict our attention to drainage flows (e.g. CO$_2$ driving out water), we confine our analysis to $s_i>s_\infty$.
To understand the different flow regimes, it is useful to introduce the stationary point saturation value $s_m$, which corresponds to the saturation at which the maximum advection velocity is achieved (e.g. see figure \ref{fig6}e). A multivalued saturation profile never develops (i.e. no shocks) for parameter values $s_m\leq s_\infty \leq s_i$, as illustrated by a yellow region in the phase diagram in figure \ref{fig7}d. Hence, in the absence of shocks, the flooding front moves at the far-field saturation speed, which is $V=V(s_\infty)$. Likewise, a shock will always develop for $s_\infty\leq s_m \leq s_i$, and the flooding front moves at the shock speed $V=V(s_s)$.

We note that \eqref{shockspeed} may result in a shock saturation value that lies outside of the range $[s_\infty,s_i]$. Therefore, in such cases \eqref{shockspeed} is replaced by the condition $s_s=s_i$, such that the shock value is simply equal to the inlet value, as illustrated with dark blue colouring in figure \ref{fig7}d.

\subsection{Viscous and capillary limits}

Now that we have summarised the Buckley-Leverett problem, the next step is to discuss the two limiting viscous and capillary cases.
In figure \ref{fig7}a,b we display a colour plot of the front velocity values for each of these limits $V_\mathrm{visc}$, $V_\mathrm{cap}$ (normalised by their maximum value) over all possible values of $s_i$, $s_\infty$. In figure \ref{fig7}c we plot the ratio between these two limits $V_\mathrm{visc}/V_\mathrm{cap}$.
Wherever the far-field saturation is larger than the stationary point $s_\infty> s_m$, viscous advection speeds dominate, whereas in regions with $s_\infty$ near zero (leading to shocks), capillary advection speeds dominate. The maximum and minimum values of the speed ratio $V_\mathrm{visc}/V_\mathrm{cap}$ are $1.44$ and $0.13$, indicating that neglecting heterogeneities at small capillary number may lead to substantial error in flooding predictions.
For modelling carbon sequestration, where $s_\infty$ is expected to be near-zero (CO$_2$ is typically injected into brine-saturated aquifers),
the implications are that in situations where the capillary number is small, heterogeneities cause an overall acceleration of the advancing front. This will play an important role in trapping mechanisms and storage efficiency.

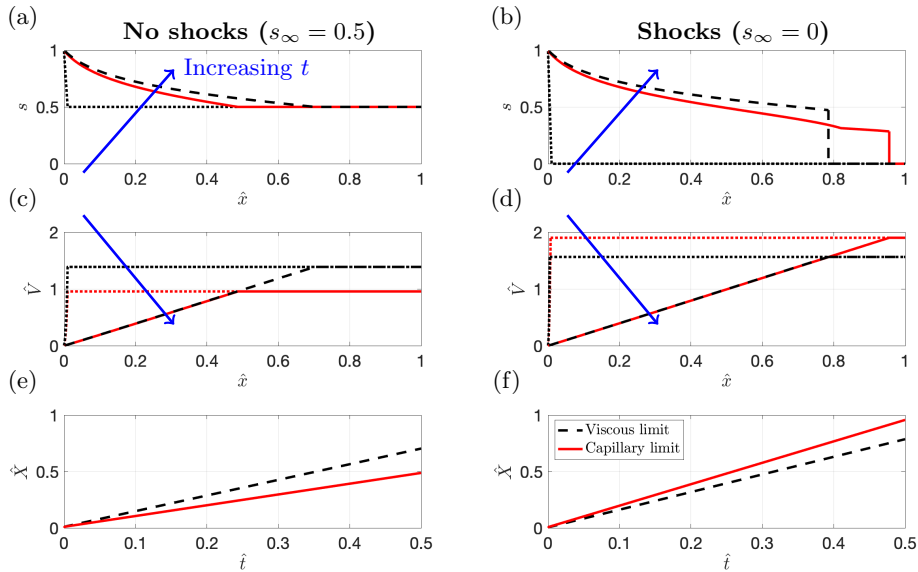
\begin{figure}
\centering
\begin{tikzpicture}[scale=0.8]
\node at (0,0) {\includegraphics[width=0.45\textwidth]{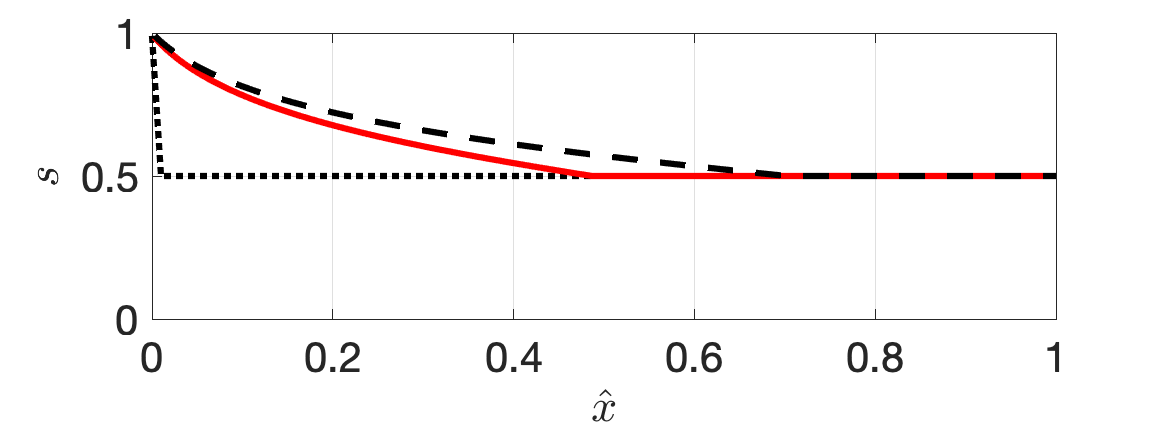}};
\node at (8,0) {\includegraphics[width=0.45\textwidth]{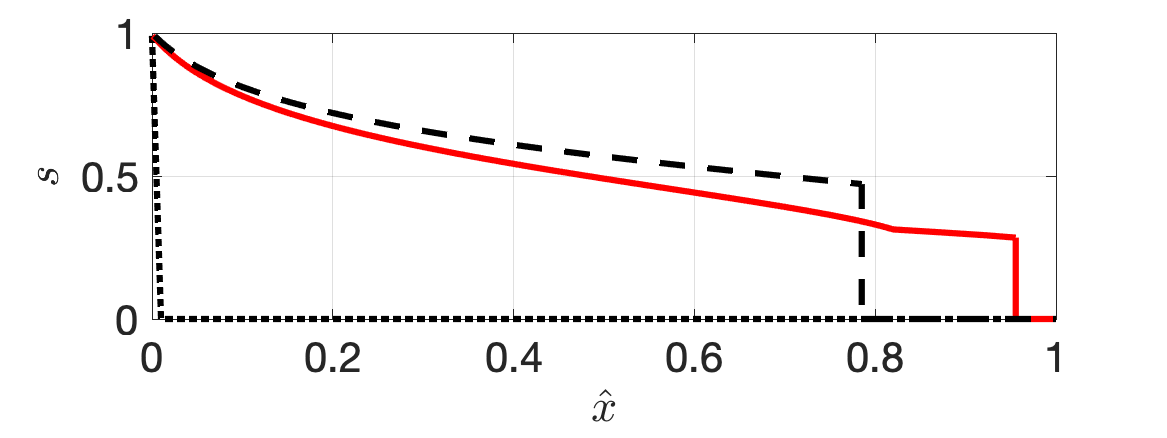}};
\node at (0,-3) {\includegraphics[width=0.45\textwidth]{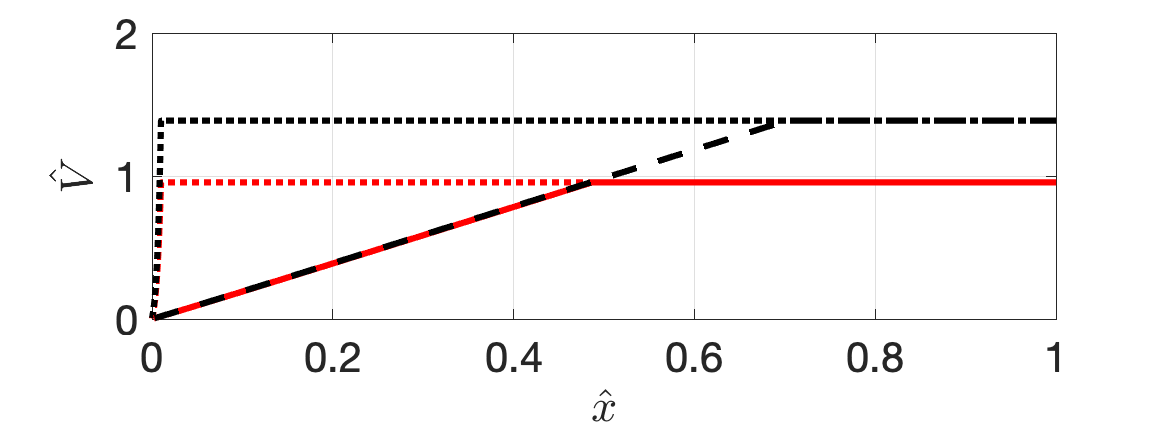}};
\node at (8,-3) {\includegraphics[width=0.45\textwidth]{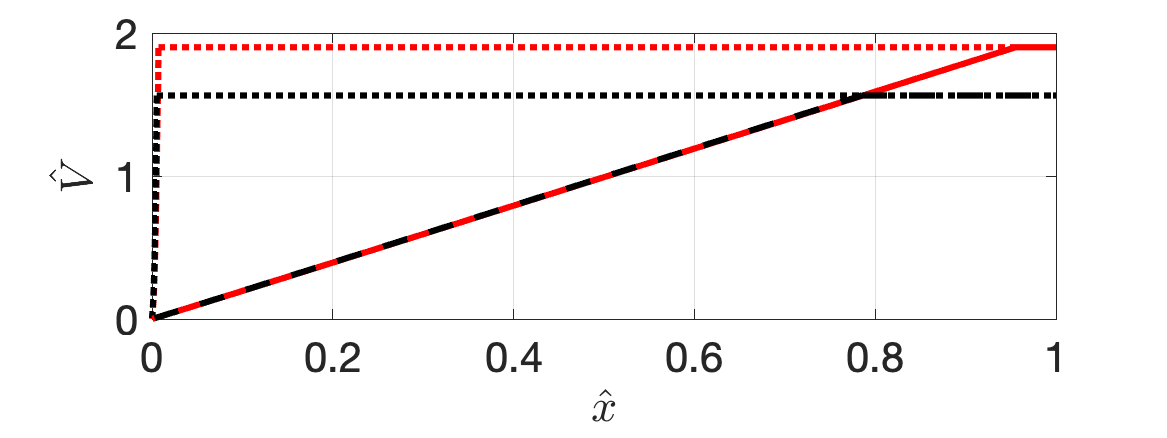}};
\node at (0,-6) {\includegraphics[width=0.45\textwidth]{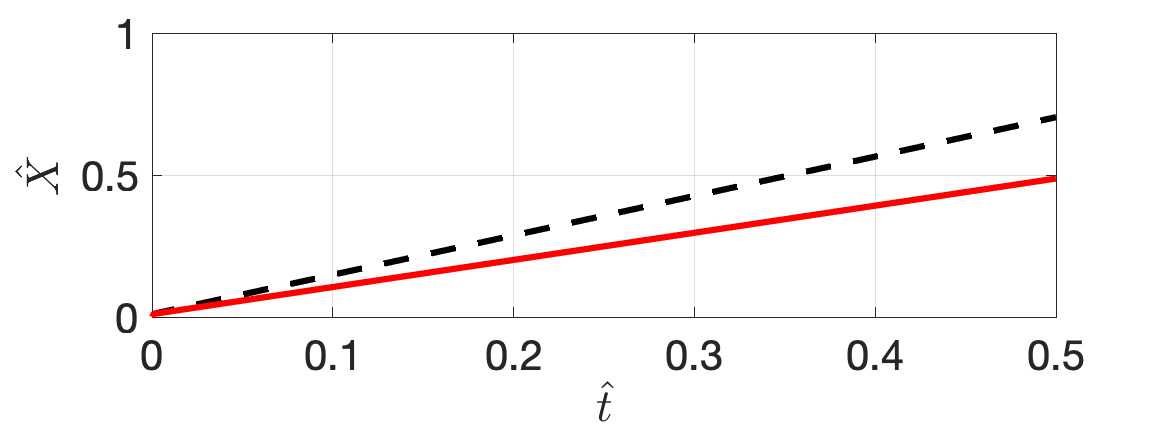}};
\node at (8,-6) {\includegraphics[width=0.45\textwidth]{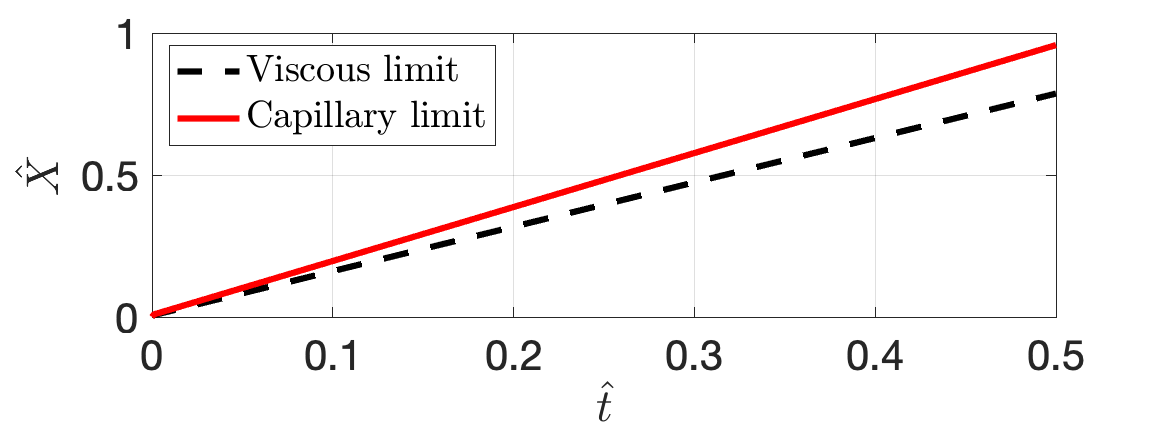}};
\node at (0.25,1.5) {\bf No shocks ($s_\infty=0.5$)};
\node at (8.25,1.5) {\bf Shocks ($s_\infty=0$)};
\draw[line width=1,->,blue] (-2.5,-0.8) -- (-1.0,0.9); 
\node[blue] at (0.2,0.9) {Increasing $t$};
\draw[line width=1,->,blue] (5.5,-0.8) -- (7.0,0.9); 
\draw[line width=1,->,blue] (5.5,-1.5) -- (7,-3.3); 
\draw[line width=1,->,blue] (-2.5,-1.5) -- (-1.0,-3.3);
\node at (-3.5,1.75) {(a)};
\node at (4.5,1.75) {(b)};
\node at (-3.5,-1.25) {(c)};
\node at (4.5,-1.25) {(d)};
\node at (-3.5,-4.25) {(e)};
\node at (4.5,-4.25) {(f)};
\end{tikzpicture}
\caption{Examples of flooding of an aquifer in capillary and viscous limits with and without shocks present. We display plots at $\hat{t}=0$ and $\hat{t}=0.5$ of (a,b) the saturation $s$ and (c,d) the advective velocity $\hat{V}$. In (e,f) we show the evolution of the front position $\hat{X}$. In both cases we set $s_i=1$. \label{fig8} }
\end{figure}

To illustrate these findings, in figure \ref{fig8} we display two solutions to the extended Buckley-Leverett problem with and without shocks. In the first case (a,c,e), we flood an aquifer which is initially saturated with a substantial fraction of gas $s_\infty=0.5$. In the second case (b,d,f), the aquifer is initially saturated with the minimum possible gas amount $s_\infty=0$ (see the markers in figure \ref{fig7}c), causing a shock to develop.
In each case we plot the saturation $s$ and velocity $V$ at both the initial time, and at a single later time, indicating both capillary (red) and viscous (black) predictions. 
We display all results in non-dimensional form, where a suitable non-dimensional timescale is
\beq
T=L/V_{tot}.\label{tscale}
\eeq
The saturation profiles are obtained by solving the characteristic equation for each $x$ value, such that
\beq
\frac{\mathrm{d}x}{\mathrm{d}t}=V(s),
\eeq
where the saturation value $s$ is conserved along characteristics. As initial conditions, we use a localised initial saturation distribution 
\beq
s(x,0)=\begin{cases}
s_i-(s_i-s_\infty){x}/{x^*}:&\quad 0\leq x \leq x^*,\\
s_\infty:&\quad x^*<x\leq L,\\
\end{cases}\label{initdata}
\eeq
where $x^*/L=10^{-3}$. In figure \ref{fig8} this initial saturation profile is advected according to either the capillary or viscous limit speed, $V_\mathrm{cap}(s)$ or $V_\mathrm{visc}(s)$ (c,d). In (e,f) we also plot the position of the leading edge of the flood $X(t)$, which increases linearly with time, with slope $V=V(s_\infty)$ or $V(s_s)$. 
The speed ratio is $V_\mathrm{visc}/V_\mathrm{cap}=1.44$ in the case without shocks, and $V_\mathrm{visc}/V_\mathrm{cap}=0.82$ in the case with shocks. For applications such as CO$_2$ sequestration, this indicates that a model which neglects the effects of heterogeneities may predict flooding speeds with nearly 50$\%$ inaccuracy.

Most flows will develop with behaviour intermediate to the viscous and capillary limits.
The flow behaviour should therefore depend on the local capillary number,
which changes with local pressure gradients according to
\beq
\mathrm{N}_c=\frac{H}{\Delta p_e} \left|\frac{\partial p_n}{\partial x}\right|,\label{newcapdef}
\eeq
where we have used the definition in terms of the non-wetting pressure gradient. 
The local pressure gradients are given by 
\beq
\frac{\partial p_n}{\partial x}=-\frac{V_{tot}\mu_w}{k_0}  \left[\frac{1 }{ M   k_{rn}+k_{rw}} \right].\label{pressgrad}
\eeq
We note that the capillary number used here \eqref{newcapdef} is defined differently to \eqref{capdef}, which was used to perform steady-state upscaling earlier. However, \eqref{newcapdef} can be interpreted as the local capillary number for a macroscopic flow description, whereas \eqref{capdef} can be interpreted as the bulk capillary number for a small-scale study. Hence, the two definitions become equivalent by zooming in or out of the aquifer appropriately. 

Since the pressure gradient \eqref{pressgrad}, and consequently the capillary number, are both functions of $s$, they are conserved along characteristics. Hence, the capillary number at the flooding front $x=X(t)$ is the same for all time (though different to the capillary number at the inlet $x=0$, for example).

To calculate the flooding speed \eqref{advective}, which depends on the capillary number via \eqref{transfun}, the nonlinear implicit equation 
\beq
\mathrm{N}_c =  \frac{ \mathcal{U}\delta}{ \sigma_P}\left[\frac{1}{ M   k_{rn_\mathrm{eq}}(s,\mathrm{N}_c)+k_{rw_\mathrm{eq}}(s,\mathrm{N}_c)}\right],
\eeq
must be solved for N$_c$, where $s$ is set as either $s_\infty$ or $s_s$ (depending on whether shocks are present). 

Interestingly, if one were to consider an axisymmetric flooding instead of two-dimensional plane flooding, the flow speed and pressure gradients would decay radially due to conservation of mass. Hence, the capillary number would also decay radially, such that different regions of the aquifer switch between viscous and capillary limits over time. 

An axisymmetric model is more realistic than the two-dimensional case for cases where injection occurs at a single point source, as is often the case in industry. 
In the context of our current modelling approach heterogeneities are below the continuum scale, and consequently the equivalent relative permeabilities derived in Section \ref{upscale} can equally be used to describe a two-dimensional (as above) or axisymmetric setting.
Hence, in the next section we extend the above analysis to axisymmetric flow.

\subsection{Axisymmetric flooding}

\begin{figure}
\centering
\begin{tikzpicture}[scale=0.8]
\node at (4,-3.5) {\includegraphics[width=0.95\textwidth]{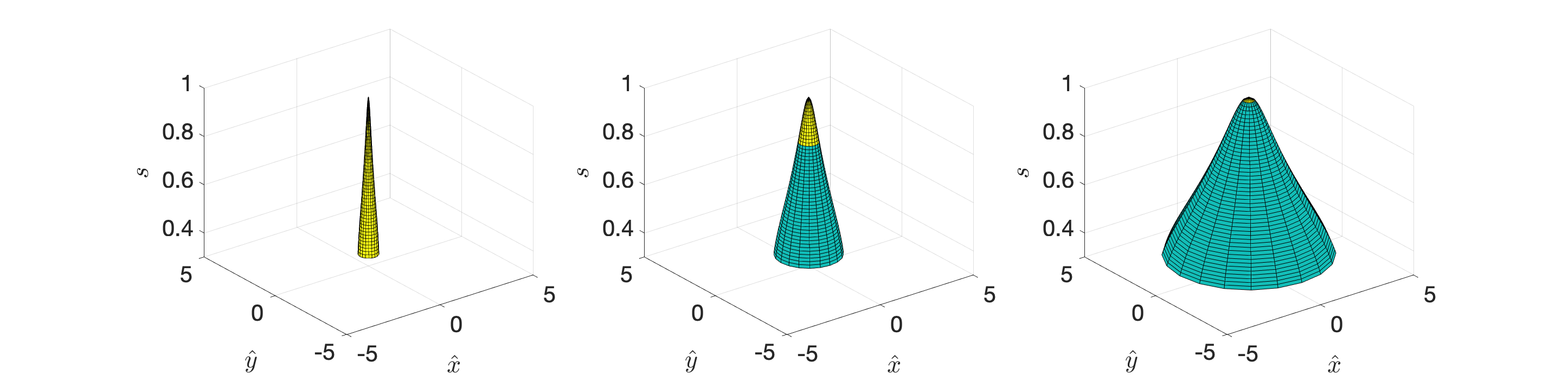}};
\node at (0,0) {\includegraphics[width=0.45\textwidth]{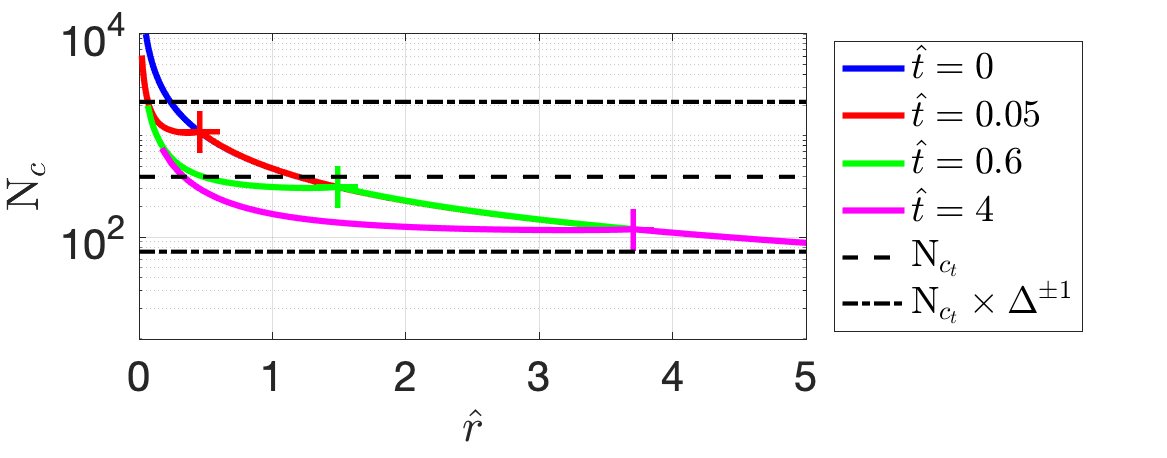}};
\node at (8,0) {\includegraphics[width=0.45\textwidth]{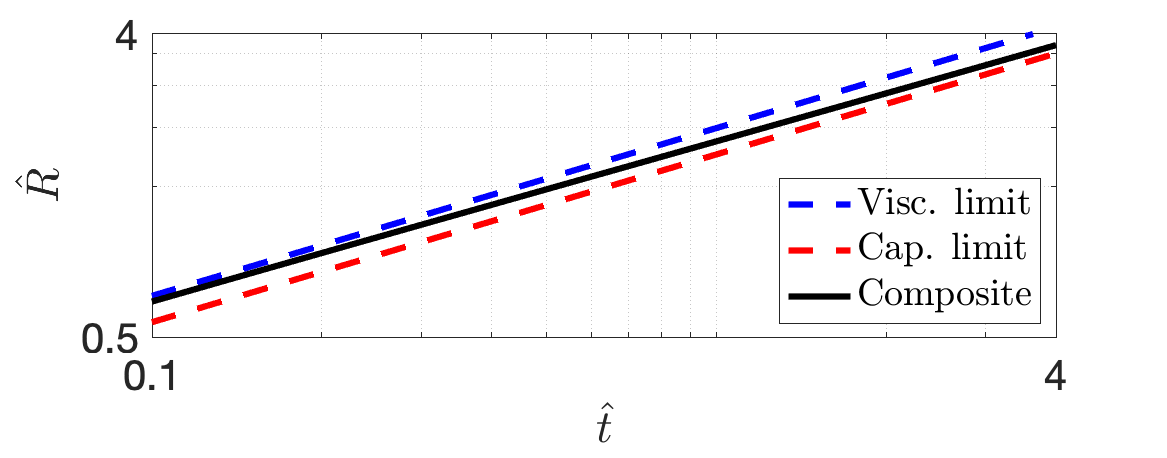}};
\node at (-0,-2) { $\boldsymbol{\hat{t}=0.05}$};
\node at (4.5,-2) { $\boldsymbol{\hat{t}=0.6}$};
\node at (9,-2) { $\boldsymbol{\hat{t}=4}$};
\node at (-3.7,1.3) {(a)};
\node at (4.4,1.3) {(b)};
\node at (-3.7,-1.5) {(c)};
\end{tikzpicture}
\caption{Axisymmetric flooding of an aquifer in the case of no shocks ($s_i=1,\,s_\infty=0.35$) using composite expressions \eqref{transfun} for the equivalent relative permeabilities. (a)  Radial variation in the capillary number at different times, illustrating the front positions as markers, and the transition capillary number N$_{c_t}$ (from Section \ref{secint}) with dotted lines. (b) Logarithmic plot of front position $\hat{R}$, evolving like the square root of time, also illustrating the viscous and capillary limits. (c) Surface plots of the axial spread of saturation at different times.   \label{fig9}}
\end{figure}
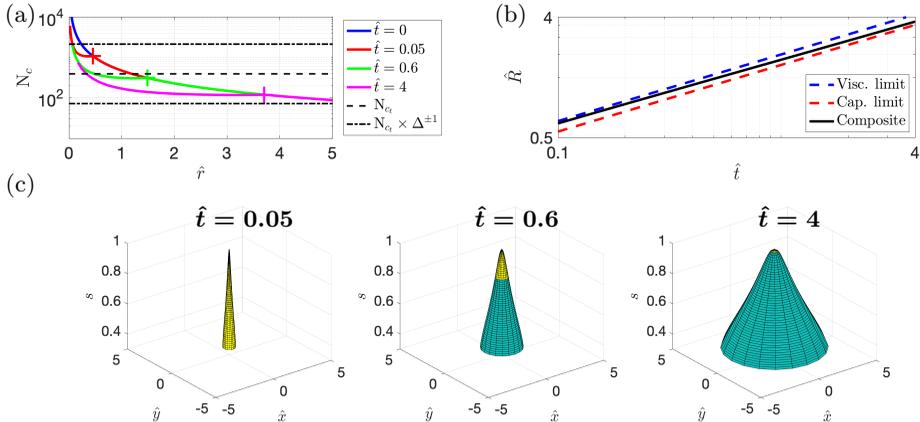

During axisymmetric flooding, the governing equation for the saturation is
\beq
\frac{\partial s}{\partial t}+\frac{Q(s)}{r}\frac{\partial s}{\partial r} = r\frac{\partial }{\partial r}\left[ \frac{K(s)}{r}\frac{\partial s}{\partial r}\right],
\eeq
where the advective and diffusive terms are the same as before \eqref{advective}-\eqref{diffusive}, except we have replaced $V(s)$ by $Q(s)$, which has an extra dimension of length. By the same argument as above, we neglect the diffusive term. In this case, the characteristic equation is
\beq
\frac{\mathrm{d}r}{\mathrm{d}t}=\frac{Q(s)}{r},
\eeq
which can be re-written as
\beq
\frac{\mathrm{d}}{\mathrm{d}t}\lb\frac{1}{2}r^2\rb=Q(s).\label{Uradeq}
\eeq
The pressure gradients are given by
\beq
\frac{\partial p_n}{\partial r}=-\frac{Q_{tot}\mu_w}{k_0} \frac{1}{r}\left[ \frac{1 }{ M   k_{rn_\mathrm{eq}}(s,\mathrm{N}_c)+k_{rw_\mathrm{eq}}(s,\mathrm{N}_c)} \right].\label{Rpressgrad}
\eeq
which are no longer constant along characteristics (since \eqref{Rpressgrad} contains $r$), and so the capillary number (now defined in terms of ${\partial p_n}/{\partial r}$) must be calculated at each radial value.
Hence, given some initial data for $s$, such as \eqref{initdata}, the solution is found by time-integrating the coupled system 
\begin{align}
\frac{\mathrm{d}}{\mathrm{d}t}\lb\frac{1}{2}r^2\rb&=Q_{tot}\frac{\partial}{\partial s} \left[ \frac{ M   k_{rn_\mathrm{eq}}(s,\mathrm{N}_c)}{ M   k_{rn_\mathrm{eq}}(s,\mathrm{N}_c)+k_{rw_\mathrm{eq}}(s,\mathrm{N}_c)} \right],\label{req1}\\
\mathrm{N}_c&=\frac{ \mathcal{Q}\delta}{ \sigma_P}  \frac{L}{r}\left[ \frac{1 }{ M   k_{rn_\mathrm{eq}}(s,\mathrm{N}_c)+k_{rw_\mathrm{eq}}(s,\mathrm{N}_c)} \right],\label{req2}
\end{align}
where $\mathcal{Q}=Q_{tot}\mu_w/k_0p_{e_0}$. In figure \ref{fig9} we display solutions to \eqref{req1}-\eqref{req2} using the parameters $s_i=1$ and $s_\infty=0.35$ (i.e. no shocks). In figure \ref{fig9}a we display the capillary number N$_c(r,t)$ at four times, which decays like $\sim1/r$ as $r\rightarrow \infty$. In figure \ref{fig9}b we display the position of the flooding front $R(t)$ for each case, also indicating the capillary and viscous limit predictions for comparison.

Unlike the two-dimensional case, here the front moves like the square root of time (instead of linearly). Also, unlike the two-dimensional case, the capillary number at the the flooding front changes over time. At early times, the entire flow is close to the viscous limit, whereas at late times, nearly all the flow is close to the capillary limit, except for a small region near the origin. At intermediate times the flow straddles between the two limits. This can be seen in figure \ref{fig9}b, where the front evolution switches between viscous-like behaviour to capillary-like behaviour over time.

We also display surface plots of the saturation at different times in figure \ref{fig9}c. The colouring in each plot is chosen as a binary value depending on whether the local capillary number is above or below the transition value N$_{c_t}$ (see also figure \ref{fig9}a, where one folding scale is illustrated). 
The result is that the flow near the source is in the viscous limit, and is consequently unaffected by heterogeneities. However, as the flood spreads through the aquifer the heterogeneities play a strong role far away from the origin. The overall effect is a deceleration, driven largely at the leading edge of the injection. Note that if we were to choose a smaller value of the far-field saturation, such as $s_\infty=0$, a shock would develop and the advection speed in the capillary limit would be faster than that of the viscous limit (as in figure \ref{fig8}b,d,f).

Similarly to the two-dimensional case, by neglecting the effects of heterogeneities, flooding speeds can be misrepresented by as much as 50$\%$. Therefore, for applications in CO$_2$ sequestration, modelling the transition of the flow between the viscous and capillary limits is critical for accurately predicting how far the injection has spread, and this is important both from safety and efficiency perspectives.

\section{Comparisons with experimental data}

In this section we compare some of our results to different sources of experimental data from other authors. Firstly, we compare the results of our steady state upscaling from Section \ref{upscale} to some X-ray CT scan experiments. Then, we compare our dynamic predictions from Section \ref{bucksec} to field measurements from a CO$_2$ injection experiment in Salt Creek, USA.

\subsection{Steady state upscaling}

\begin{figure}
\centering
\begin{tikzpicture}[scale=0.8]
\node at (0,0) {\includegraphics[width=0.45\textwidth]{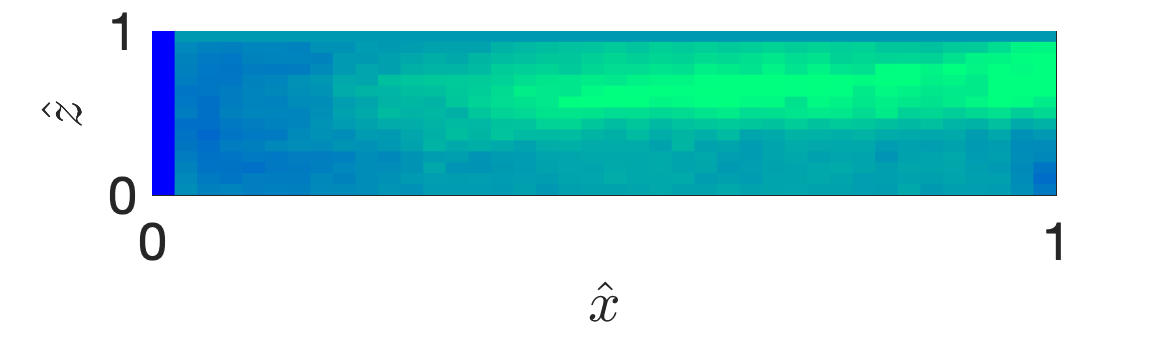}};
\node at (8,0) {\includegraphics[width=0.45\textwidth]{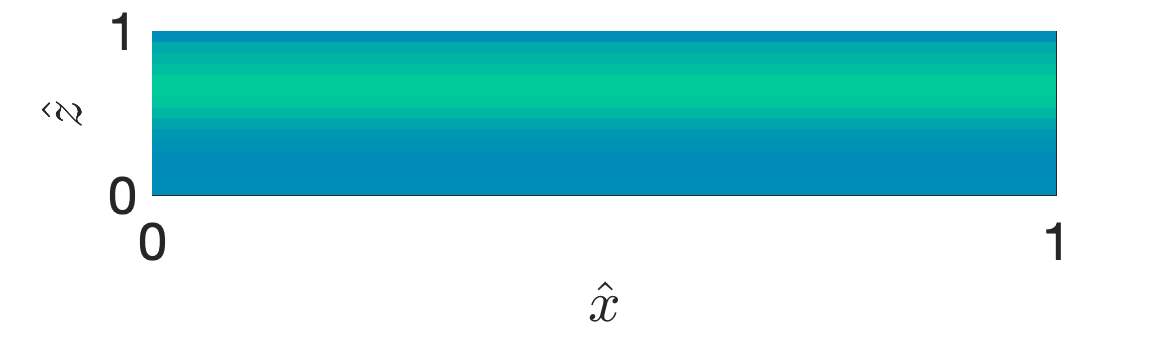}};
\node at (4,-1.5) {\includegraphics[width=0.4\textwidth]{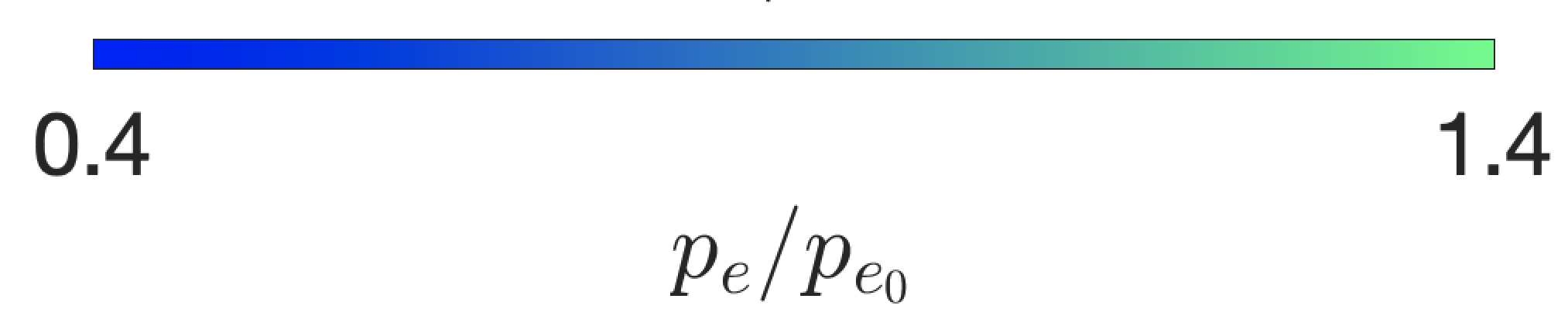}};
\node at (0.2,1.2) {Bentheimer sandstone};
\node at (8.2,1.2) {Transverse/vertical average};
\node at (-4,1) {(a)};
\node at (4,1) {(b)};
\node at (0,-4.75) {\includegraphics[width=0.45\textwidth]{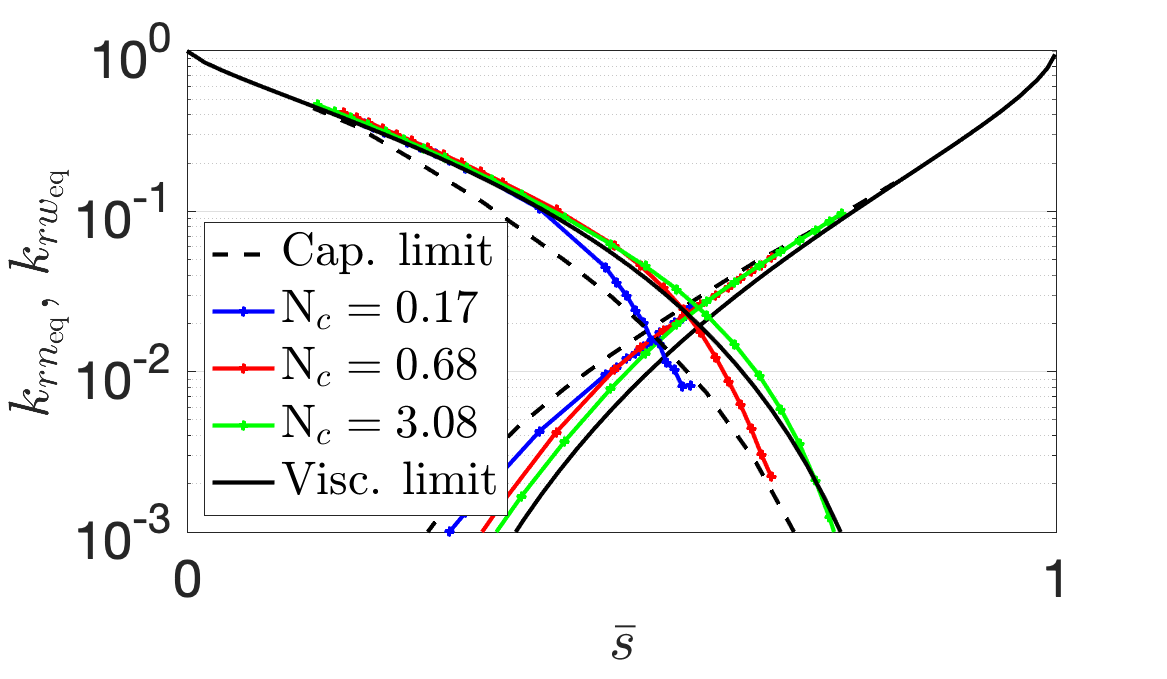}};
\node at (8,-4.75) {\includegraphics[width=0.45\textwidth]{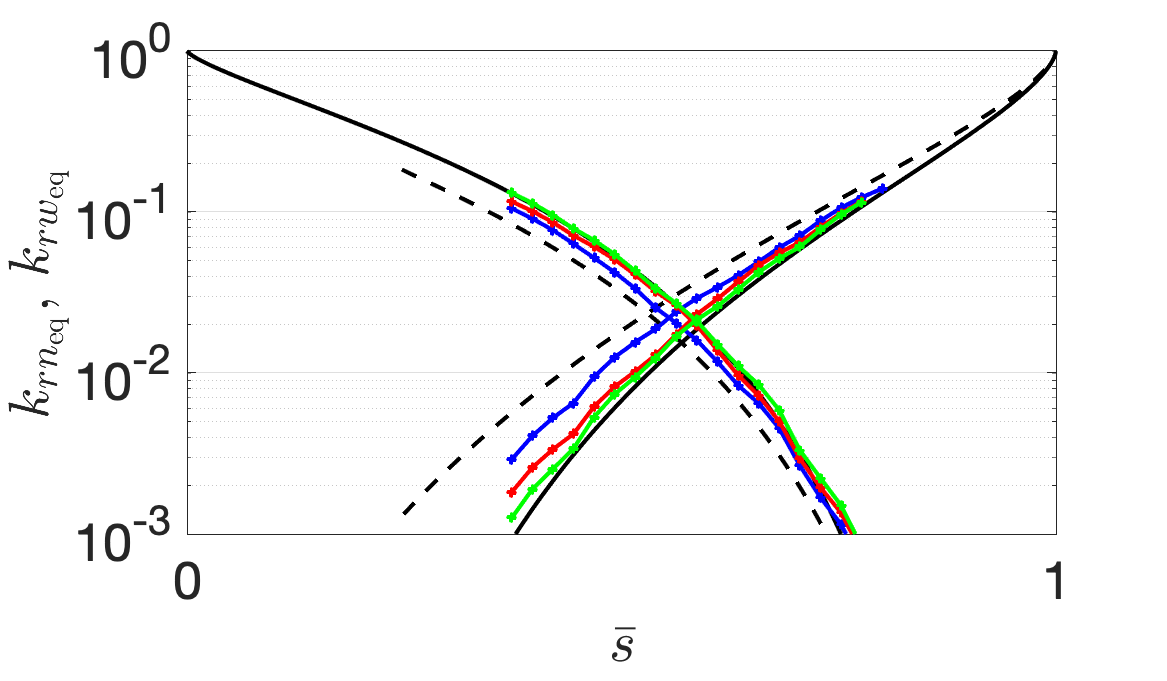}};
\node at (0.25,-2.5) {Jackson \textit{et al.} (2018)};
\node at (8.25,-2.5) {Present study};
\node at (-4,-2.75) {(c)};
\node at (4,-2.75) {(d)};
\end{tikzpicture}
\caption{(a) Colour map of a two-dimensional slice of the capillary heterogeneity $p_e(x,y,z)$, derived by \citet{jackson2018characterizing}, for a core of Bentheimer sandstone. (b) Transverse/vertical average of (a) $p_e(z)$. (c,d) Comparison of the equivalent relative permeabilities $k_{rn_\mathrm{eq}}$, $k_{rw_\mathrm{eq}}$ over a range of capillary numbers, also showing the viscous and capillary limits. \label{bentheimer}}
\end{figure}
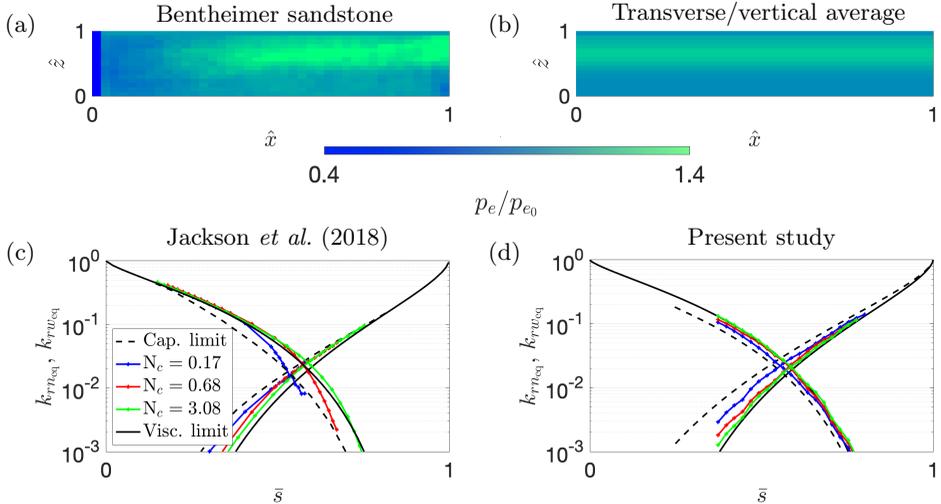

We now quantitatively compare our results to data taken from core flooding experiments. 
The recent study of \citet{jackson2018characterizing} calculates equivalent relative permeabilities using X-ray CT scans of Bentheimer sandstone with parallel layers \citep{peksa2015bentheimer}. 
Their analysis provides a three-dimensional map of the pore entry pressure in a rock core, a two-dimensional slice of which is illustrated in figure \ref{bentheimer}a. 
To upscale the observed heterogeneities, the intrinsic relative permeabilities $k_{ri}$ were first approximated by fitting the empirical relationship proposed by \citet{chierici1984novel}, which is given explicitly by \eqref{cher1}-\eqref{cher2}, to CT scans at very high capillary number. Then, a set of experiments at very low capillary number were used to iteratively fit a numerical model of the core to experimentally observed saturation data. 
A full list of the parameter values is given in Appendix \ref{appA}.

Unlike their three-dimensional data, heterogeneities discussed here depend on the vertical dimension alone. Therefore, we take an average of the experimental data, $p_e(z)=\int\int p_{e_\mathrm{exp}}(\hat{x},\hat{y},\hat{z})\,\mathrm{d}\hat{x}\mathrm{d}\hat{y}$, which is illustrated in figure \ref{bentheimer}b. 
Evidently, the experimental rock core has some longitudinal variation, so we do not expect our comparison to be perfect. However, a good approximation should be attained, since the layering is predominantly parallel to the flow.

To compare with these experiments, we start with the two viscous and capillary limiting cases, since all other cases must lie between these. The capillary and viscous limits derived by Jackson \textit{et al.} are displayed in figure \ref{bentheimer}c. 
Spatial variation in the permeability $k$ is not provided, so we fit the our power law relationship \eqref{powerlaw} against their capillary limit data, giving $B=1/10$, with a mean relative error of $23\%$, which is most likely attributed to our approximation of heterogeneity by a simple vertical variation. 
For each of the pore entry pressure and permeability, we calculate the standard deviation divided by the mean, giving $\sigma(p_e)/\mu(p_e)=0.16$ (which is the same as quoted by Jackson \textit{et al.}) and $\sigma(k)/\mu(k)=0.74$ (which is similar to field observations from Salt Creek, discussed later).

The next step is to compare equivalent relative permeabilities for intermediate capillary numbers.
To do so, we use our numerical simulations, as described earlier.
Our calculated equivalent relative permeabilities are shown in figure \ref{bentheimer}d, compared against the data of \citet{jackson2018characterizing} in figure \ref{bentheimer}c. Each coloured line on the plot has the same  value of the total Darcy flow $U_{tot}=U_n+U_w$ and different values of the flow fraction $ f_0 =U_w/U_n$. 
Consequently, the capillary number varies greatly over one value of $U_{tot}$ and so, following Jackson \textit{et al.}, we quote the value at $ f_0 =0.5$.
To ensure that the quoted capillary numbers are the same, we use the same definition as Jackson \textit{et al.} for the capillary number, where the pressure change in \eqref{capdef} is over the whole core. 
Overall, the comparison is good, with our data points varying between the viscous and capillary limits in a similar manner to Jackson \textit{et al.} However, the slight differences in the curve shapes are most likely attributed to our one-dimensional approximation of the heterogeneity.


\subsection{Dynamic flooding}

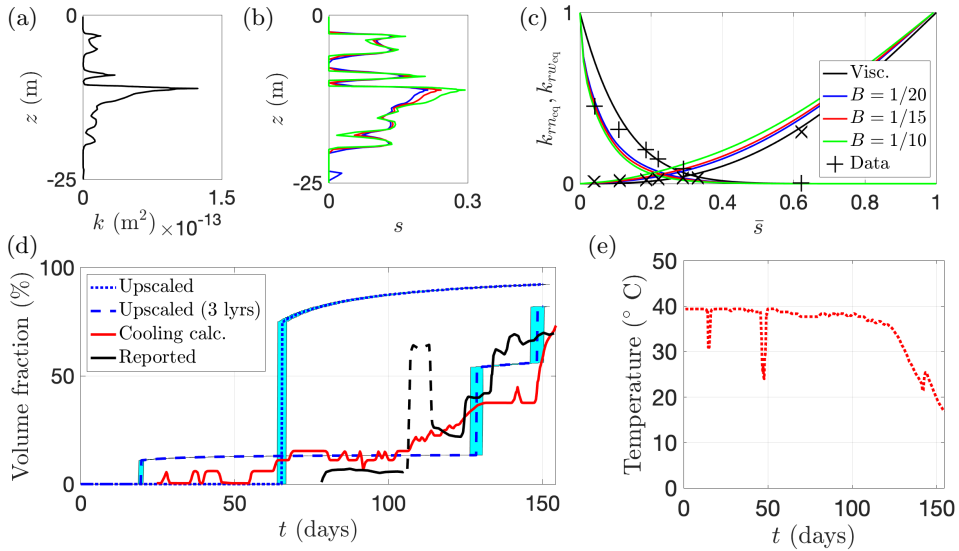
\begin{figure}
\centering
\begin{tikzpicture}[scale=1.3]
\node at (-1.3,0.3) {\includegraphics[width=0.22\textwidth]{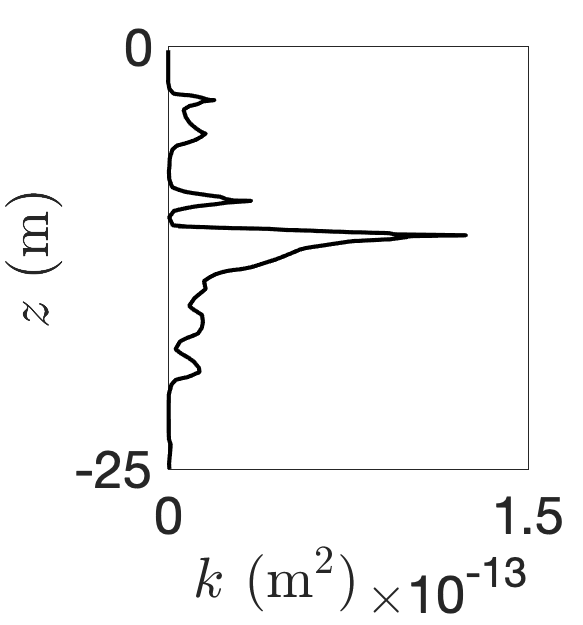}};
\node at (1.2,0.3) {\includegraphics[width=0.22\textwidth]{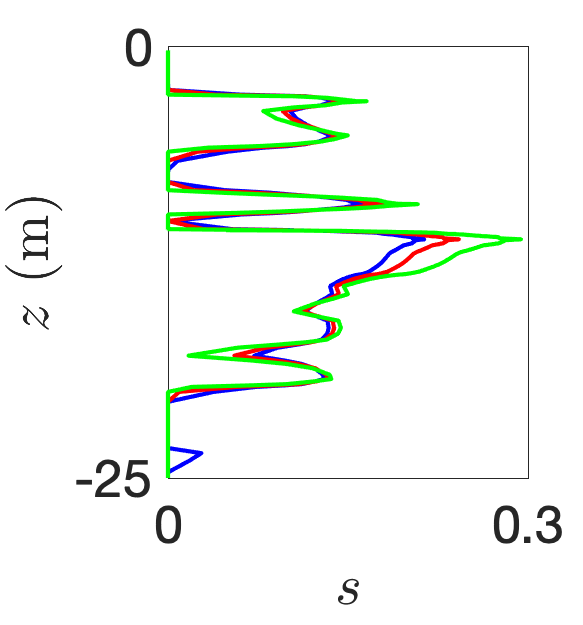}};
\node at (5,0.3) {\includegraphics[width=0.45\textwidth]{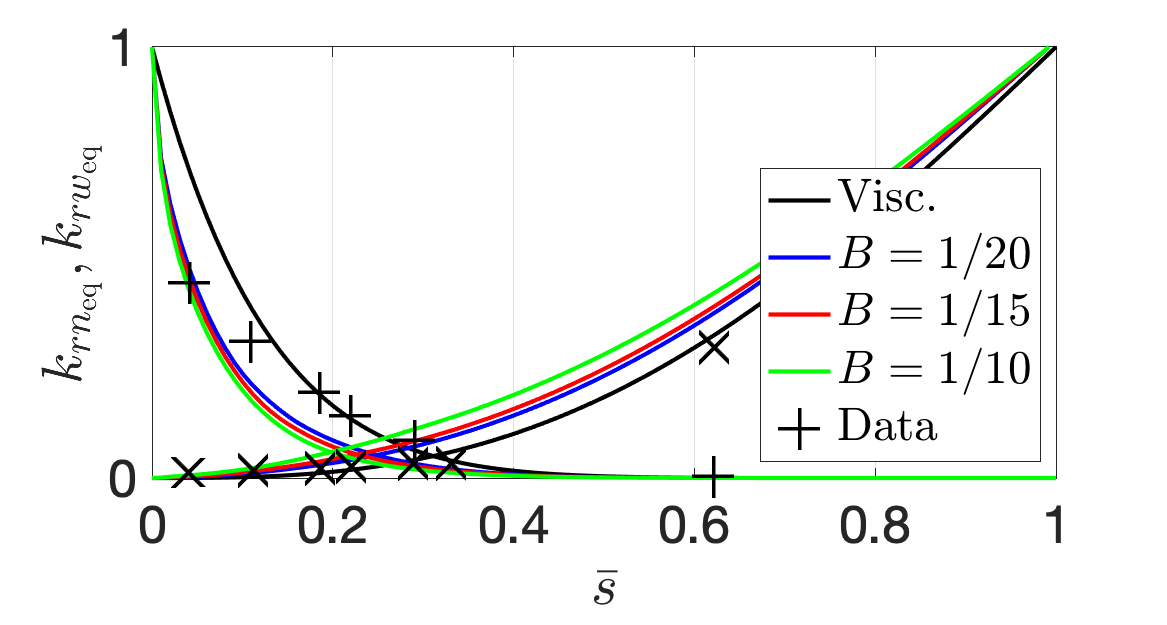}};
\node at (0.5,-2.5) {\includegraphics[width=0.6\textwidth]{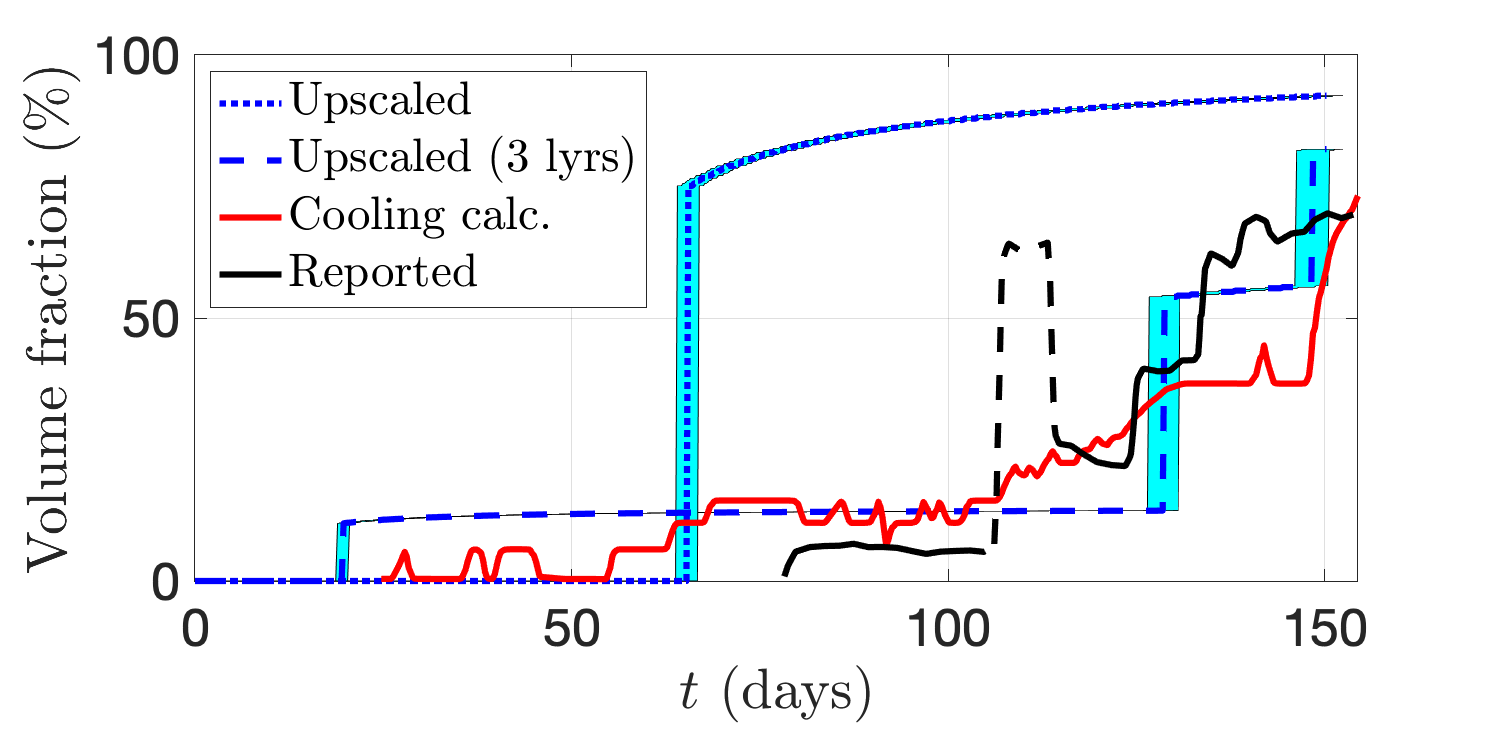}};
\node at (5.5,-2.5) {\includegraphics[width=0.35\textwidth]{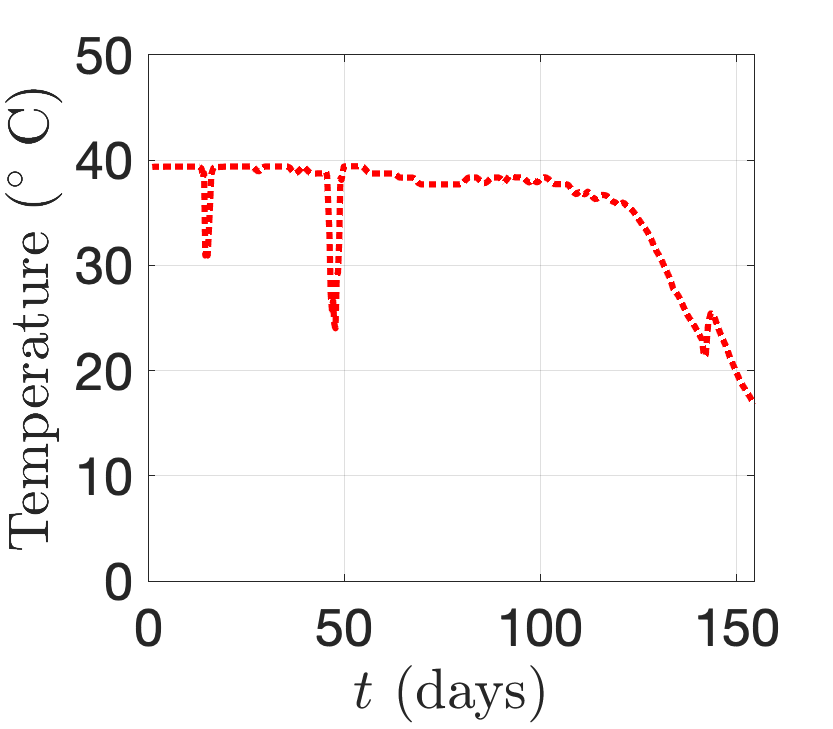}};
\node at (-2.4,1.3) {(a)};
\node at (0.,1.3) {(b)};
\node at (2.8,1.3) {(c)};
\node at (-2.4,-1.) {(d)};
\node at (3.5,-1.) {(e)};
\end{tikzpicture}
\caption{ Case study of CO$_2$ injection at Salt Creek. (a) Vertical permeability profile inferred from downhole porosity measurements \citep{bickle2017rapid}. (b) Vertical capillary limit saturation profiles for different values of the power law $B$ \eqref{powerlaw}. (c) Corresponding equivalent relative permeability curves (Experimental data taken from \citet{krevor2012relative} for Paaratte sandstone in the viscous limit). 
(d) Upscaled predictions of the volume fraction of CO$_2$ at the observation well \eqref{volumefrac}, compared with field measurements. The CO$_2$ volume fraction of the produced fluids (red solid curve) is calculated from the the temperature (e), assuming adiabatic cooling, given the variation of density and coefficient of thermal expansion of CO$_2$ with pressure and temperature from \citet{dubacq2013activity} and specific heats of CO$_2$ and water from \citet{holland2011improved}. Temperature drops at days 15, 47 - 48 and 143 - 144 are related to reductions in production rates. High reported volumes of produced CO$_2$ between days 107 - 113 do not coincide with any changes in production rate or temperature fluctuations and are disregarded.  \label{fig5}}
\end{figure}

To compare our extension to the Buckley-Leverett problem for heterogeneous media to field data, we use the Salt Creek CO$_2$ injection experiments from 2010, as detailed by \citet{bickle2017rapid}. CO$_2$ was injected into a sandstone aquifer with vertical permeability structure as shown in figure \ref{fig5}a, and aspect ratio $\delta\approx 25\un{m}/200\un{m}$. 
Injection was performed in several rows of wells, so that a two-dimensional model is probably more accurate than a radially symmetric one.
Variations in the topography are neglected. 

Relative permeability curves are not available for this sandstone, so to model this case study we use the curves of a similar sandstone called the \textit{Paaratte} formation located in SE Australia, as detailed by \citet{krevor2012relative}. We display the empirical relationships \eqref{krevor1}-\eqref{krevor2} in Appendix \ref{appA}. Likewise, pore entry pressure variation is not available, so we try using several different values of the power law $B$ \eqref{powerlaw}. We display the equivalent relative permeability curves for both the viscous limit, and the capillary limit (for several different values of $B$) in figure \ref{fig5}c. Power laws $1/20\leq B\leq 1/10$ seem to give reasonable results. Moreover, for these $B$ values, the value of the ratio between the pore entry pressure standard deviation and mean is $\sigma(p_e)/\mu(p_e)\in[0.1,0.2]$, as compared to the Bentheimer sandstone of \citet{jackson2018characterizing} which has $\sigma(p_e)/\mu(p_e)=0.16$. For such pore entry pressure distributions, we display the corresponding capillary limit saturation distributions \eqref{leads2} in figure \ref{fig5}b. For comparison with field data, we use a mid-range value of $B=1/15$.

Following our extension to the Buckley-Leverett problem (ignoring diffusion), we use \eqref{sgoveq} to describe the temporal evolution of an injection of CO$_2$, with $s_i=1$, $s_\infty=0$.
We use \eqref{transfun} for the equivalent relative permeabilities with N$_{c_t}=394$ and $\Delta=5.5$, as before.
We choose a driving flow of $V_{tot}=1.6\times 10^{-6}\un{m/s}$ which results in a pressure drop across the aquifer between $4-8\un{MPa}$, which is consistent with field measurements.
The full list of parameter values for this problem is given in Appendix \ref{appB}. 

Using all of the above information, we can compare our model predictions to field measurements. 
One useful metric for comparison is the volume fraction of CO$_2$ at the observation well, for which field data is available. The predicted volume fraction of CO$_2$ at any given saturation value and capillary number is
\beq
J(s,\mathrm{N}_c)=\frac{u_n}{u_n+u_w}=\frac{Mk_{{rn}_\mathrm{eq}}(s,\mathrm{N}_c)}{Mk_{{rn}_\mathrm{eq}}(s,\mathrm{N}_c)+k_{{rw}_\mathrm{eq}}(s,\mathrm{N}_c)},\label{volumefrac}
\eeq
which we calculate at observation well 28WC2NW05 (200 m from the injection well) and plot in figure \ref{fig5}d (dotted blue curve). 
Due to the small far-field saturation value, a shock develops, creating a sharp advection front which moves at constant velocity through the aquifer, such that arrival at the observation well manifests as a discontinuous jump in CO$_2$ volume fraction. 
The diffusion term \eqref{diffusive} which we neglected would smooth out the saturation profile near the shock in a diffusive boundary layer of growing width $\ell\propto(t/$Pe$)^{1/2}$. However, since the Peclet number for this flow is so large, this manifests in a very small error margin, as illustrated with blue shading in figure \ref{fig5}d.

In figure \ref{fig5}d we compare these predictions to field measurements of the volume fraction of CO$_2$ in the produced fluids. We consider that the volume fraction given at reservoir temperature and pressure (red solid curve), which is calculated from the temperature (figure \ref{fig5}e) of the produced fluids (assuming adiabatic cooling), is more reliable than the reported CO$_2$ production based on spot measurements (black curve).

Our modelling predicts breakthrough of CO$_2$ at volume fraction $J\approx 75\%$, after 66 days, whereas the observations suggest significant CO$_2$ ($J\approx 10-20\%$) arriving between 65 and 86 days after the start of injection. The breakthrough times for the capillary and viscous limits (which we plot in figure \ref{threelayer} in Appendix \ref{appB}) are 50 and 83 days, indicating a significant effect of heterogeneities.

It should be noted that, whilst the field measurements only detected significant CO$_2$ breakthrough after $\sim$65 days, small quantities of noble gas tracers ($^3$He \& $^{129}$Xe) added to the CO$_2$ stream at the start of injection were detected only 10 days later. 
This suggests that regions of the aquifer, such as the high permeability zone at mid-depth, may advect CO$_2$ at much greater velocity than the bulk. 
This would also explain why the field data has a much lower, more spread out volume fraction than our predicted curve. Therefore, this motivates a slightly more resolved upscaled model that breaks up the aquifer into smaller regions. We discuss this and other questions regarding the choice of length scales in the next section.

\subsection{A note on the choice of length scales}
\label{seclength}

\begin{figure}
\centering
\begin{tikzpicture}[scale=1]
\node at (0.5,-0.1) {\includegraphics[height=0.4\textwidth]{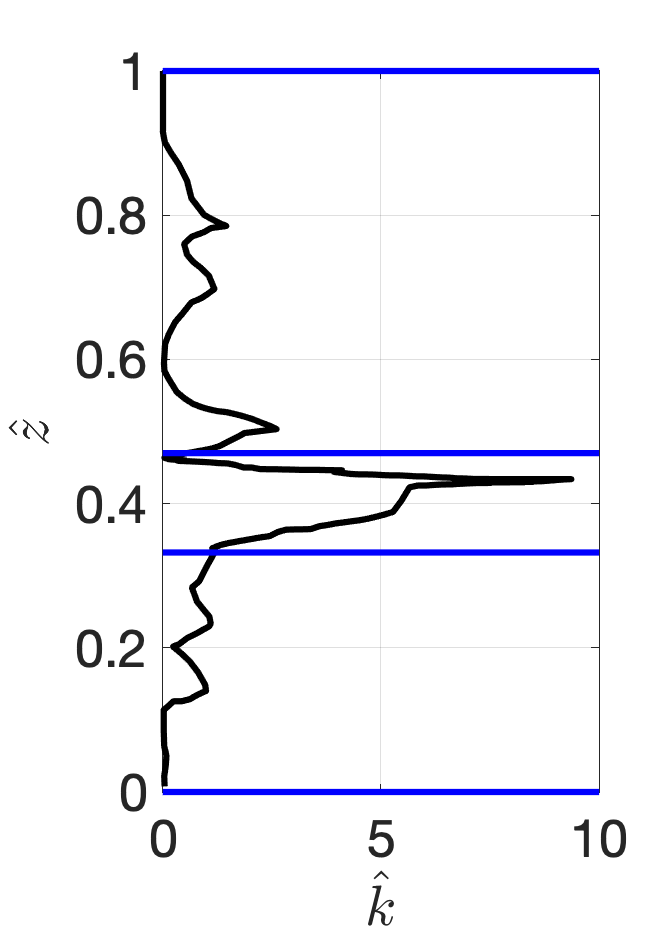}};
\node at (3.5,1.25) {\includegraphics[height=0.15\textwidth]{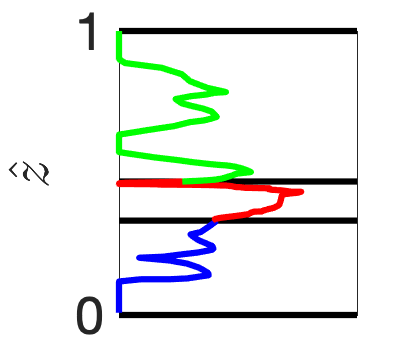}};
\node at (3.5,-1.5) {\includegraphics[height=0.175\textwidth]{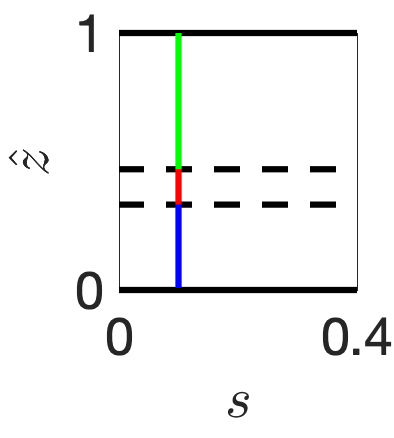}};
\node at (7.75,1.02) {\includegraphics[height=0.175\textwidth]{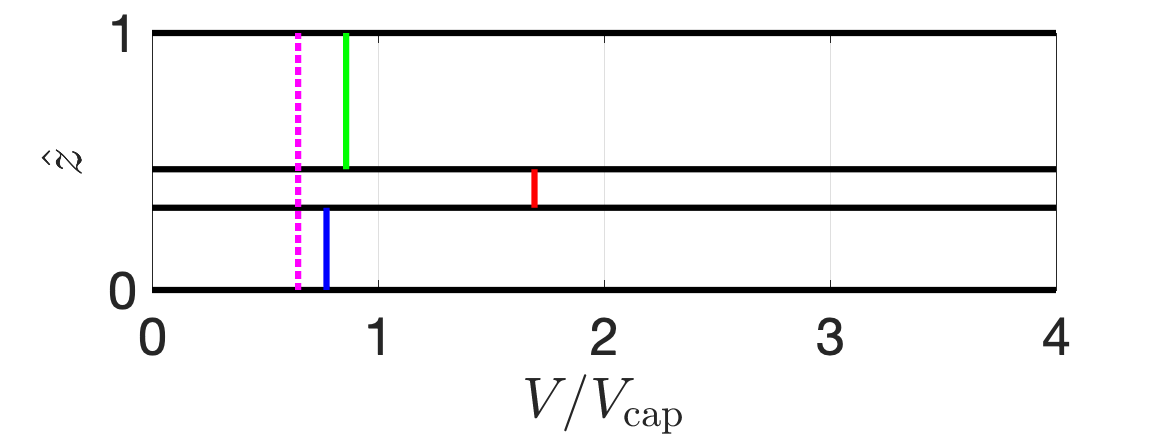}};
\node at (7.75,-1.48) {\includegraphics[height=0.175\textwidth]{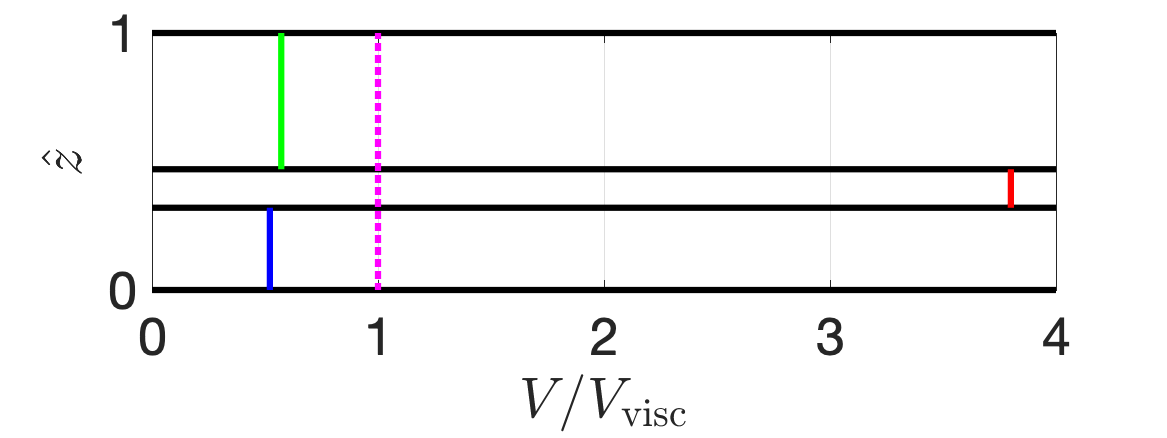}};
\draw[line width =1,->,green] (5.45,1.6) -- (6.45,1.6);
\draw[line width =1,->,red] (5.45,1.2) -- (7.45,1.2);
\draw[line width =1,->,blue] (5.45,0.85) -- (6.35,0.85);
\draw[line width =1,->,cyan,dashed] (5.45,2.2) -- (6.25,2.2);
\draw[line width =1,->,green] (5.45,-0.9) -- (6.1,-0.9);
\draw[line width =1,->,red] (5.45,-1.3) -- (10,-1.3);
\draw[line width =1,->,blue] (5.45,-1.65) -- (6.05,-1.65);
\draw[line width =1,->,cyan,dashed] (5.45,-0.3) -- (6.7,-0.3);
\node[cyan] at (7.3,2.3) {$\bar{V}_\mathrm{cap}<{V}_\mathrm{cap}$};
\node[cyan] at (7.6,-0.25) {$\bar{V}_\mathrm{visc}={V}_\mathrm{visc}$};
\node at (-1.2,2.) {(a)};
\node at (2.4,2.) {(b)};
\node at (4.8,2.) {(c)};
\node at (2.4,-0.7) {(d)};
\node at (4.8,-0.7) {(e)};
\node at (4,-0.) {\bf Viscous limit};
\node at (4,2.5) {\bf Capillary limit};
\node[green] at (1.5,1.2) {\bf (i)};
\node[red] at (1.5,-0.35) {\bf (ii)};
\node[blue] at (1.5,-1.5) {\bf (iii)};
\end{tikzpicture}
\caption{The effect of dividing the Salt Creek vertical heterogeneity into three different regions, each upscaled separately (a). With a mean saturation of  $\bar{s}=0.2$, the saturation distribution is illustrated in (b,d). After upscaling the heterogeneities within each of the three regions, the corresponding upscaled advection velocities ${V}$ in each region are illustrated in (c,e). We also illustrate the standard upscaled velocities  ${V}_\mathrm{cap}$, ${V}_\mathrm{visc}$, as well as the average velocity after upscaling the three regions independently ($\bar{V}_\mathrm{cap}$, $\bar{V}_\mathrm{visc}$).
\label{saltcreek2}}
\end{figure}
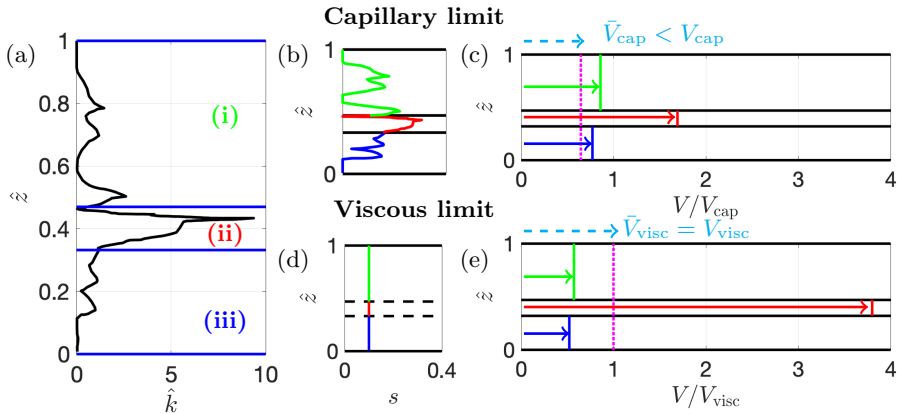

One of the key difficulties, and still an open question in the process of upscaling, is the choice of length scales. We demonstrated this earlier in figure \ref{numcol} by showing that the boundary layer thickness depends on the aspect ratio of the upscaling domain, independently of the capillary number. Therefore, the viscous-capillary transition, characterised by the parameter N$_{c_t}$ clearly depends on the domain over which upscaling is performed.
However, as illustrated with regions (i)-(iii) in figure \ref{saltcreek2}a 
for the Salt Creek permeability data, the aquifer can sometimes be naturally divided into subdomains. For the Salt Creek site, there is clearly a mid-depth region of very high permeability between regions of relatively low permeability. 
As the field data in figure \ref{fig5}d suggests, this may be responsible for a more distributed arrival of CO$_2$ at lower volume fraction than predicted by our upscaled model.
Hence, it is not obvious whether it is more accurate to think of the aquifer as a single medium or three vertically stacked media, each to be upscaled separately.

In figure \ref{saltcreek2}b,d we illustrate how the saturation of CO$_2$ would be distributed vertically in the aquifer in each of the capillary and viscous limits (for a mean value of $\bar{s}=0.2$). The viscous limit has a uniform distribution, whereas the capillary limit is given by \eqref{leads2}, leading to a focusing of CO$_2$ in the high permeability region, and mean saturation values within each of the three subdomains as $\bar{s}=0.082, 0.225$ and $0.081$. Now, if we upscale each of the three subdomains separately, we get three sets of equivalent flow properties $k_{ri_\mathrm{eq}}$, and three different advection coefficients $\hat{V}$ in the Buckley-Leverett problem. In figure \ref{saltcreek2}c,e we illustrate how each of the three individual upscaled advection speeds would vary between subdomains, compared to the original viscous and capillary limits for the whole domain. The high-permeability region has a high-speed finger of CO$_2$ which precedes the low-permeability regions on either side, which is consistent with field observations \citep{bickle2017rapid}. Indeed this CO$_2$ finger may travel at almost double the speed of the bulk in the case of the capillary limit, and at almost quadruple the speed of the bulk in the viscous limit.
In the viscous case, the mean advection speed of the three upscaled regions $\bar{V}_\mathrm{visc}$ is equal to the upscaled speed of the whole region ${V}_\mathrm{visc}$, as expected. By contrast, this is not the case for the capillary limit, with $\bar{V}_\mathrm{cap}$ being about $65\%$ of the original upscaled advection speed ${V}_\mathrm{cap}$. 

We compare this three-layered upscaling approach to the Salt Creek field data in figure \ref{fig5}d (blue dashed curve). Now instead of a single bulk arrival of CO$_2$, we see more of a staircase structure, with the mid-depth region of the aquifer delivering a CO$_2$ volume fraction of 12$\%$ at 20 days after injection, followed by the other two regions at 128 and 148 days. This gives much better comparison with the field observations, indicating that a three-layered model is more appropriate than a single-layered model
if one is interested in predicting the first arrival of CO$_2$ (e.g. the first tracers of CO$_2$ were detected at Salt Creek 10 days after injection) and the arrival distribution, but less useful if one is only interested in predicting the breakthrough of the bulk CO$_2$ quantity (65 days).
More generally, there is an interesting question about how many upscaled layers are needed to accurately capture the CO$_2$ injection in a given aquifer. By breaking the aquifer up into smaller and smaller subdomains, we can achieve better and better comparison with field data, but at some point this defeats the point of upscaling, since our original objective was to avoid resolving all the heterogeneities.

The main implications from the comparison with Salt Creek are threefold: 
Firstly, we have shown that bulk CO$_2$ breakthrough times can be reasonably well predicted by our single-layered upscaling approach, though with an over-predicted volume fraction. Secondly, we illustrated that by breaking the aquifer into three subdomains, a much better comparison with field data is achieved, including realistic predictions of CO$_2$ volume fraction at the observation well. Finally, we have shown that there is clearly significant difference between capillary and viscous limit predictions, indicating that an accurate flow description requires careful modelling of the heterogeneities. In particular, we have shown that treating the transition between viscous and capillary limits using \eqref{transfun} gives good agreement with the field data.

\section{Concluding remarks}

We have studied the effect of a vertical heterogeneity in a porous medium on the overall flow properties by way of upscaling. This is characterised by the two limiting cases of large capillary number (viscous limit), where heterogeneities play a weak role, small capillary number (capillary limit), where heterogeneities play a dominant role, and intermediate capillary number, for which a balance is sustained.
In the former limiting cases we derived analytical expressions for the upscaled equivalent relative permeabilities using asymptotic analysis. For intermediate capillary numbers we used numerical simulations to suggest a composite (heuristic) form for the equivalent relative permeabilities that remains accurate across all flow regimes. The CT scan experiments of  \citet{jackson2018characterizing} were used for comparison with some of these upscaling results.

Using an analysis that stemmed from the classic Buckley-Leverett problem \citep{buckley1942mechanism}, we applied the upscaled quantities to describe the flooding of an aquifer with heterogeneities. We illustrated how and when heterogeneities accelerate/decelerate the dynamic flow. By extending this analysis to the case of a radially symmetric injection, we illustrated how the capillary number at the flooding front changes over time. At early times, near the source, the front is in the viscous limit regime (where heterogeneities are unimportant), whereas later on, far away from the source, it is in the capillary limit regime (where heterogeneities dominate the flooding speed). The implications for CO$_2$ sequestration are that heterogeneities can alter advection of CO$_2$ by as much as 50$\%$, indicating the need for modelling such effects, as illustrated by our comparisons with field data from the injection experiments at Salt Creek, Wyoming. Finally, we illustrated how the choice of length scales for upscaling significantly affects predictions, underlining one of the key outstanding challenges in this field.

For future work,
the effects of a dynamic flow on the equivalent properties could be investigated (i.e. instead of steady-state upscaling), using some canonical time-dependent case studies.
This would be particularly useful for understanding when steady-state upscaling is an accurate approach, and when more detailed models are necessary.
In addition, this analysis could be extended to the case of a gravity current beneath an impermeable cap rock. This is  relevant to CO$_2$ storage applications, in which such trapping mechanisms are of key importance.
In particular, recent studies have shown how vertical heterogeneities can alter the flow in the case of a gravity current of miscible fluids \citep{hinton2018buoyancy}. It would be interesting to compare and contrast such results to the case of an immiscible gravity current.

Another common challenge in hydrology applications is estimating rock heterogeneities, where it is often only possible to obtain very sparse measurements. It would be interesting to use our analysis here to explore the inverse problem of estimating rock heterogeneities from a small number of data points of the equivalent properties of the flow (and mean saturation). This would be easiest in the case of small capillary number, where one could use the function $p_e(z)$ and the power law $B$ to fit the equivalent relative permeability curves to measurements. This approach is unlikely to be well-posed, since multiple types of rock heterogeneity may give the same upscaled properties, but still one could develop an ensemble of likely heterogeneity profiles as an informative tool for geoscientists.

\acknowledgements{
This research is funded in part by the GeoCquest consortium, a BHP-supported collaborative project between Cambridge, Stanford and Melbourne Universities, and by a NERC consortium grant ``Migration of CO$_2$ through North Sea Geological Carbon Storage Sites'' (grant no. NE/N016084/1).
}\\

Declaration of Interests. The authors report no conflict of interest.

\appendix
\section{Empirical relationships for the relative permeabilities\label{appA}}

Here we give the explicit relationships for the intrinsic relative permeabilities of various rock types, as discussed in the main text. In all of the following cases the Brooks-Corey relationship is used to model the capillary pressure with different values of $\lambda$, $p_{e_0}$ and $S_{wi}$.

Firstly, the model of \citet{corey1954interrelation} used by \citet{golding2011two} for Ellerslie sandstone is given by
\begin{align}
k_{rn}&=k_{rn_0}{s}^\alpha,\label{corey1}\\
k_{rw}&=(1-{s})^\beta,\label{corey2}
\end{align}
where the parameters are given by
$k_{rn_0}=0.116$,
$\alpha=2$,
$\beta=2$,
$S_{wi}=0.651$,
$\lambda=1$.
The value of $p_{e_0}$ is not given.

Secondly, the model of \citet{chierici1984novel} used by \citet{jackson2018characterizing} for Bentheimer sandstone is given by
\begin{align}
{k}_{rn}&=e^{-B\lb \frac{1-{s}}{s}\rb^M},\label{cher1}\\
{k}_{rw}&=e^{-A\lb \frac{{s}}{1-s}\rb^L},\label{cher2}
\end{align}
where the parameters are given by
$M=0.65$,
$L=0.75$,
$A=3$,
$B=5$,
$S_{wi}=0.081$,
$\lambda=2.3$,
and
$p_{e_0}=3.51\un{kPa}$.

Finally, the Brooks-Corey model \citep{dullien2012porous} used by \citet{krevor2012relative} for the Paaratte sandstone is given by
\begin{align}
k_{rn}&=k_{rn_0} s^2 (1 - (1 - s)^\alpha),\label{krevor1}\\
k_{rw}&=(1-{s})^\beta,\label{krevor2}
\end{align}
where the parameters are given by
$k_{rn_0}=0.95$,
$\alpha=2$,
$\beta=8$,
$S_{wi}=0.05$,
$\lambda=0.9$,
and
$p_{e_0}=2.1\un{kPa}$.

\section{Parameter values and extra plots for the Salt Creek case study\label{appB}}
\begin{table}
\centering
\begin{tabular}{|c|c|c|c|}
\hline
Parameter & Description & Value & Units\\
\hline
$H$ & Aquifer depth & 25 & m\\
$L$ & Aquifer length & 200 & m\\
$\mu_w$ & Viscosity of water & $6\times10^{-4}$ & Pa$\cdot$s\\
$\mu_n$ & Viscosity of CO$_2$ &$2\times10^{-5}$ & Pa$\cdot$s\\
$p_{e_0}$ & Base level pore entry pressure & $2.1\times10^{3}$ & Pa\\
$k_0$ & Mean permeability & $4.3\times10^{-14}$ & m$^2$\\
$V_{tot}$ & Total Darcy flow & $1.6\times10^{-6}$ & m/s\\
$\phi_0$ & Mean porosity  & 0.22 & $\sim$\\
$S_{wi}$ & Irreducible water saturation & 0.05 & $\sim$\\
\hline
\end{tabular}
\caption{Table of parameter values for the Salt Creek case study. \label{table1}}
\end{table}

\begin{figure}
\centering
\begin{tikzpicture}[scale=1]
\node at (0,0) {\includegraphics[width=0.45\textwidth]{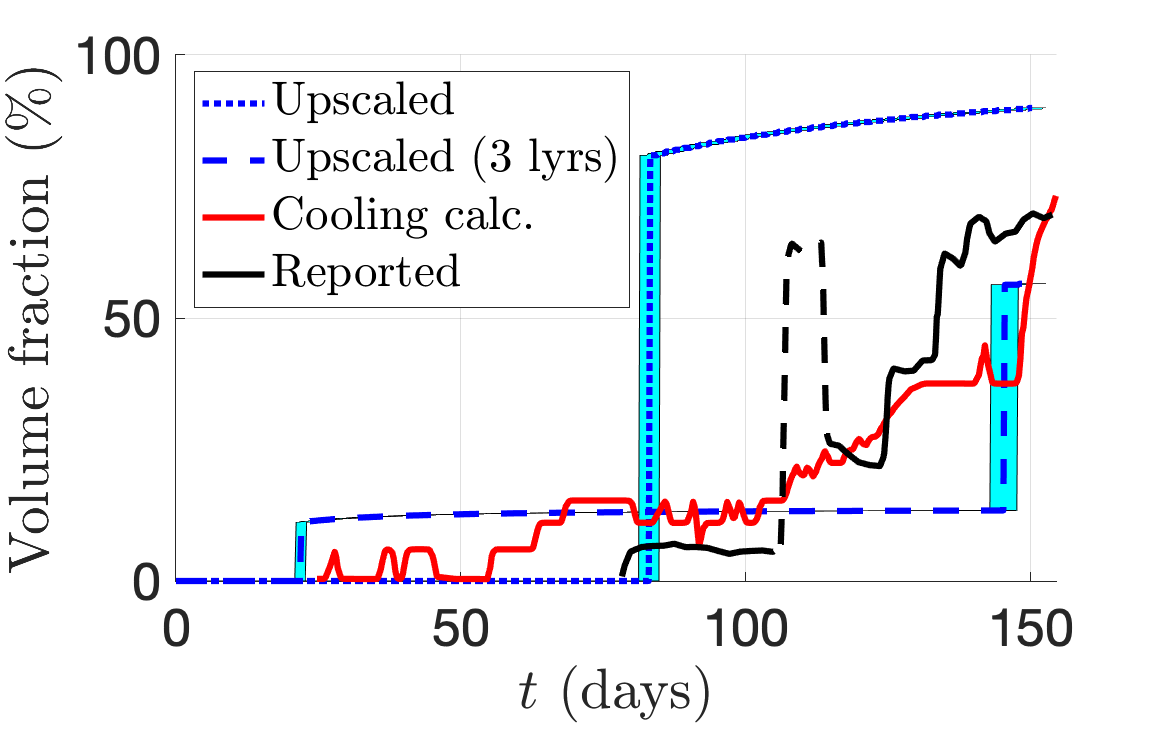}};
\node at (7,0) {\includegraphics[width=0.45\textwidth]{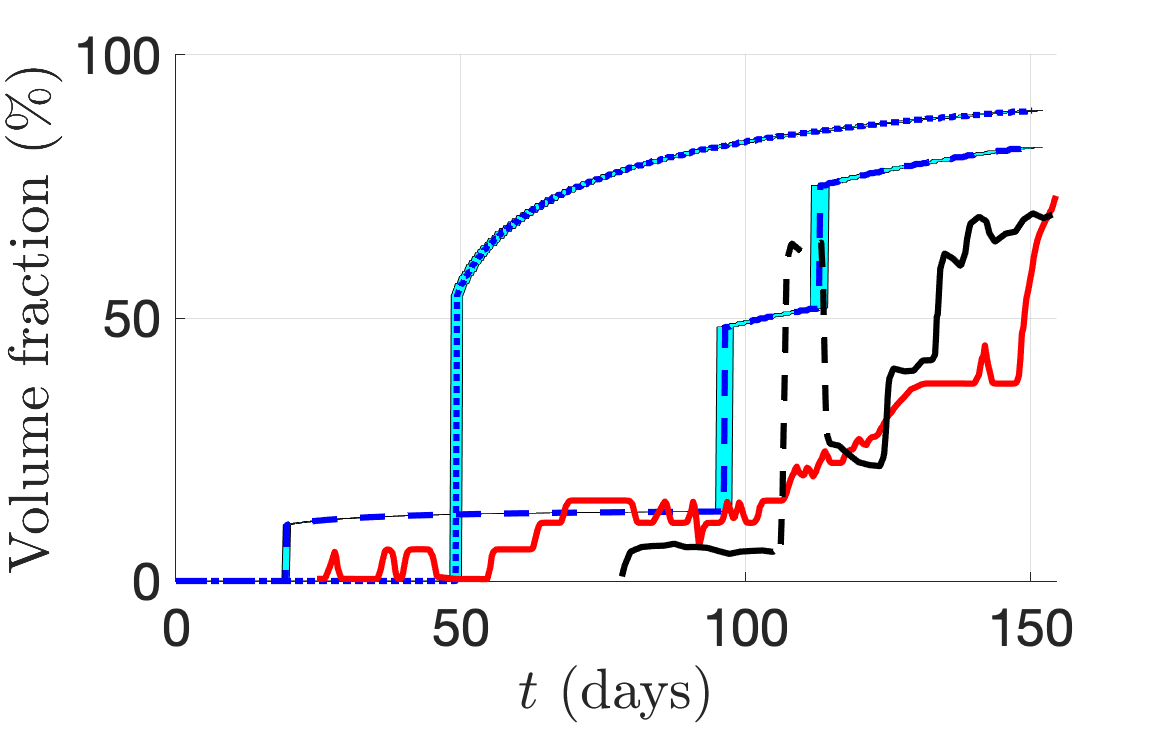}};
\node at (-3.5,1.3) {(a)};
\node at (3.5,1.3) {(b)};
\end{tikzpicture}
\caption{Upscaled (a) viscous limit and (b) capillary limit predictions for the volume fraction of CO$_2$ \eqref{volumefrac} at the observation well in Salt Creek, compared to field measurements (see figure \ref{fig5}).  \label{threelayer}}
\end{figure}

\bibliographystyle{jfm}
\bibliography{bibfile.bib}

\end{document}